\PassOptionsToPackage{table}{xcolor}
\documentclass[]{article}

\usepackage[T1]{fontenc}
\usepackage[utf8]{inputenc}
\usepackage{textcomp} 
\usepackage{lmodern}  

\usepackage{amsmath, amssymb, amsthm}
\usepackage{siunitx}
\usepackage{bbm} 

\newtheorem{Theorem}{Theorem}
\newtheorem{cor}{Corollary}

\newtheorem{definition}{Definition}
\theoremstyle{definition}
\newtheorem{exmp}{Example}[section]
\newtheorem{remark}{Remark}

\DeclareMathOperator*{\argmin}{arg\,min}
\newcommand{\ind}[1]{\mathbbm{1}_{\lrb{#1}}}
\newcommand{\lrb}[1]{\left\{#1\right\}}

\usepackage{graphicx}
\usepackage{tikz}
\usepackage{pgf}
\usepackage{float}
\usepackage{xcolor}

\usepackage{booktabs}
\usepackage{multirow}
\usepackage{collcell}

\PassOptionsToPackage{hyphens}{url}
\usepackage[unicode]{hyperref}
\hypersetup{
	citecolor  = {blue},
	colorlinks = true,
	linkcolor  = blue,
	urlcolor   = blue
}    

\usepackage[round]{natbib}

\usepackage[margin=1in]{geometry}
\usepackage{titlesec}
\titlelabel{\thetitle.\quad}
\usepackage{enumitem}
\usepackage{authblk} 

\usepackage[]{microtype}
\UseMicrotypeSet[protrusion]{basicmath}

\title{A kernel-based framework for covariate significance tests in nonparametric regression}
\author[1]{Daniel Diz-Castro\thanks{Corresponding author. E-mail address: \href{mailto:danieldiz.castro@usc.es}{danieldiz.castro@usc.es}. The authors acknowledge support from project PID2020-116587GB-I00 funded by MICIU/AEI/10.13039/501100011033.}}
\author[1]{Manuel Febrero-Bande}
\author[1]{Wenceslao Gonz\'alez-Manteiga}
\affil[1]{Department of Statistics, Mathematical Analysis and Optimization. University of Santiago de Compostela, Spain.}
\date{}

\begin{document}
	\maketitle

\begin{abstract}
	It is well known that nonparametric regression estimation and inference procedures are subject to the curse of dimensionality. Moreover, model interpretability usually decreases with the data dimension. Therefore, model-free variable selection procedures and, in particular, covariate significance tests, are invaluable tools for regression modelling as they help to remove irrelevant covariates. In this contribution, we provide a general framework, based on recent developments in the theory of kernel-based characterizations of null conditional expectations, for testing the significance of a subgroup of Hilbert space-valued covariates in a nonparametric regression model. Moreover, we propose a test designed to be robust against the curse of dimensionality and we provide some asymptotic results regarding the distribution of the test statistic under the null hypothesis of non-significant covariates as well as under fixed and local alternatives. Regarding the test calibration, we present and prove the consistency of a multiplier bootstrap scheme. An extensive simulation study is conducted to assess the finite sample performance of the test. We also apply our test in a real data scenario.
\end{abstract}

\bigskip
\section{Introduction}
\label{Sec1}

In the last decades, there has been a surge of interest in statistics with high-dimensional and functional data, driven by the growth in computational power, the ever-increasing volume of generated and stored data and the advancement of the statistical theory in infinite-dimensional spaces. For an introduction to the statistical analysis of functional data, one may refer to the monographs by \citeauthor{Ramsay2002} (\citeyear{Ramsay2002}, \citeyear{Ramsay2005}), \cite{Ferraty2006} and \cite{Horvath2012}, among others. Simultaneously, variable selection mechanisms have received considerable attention, particularly in the regression context, to reduce the complexity of the proposed models. On the one hand, the removal of redundant covariates allows for more parsimonious models and thus easier to interpret. On the other hand, it is well known that nonparametric estimation and inference procedures tend to suffer as the dimensionality of the data increases (see e.g. \citeauthor{Stone1980}, \citeyear{Stone1980}), particularly under codependence between covariates. There are various approaches to reducing the dimension of covariates in regression: regularization techniques for overparametrized models (see \citeauthor{Freijeiro‐Gonzalez2021}, \citeyear{Freijeiro‐Gonzalez2021} for a comprehensive review on LASSO regularization), screening methods (see e.g. \citeauthor{Zhu2011}, \citeyear{Zhu2011} and \citeauthor{Li2012}, \citeyear{Li2012}) and covariate significance tests. The latter are particularly useful as they provide, by construction, hierarchical variable selection mechanisms with a controlled type I error. These types of tests are also known as conditional mean independence tests or omitted variable tests.

In this paper, we are concerned with testing the significance of a subset of covariates in nonparametric regression models that may involve some functional variables. In particular, given a response $Y\in \mathcal{Y}$ and a pair of covariates $Z=(X,W)\in \mathcal{X}\times \mathcal{W} = \mathcal{Z}$, where $\mathcal{Y}, \mathcal{X}$ and $\mathcal{W}$ are separable Hilbert spaces, we are 
interested in the hypothesis test
\begin{equation}\label{1.1}\tag{1.1}
	\mathcal{H}_0: \ \ g(Z) = m(X) \ \text{a.s.} \quad \text{vs.} \quad \mathcal{H}_1: \ \ \mathbb{P}(g(Z) \neq m(X))>0, 
\end{equation}
where a.s. means almost surely, $m(X)=\mathbb{E}[Y|X]$ and $g(Z)=\mathbb{E}[Y|Z]$ are both unspecified and $W$ cannot be obtained as a measurable function of $X$; that is, in terms of $\sigma$-algebras, $\sigma(W)\nsubseteq\sigma(X)$.

The statistical literature on significance tests is extensive for regression models with finite-dimensional (and, particularly, low-dimensional) covariates and responses. Many proposals, such as the ones in \cite{Fan1996}, \cite{Lavergne2000}, \cite{Lavergne2015} and \cite{Zhu2018} are smoothing-based tests that require a kernel estimator of the regression function under the null hypothesis, $\hat{m}(X)$, and another kernel estimator of some weighted second order moment of either $\mathbb{E}\left[Y-\hat{m}(X)|Z\right]$ or other, similar quantities. There are, however, a few tests following completely different approaches. \cite{Racine1997} developped a test based on the nonparametric estimation of the derivatives of $\mathbb{E}[Y|Z]$ with respect to the components of a vector-valued $Z$, whereas the tests in \cite{Chen1999} and \cite{Delgado2001} rely on the empirical process ideas of \cite{Stute1997} associated with the integrated regression function. Moreover, in recent years, new contributions related to the field of machine learning have arised, dealing with the problem of covariate significance testing for big data. See for example \cite{Williamson2022} and \cite{Dai2024}, both based on the comparison of nonparametric estimators under the null and alternative hypothesis with an embedded sample splitting procedure, which can heavily impact the power performance of the tests. In any case, the aforementioned proposals are restricted to the case of scalar response and vector covariates.

The theory behind significance tests becomes more challenging when any of the variables involved in a regression model is infinite-dimensional. This is probably the reason behind the little literature available on this specific subject. Some examples of no-effect tests of covariates in nonparametric functional regression can be found in \cite{Delsol2012}, \cite{Patilea2016}, \cite{Patilea2020} \cite{Lee2020} and \cite{Lai2021}. However, in the context of testing the conditional mean independence of a response on a covariate, one of them being functional, given another covariate, we are only aware of the proposals in \cite{Maistre2020}, although the asymptotic theory for the completely nonparametric case is not developed, and in \cite{Delsol2011}, which formally requires to know the distribution function of $Z$. Consequently, this area of research remains relatively unexplored. 

The main contribution of this paper is a new significance test based on recent developments in the theory of kernel-based characterizations of null conditional expectations. This theory is adapted, in Section \ref{Sec2}, to the context of the hypothesis test in (\ref{1.1}) and leads to a procedure which, according to the results in Section \ref{Sec3}, is robust against the curse of dimensionality and it is able to handle functional data, thus helping to fill one of the many gaps still existing in the functional data literature. In \cite{Escanciano2024} it was established that quadratic functionals of the empirical process-based test statistics used in \cite{Chen1999} and \cite{Delgado2001} can be regarded as a particular class of kernel-based test statistics. Therefore, it is not surprising that our proposal preserves the good theoretical properties of the empirical-process approach, namely the ability to detect local alternative hypothesis converging to the null at the rate of $n^{-1/2}$, independently of the presence of functional variables in the regression model, at the cost of a non-pivotal limiting distribution of the test statistic under $\mathcal{H}_0$. The latter issue motivates the consistent resampling method proposed in Section \ref{Sec4}, which can be implemented in practice as a fast bootstrap scheme, enhancing the computational efficiency of our test.

The theoretical results in this document focus on the case of an absolutely continuous, vector valued $X\in \mathbb{R}^d$ for some $d\in \mathbb{N}$. The extension to models involving discrete variables as well as the posibility to include a Hilbert space-valued $X$ is discussed in Section \ref{Sec5}. Practical issues and implementation suggestions are addressed in Section \ref{Sec6} whereas Sections \ref{Sec7} and \ref{Sec8} contain a simulation study and an application to real data, respectively. Finally, conclusions are presented in Section \ref{Sec9}. The proofs of theoretical results can be found in the supplementary material.

\section{Theoretical background}
\label{Sec2}

The null hypothesis in (\ref{1.1}) holds if and only if $\mathbb{E}[Y-m(X)|Z] = 0$ a.s. Therefore, testing the significance of $W$ in the regression model of $Y$ on $Z$ is equivalent to testing the conditional mean independence of $Y-m(X)$ on $Z$. In this section we introduce the fundamental ideas behind kernel-based conditional mean independence tests and we adapt them to account for the unknown, unspecified $m$ issue. These ideas are closely related to the characterization of the hypothesis $\mathbb{E}[Y-m(X)|Z] = 0$ a.s., used in empirical process based tests, given by
$$ \mathbb{E}[Y-m(X)|Z] = 0 \ \text{a.s.}  \iff \mathbb{E}[l(Z,\theta)(Y-m(X))] = 0 \ \text{for all}\ \theta \in \Theta,$$
where $l(z,\theta)$ is an adequate complex or real-valued function for all $(z,\theta)\in \mathbb{Z}\times \Theta$ and $\Theta$ is some set. For example, if $\mathcal{Y} = \mathbb{R}$ and $\mathcal{Z}=\mathbb{R}^{d_Z}$ for some $d_Z>d$, it is possible to choose $\Theta = \mathbb{R}^{d_Z}$ and either $l(z,\theta) = \exp(i\langle z, \theta\rangle_{\mathbb{R}^{d_{\scalebox{0.5}{$\scriptstyle Z$}}}})$ (\citeauthor{Bierens1982}, \citeyear{Bierens1982}) or $l(z,\theta) = \ind{z\le \theta}$ (\citeauthor{Stute1997}, \citeyear{Stute1997}). Moreover, a Gaussian process motivation for kernel-based tests involving euclidean data can be found in \cite{Escanciano2024}.

\subsection{The kernel approach to assess conditional mean independence}
\label{Sec2.1}

Let $c:\mathcal{Z}\times \mathcal{Z}\rightarrow \mathbb{R}$ be a continuous, positive definite kernel and denote by $\mathcal{C}$ 
the reproducing kernel Hilbert space (RKHS) generated by $c$; that is, $\mathcal{C}$ is the completion of the linear span of $\{c(z,\cdot): z\in \mathcal{Z}\}$, where the following reproducing property holds: $\langle c(z,\cdot), \zeta\rangle_{\mathcal{C}} = \zeta(z)$ for all $\zeta\in \mathcal{C}$. Note that, if $\mathcal{Z}$ is separable and $c$ is continuous, $\mathcal{C}$ is also separable (see e.g. Theorem 7 in \citeauthor{Hein2004}, \citeyear{Hein2004}).
Denote by $\mathcal{HS}= \mathcal{C} \otimes \mathcal{Y}$ the (also separable) tensor product Hilbert space, defined as the completion of the set of all finite linear combinations of tensor products of the form $\zeta\otimes y:\mathcal{Y} \rightarrow \mathcal{C}$, where $\zeta\in \mathcal{C}$ and $y\in \mathcal{Y}$, such that $\zeta\otimes y(y') = \zeta\langle y,y' \rangle_{\mathcal{Y}}$, endowed with the following inner product structure: $\langle \zeta\otimes y, \zeta'\otimes y'\rangle_{\mathcal{HS}} = \langle \zeta\otimes \zeta'\rangle_{\mathcal{C}}\langle y\otimes y'\rangle_{\mathcal{Y}}$. It is easy to verify that
$$\mathbb{E}\left[c(Z,\cdot)\otimes (Y-m(X))\right] = \mathbb{E}\left[c(Z,\cdot)\otimes \mathbb{E}[Y-m(X)|Z]\right],$$
assuming such expected values are well defined. Therefore, $\mathbb{E}\left[c(Z,\cdot)\otimes (Y-m(X))\right] = 0$ under $\mathcal{H}_0$. Equivalently,
$$\mathcal{H}_0\implies  0 = \left|\left| \mathbb{E}\left[c(Z,\cdot)\otimes (Y-m(X))\right]\right|\right|_{\mathcal{HS}}^2 = \mathbb{E}\left[c(Z,Z')\langle Y-m(X), Y'-m(X')\rangle_{\mathcal{Y}}\right],$$
given two i.i.d (independent and identically distributed) copies $(Z,Y)$, $(Z',Y')$. The reciprocal also holds under some conditions on the kernel $c$. Considering $c(z,z') = ||z-z'||_{\mathcal{Z}}$ leads to the \emph{Martingale Difference Divergence} (MDD) introduced in \cite{Shao2014} and later extended to the context of functional data in \cite{Lee2020}. Furthermore, given the results in \cite{Lai2021}, it suffices to choose any bounded characteristic kernel $c$ to completely characterize the null hypothesis.

\begin{definition}
	Let $\mathcal{S}$ be a separable Hilbert space and let $\mathcal{M}$ denote the set of all finite signed measures $\mu$ defined on the measurable space $(\mathcal{S},B_{\mathcal{S}})$, where $B_{\mathcal{S}}$ denotes the Borel $\sigma$-algebra on $\mathcal{S}$. In addition, let $\mathcal{M}_1$ denote the set of probability measures on $(\mathcal{S},B_{\mathcal{S}})$. We say that a positive definite kernel $c_{\mathcal{S}}:\mathcal{S}\times \mathcal{S}\rightarrow \mathbb{R}$, with associated RKHS $\mathcal{C}_{\mathcal{S}}$, is characteristic if the kernel mean embedding $\Pi:\mathcal{M}\rightarrow \mathcal{C}_{\mathcal{S}}$ defined as 
	$$ \Pi(\mu) = \int c_{\mathcal{S}}(s,\cdot) d\mu(s),$$
	for all $\mu\in \mathcal{M}$, is injective in $\mathcal{M}_1$.
\end{definition}

There are alternative characterizations of the characteristic property for positive definite kernels (see e.g. \citeauthor{Sriperumbudur2010}, \citeyear{Sriperumbudur2010} and \citeauthor{Ziegel2024}, \citeyear{Ziegel2024}). Moreover, non-characteristic kernels may be unable to detect some alternative hypothesis but, at the same time, can be the optimal choice to test against a particular family of alternatives.

We define the \emph{Kernel-based Conditional Mean Dependence} measure (KCMD) as follows.
 
\newpage
\begin{definition}\label{Def2}
	Let $\mathcal{S}_1$ and $\mathcal{S}_2$ be any separable Hilbert spaces and let $c_{\mathcal{S}_2}:\mathcal{S}_2\times \mathcal{S}_2\rightarrow \mathbb{R}$ be a continuous, positive definite kernel. Given two i.i.d. copies of a pair of random elements $(S_1,S_2),(S'_1,S'_2)\in \mathcal{S}_1\times \mathcal{S}_2$, we define the KCMD measure of conditional mean independence of $S_1$ on $S_2$, based on kernel $c_{\mathcal{S}_2}$, as 
	$$\text{KCMD}(S_1|S_2;c_{\mathcal{S}_2}) = \mathbb{E}\left[c_{\mathcal{S}_2}(S_2,S'_2)\langle S_1, S'_1\rangle_{\mathcal{S}_1} \right]\ge 0.$$
\end{definition}

For notational simplicity, we will omit the reference to the kernel $c_{\mathcal{S}_2}$ in Definition \ref{Def2} and simply denote $\text{KCMD}(S_1|S_2;c_{\mathcal{S}_2})$ by $\text{KCMD}(S_1|S_2)$. Then, by previous arguments, $\text{KCMD}(Y-m(X)|Z)$ characterizes the null hypothesis in (\ref{1.1}) for every bounded characteristic kernel $c$ and, thus, it is reasonable to suggest rejecting $\mathcal{H}_0$ whenever an empirical estimator of $\text{KCMD}(Y-m(X)|Z)$ surpasses a suitable threshold.

\subsection{Plug-in KCMD estimator}

To use the $\text{KCMD}$ to test $\mathcal{H}_0$ in practice, since we are assuming that $m$ is unknown, we need an estimator $\hat{m}$ which, in order to avoid imposing a particular structure on the regression model of $Y$ on $X$, must be nonparametric. In the following we assume that $\mathcal{X} = \mathbb{R}^d$ for some $d\in \mathbb{N}$; that is, $X$ is vector-valued. Moreover, we also assume that $X$ is absolutely continuous with density function $f$. The case of discrete and mixed vector-valued $X$ is discussed in Section \ref{Sec5.1}. The general case of a Hilbert space-valued $X$ is more complicated, although an elegant alternative procedure is proposed in Section \ref{Sec5.2} under the assumption of independence between $X$ and $W$.

To simplify both the exposition and the derivation of theoretical results, we consider the Nadaraya-Watson estimator of $m$ (\citeauthor{Nadaraya1964}, \citeyear{Nadaraya1964}), although the methodology presented in this section can be adapted to fit other types of nonparametric regression estimators. Given an i.i.d. sample $\{(Z_i,Y_i)\}_{i=1}^n$  and a product smoothing kernel $K$ such that $K(x) = \prod_{j=1}^{d}K^u(x^j)$ for all $x = (x^1,\ldots, x^d)\in \mathbb{R}^d$, where $K^u$ stands for some symmetric univariate kernel, the Nadaraya-Watson estimator corresponds to the formula
$$\hat{m}_h(x) = \sum\limits_{k=1}^n\frac{K_h(x-X_k)}{n\hat{f}_h(x)}Y_k,$$
where $\hat{f}_h(x) = \sum_{k=1}^nK_h(x-X_k)/n$ is the Parzen-Rosenblatt estimator of $f$ at $x$ (\citeauthor{Rosenblatt1956}, \citeyear{Rosenblatt1956}), and $K_h$ is the scaled kernel defined as $K_h(x) = K(x/h)/h^d$ given the bandwith parameter $h>0$. Intuition leads to plug $\{Y_i-\hat{m}_h(X_i)\}_{i=1}^n$ in the empirical estimator of $\text{KCMD}(Y-m(X)|Z)$. However, to avoid the issue of random denominators, we rely on the equivalent formulation of the test in (\ref{1.1}), already used in many previous proposals, given by
\begin{equation}\label{2.1}\tag{2.1}
	\mathcal{H}_0 : \ \ \mathbb{E}[f(X)(Y -m(X))|Z] = 0  \ \text{a.s.} \ \ \ \text{vs.} \ \ \ \mathcal{H}_1: \ \ \mathbb{P}(\mathbb{E}[f(X)(Y -m(X))|Z] \neq 0) >0,
\end{equation}
and the corresponding modification of the KCMD measure: $\text{KCMD}(f(X)(Y-m(X))|Z)$. Then, the empirical plug-in estimator of $\text{KCMD}(f(X)(Y-m(X))|Z)$ is defined as
$$V_n = \frac{1}{n^2}\sum\limits_{i,j}c(Z_i,Z_j)\langle \hat{f}_h(X_i)(Y_i-\hat{m}_h(X_i)), \hat{f}_h(X_j)(Y_j-\hat{m}_h(X_j))\rangle_\mathcal{Y}.$$
Let the bandwidth parameter-dependent (and, thus, sample size-dependent) function $\psi_n:(\mathcal{Z},\mathcal{Y})^2 \rightarrow \mathcal{HS}$ be defined as
$$ \psi_n((z,y),(z',y')) = \frac{1}{2}(c(z,\cdot)-c(z',\cdot))\otimes K_h(x-x')(y-y'),$$
for all $(z,y),(z',y')$ in the support of $(Z,Y)$. Expanding the expression of $V_n$ in terms of the sums implicit in the definition of the Parzen-Rosenblatt and Nadaraya-Watson estimators and taking into account the inner product stucture of the Hilbert space $\mathcal{HS}$ defined in Section \ref{Sec2.1}, as well as the symmetry of the product kernel $K$ around $0$, it is trivial to show that
\begin{align*}
	V_n &= \frac{1}{n^4}\sum\limits_{i,j,k,l}c(Z_i,Z_j)\langle K_h(X_i-X_k)(Y_i-Y_k), K_h(X_j-X_l)(Y_k-Y_l)\rangle_{\mathcal{Y}}\\
	&= \left|\left| \frac{1}{n^2}\sum\limits_{i\neq k} c(Z_i,\cdot)\otimes K_h(X_i-X_k)(Y_i-Y_k)  \right|\right|^2_{\mathcal{HS}}\\
	&= \left|\left| \frac{n-1}{n}\frac{1}{{n\choose 2}}\sum\limits_{i<k} \psi_n((Z_i,Y_i),(Z_i,Y_k))  \right|\right|^2_{\mathcal{HS}}. 
\end{align*}
Therefore, $V_n$ can be interpreted as the squared norm of a scaled $\mathcal{HS}$-valued U-statistic with kernel $\psi_n$. Despite this convenient interpretation, we suggest an asymptotically unbiased modification of $V_n$, defined as
\begin{align*}\label{2.2}\tag{2.2}
	 U_n &= \frac{1}{n(n-1)(n-2)(n-3)}\sum\limits_{i\neq j\neq k\neq l}c(Z_i,Z_j)\langle K_h(X_i-X_k)(Y_i-Y_k), K_h(X_j-X_l)(Y_k-Y_l)\rangle_{\mathcal{Y}}\\
	 &= \frac{1}{n(n-1)(n-2)(n-3)}\sum\limits_{i\neq j\neq k\neq l}\langle\psi_n((Z_i,Y_i),(Z_k,Y_k)),\psi_n((Z_j,Y_j),(Z_l,Y_l)) \rangle_{\mathcal{HS}}.
\end{align*}	
Theoretical results for $U_n$ (and also for $V_n$) can be derived by adapting the theory of Hilbert space-valued U-statistics to account for the bandwidth parameter $h$ which, as we discuss in the following sections, has a considerable impact on the asymptotic behavior of the statistic.

\begin{remark}
	If, instead of $c$, we were to use another smoothing kernel depending on a bandwidth parameter which tends to zero as the sample size increases, we would have the same type of test statistic used in \cite{Fan1996}, \cite{Lavergne2000}, \cite{Lavergne2015} and \cite{Maistre2020}. However, it is precisely the usage of characteristic kernels what, as we show in the following sections, allows our test to be omnibus (able to detect any type of fixed alternatives) and to obtain better asymptotic power properties than the aforementioned authors proposals, at the cost of a non-pivotal limiting distribution under $\mathcal{H}_0$.
\end{remark}

\renewcommand{\labelenumi}{(A\arabic{enumi})}

\section{Asymptotic behavior of the KCMD estimator}
\label{Sec3}

We provide some results in Section \ref{Sec3.2} concerning the asymptotic behavior of $U_n$ under the null hypothesis $\mathcal{H}_0$ and the fixed alternative $\mathcal{H}_1$, both defined in (\ref{2.1}). In particular, we assume that $Y$ and $Z$ verify the regression model
\begin{equation}\tag{3.1}\label{3.1}
	Y=m(X) + \tilde{g}(Z) + \mathcal{E},
\end{equation}	
where $\mathbb{E}[\mathcal{E}|Z]=0$ a.s. and $\mathbb{E}[\tilde{g}(Z)|X]= 0$ a.s. so that $g(Z) = m(X) + \tilde{g}(Z)$ a.s. It is clear that $\mathcal{H}_0$ holds if $\tilde{g}(Z)=0$ a.s. whereas $\mathcal{H}_1$ holds if $\mathbb{P}(\tilde{g}(Z)\neq 0)>0$. 

We also explore in Section \ref{Sec3.3} the properties of $U_n$ under local alternatives in Pitman's sense by considering a sequence of alternative hypotheses $\{\mathcal{H}_{1n}\}_{n\in \mathbb{N}}$ defined as
\begin{equation}\tag{3.2}\label{3.2}
	\mathcal{H}_{1n}:\quad Y = m(X) + \frac{\tilde{g}(Z)}{n^\gamma} + \mathcal{E},
\end{equation}	
where $\mathbb{E}[\tilde{g}(Z)|X] = 0$ a.s., $\mathbb{P}(\tilde{g}(Z)\neq 0)>0$ and $\gamma>0$, which converge to $\mathcal{H}_0$ as the sample size $n$ increases.

The assumptions and technical conditions required to derive the theoretical results in Sections \ref{Sec3.2} and \ref{Sec3.3} are presented and briefly discussed in Section \ref{Sec3.1}.

\subsection{Assumptions}
\label{Sec3.1}

We make the following
assumptions about the user-controlled functions and parameters involved in the definition of $U_n$:

\begin{enumerate}
\item Let $K^u$ be a bounded function with support in $[-1,1]$ and symmetric with respect to $0$ such that
$$\int K^u(t)t^idt = \ind{i=0}, \quad 0\le i \le \nu-1,$$
for some $\nu\in \mathbb{N}$ and $\nu \ge 2$. $K$ is a product kernel based on $K^u$; that is, $K(x) = \prod_{j=1}^{d}K^u(x^j)$ for all $x=(x^1,\ldots, x^d)$.
\item $\{h(n)\}_{n=1}^\infty$, where $h(n)$ is simply denoted as $h$, is a deterministic sequence of positive bandwidths, such that $h\rightarrow0$ and $nh^d\rightarrow \infty$ as $n\rightarrow \infty$. 
\item $c$ is a continuous, symmetric and bounded characteristic kernel. Without loss of generality, we assume $||c||_\infty \le 1$.
\end{enumerate}

It is not uncommon for assumptions such as (A1) and (A2) to be made in the context of specification tests based on kernel smoothing. The bounded support assumption for the univariate kernel $K^u$ is relatively unrestrictive in practice since it is verified by many of the most used kernels. Furthermore, it is well known that the choice of kernel has minimal impact on the behavior of smoothers and, ultimately, on the performance of kernel-based tests. It should be noted that $||K||_\infty$ is finite (which follows from the compact support and the boundedness of $K^u$). Assumption (A3) ensures that $c$ is a  kernel that characterizes the null hypothesis through the KCMD. A notoriuous example is the Gaussian kernel, defined as $c(z,z') = \exp\left(-||z-z'||_\mathcal{Z}^2/\sigma_c^2\right)$ for some $\sigma_c^2>0$. 
	
Let $\text{supp}(\mathbb{P}_X)$ and $\partial \text{supp}(\mathbb{P}_X)$ denote the support of $X$ and its boundary, respectively. Moreover, denote by $d(x,\partial \text{supp}(\mathbb{P}_X))$ the distance from $x\in \text{supp}(\mathbb{P}_X)$ to $\partial \text{supp}(\mathbb{P}_X)$.
 With this notation, we also assume that the following technical conditions are verified for the regression model in (\ref{3.1}) and also for the sequence of local alternatives $\{\mathcal{H}_{1n}\}_{n\in \mathbb{N}}$ defined as in (\ref{3.2}):
 
\renewcommand{\labelenumi}{(T\arabic{enumi})}
\begin{enumerate}
\item $\mathbb{E}\left[||m(X)||_\mathcal{Y}^4 + ||\tilde{g}(Z)||_\mathcal{Y}^2 + ||\mathcal{E}||_\mathcal{Y}^4 \right]<\infty$.
\item Both the density $f$ and the regression function $m$ are $\nu+1$ times continuously Fréchet differentiable in the interior of $\text{supp}(\mathbb{P}_X)$. Furthermore, there will be some  $h_1>0$ such that, for $a\in \{m,f\}$,
\begin{itemize}
	\item $\sum\limits_{k_1,\ldots, k_j = 1}^d \mathbb{E}\left[\left|\left|\dfrac{\partial^j a(X)}{\partial x^{k_1}\cdots\partial x^{k_j}}\right|\right|^4\right]<\infty$ for $j=1,\ldots,\nu$.
	\item $\sum\limits_{k_1,\ldots, k_{\nu + 1} = 1}^d \mathbb{E}\left[\sup\limits_{||x_0||\le h_1}\left|\left|\dfrac{\partial^{\nu + 1} a(X+x_0)}{\partial x^{k_1}\cdots\partial x^{k_{\nu + 1}}}\right|\right|^4\ind{X + x_0 \in \text{supp}(\mathbb{P}_X)}\right]<\infty$.
\end{itemize}
\item For every  $z'\in \mathcal{Z}$, $b(x,z') = \mathbb{E}[c(Z,z')|X=x]$ is continuously differentiable with respect to $x$ in the interior of $\text{supp}(\mathbb{P}_X)$. Moreover, there will be some $h_2>0$ such that
\begin{itemize}
		\item $\sum\limits_{k=1}^d \mathbb{E}\left[\sup\limits_{||x_0||\le h_2}\left|\dfrac{\partial b(X+x_0,Z)}{\partial x^{k}}\right|^2\ind{X + x_0 \in \text{supp}(\mathbb{P}_X)}\right]<\infty$.
\end{itemize} 
\item If $\partial \text{supp}(\mathbb{P}_X)$ is not empty, we have, as $\epsilon>0$ goes to zero, 
\begin{itemize}
	\item $\sup\limits_{d(x,\partial \text{supp}(\mathbb{P}_X))\le \epsilon} f(x) = O(\epsilon^{\nu})$.
\end{itemize} 
Moreover, there is some $h_3>0$ such that 
\begin{itemize}
	\item $\mathbb{E}\left[\sup\limits_{||x_0||\le h_3} ||m(X + x_0)||_\mathcal{Y}^2\ind{d(X,\partial \text{supp}(\mathbb{P}_X))\le h_3, \ X + x_0 \in \text{supp}(\mathbb{P}_X)}\right] = O(1)$.
\end{itemize} 
\item The density function $f$ is bounded: $||f||_{\infty} <\infty$.
\end{enumerate}

Condition (T1) imposes some finite moments restrictions on the regression errors and the conditional expectations of $Y$. If (T1) holds, $\mathbb{E}\left[||Y||_\mathcal{Y}^2\right]<\infty$ under $\mathcal{H}_1$, $\mathbb{E}\left[||Y||_\mathcal{Y}^4\right]<\infty$ under $\mathcal{H}_0$ and $\sup_{n}\mathbb{E}\left[||Y||_\mathcal{Y}^2\right]<\infty$ under $\{\mathcal{H}_{1n}\}_{n\in \mathbb{N}}$.
 Conditions (T2)-(T3) allow us to bound the moments of Taylor expansions of $m$, $f$ and $b$. In particular, in condition (T2) we require the $\mathcal{Y}$-valued function $m$ to be Fréchet differentiable (see e.g. \citeauthor{Cartan1971}, \citeyear{Cartan1971}) and the partial Gateaux derivatives of $m$ to have finite fourth order moments. In addition, by the definition of conditional expectations for Hilbert space-valued random variables, we can identify $\mathbb{E}[c(Z,\cdot)|X=x]\in \mathcal{C}$ a.s. with $b(x,\cdot)$ since, by the reproducing property,
$$ \mathbb{E}[c(Z,\cdot)|X=x](z') = \langle \mathbb{E}[c(Z,\cdot)|X=x], c(z',\cdot)\rangle = \mathbb{E}[\langle c(Z,\cdot), c(z',\cdot)\rangle|X=x] = \mathbb{E}[c(Z,z')|X=x] = b(x,z').$$
Condition (T4) controls the boundary effect on the nonparametric kernel smoothers. It holds, for instance, if $f$ and its partial derivatives up to order $\nu$ vanish on $\partial \text{supp}(\mathbb{P}_X)$ and $m$ is bounded next to the boundary of the support of $X$. If (T4) does not hold but there is a smooth map $\Phi:\text{supp}(\mathbb{P}_X) \rightarrow \mathbb{R}^d$ such that $\text{supp}(\mathbb{P}_{\Phi(X)}) = \mathbb{R}^d$ and $\sigma(X) = \sigma(\Phi(X))$, we can consider $\Phi(X)$ as the covariate under the null instead of $X$ and compute $U_n$ based on $\Phi(X)$.
Condition (T5) is analytically convenient to simplify the derivation of theoretical results but it can be replaced by the finiteness of $\mathbb{E}[|f(X)|^4]$ and $\mathbb{E}\left[||\tilde{g}(Z)||_\mathcal{Y}^4\right]$. Moreover, the boundedness assumption in (A3), which also helps to simplify the derivation of theoretical results, can be removed at the cost of more restrictive moment conditions in (T1)-(T3). For example, it would be posible to consider $c(z,z') = -|||z-z'||_\mathcal{Z}$ and adapt the theory in \cite{Lee2020} to test $\mathcal{H}_0$.

\subsection{Null hypothesis and fixed alternatives}
\label{Sec3.2}
Let $\{(Z_i, Y_i)\}_{i=1}^n$ be an i.i.d. sample with the same distribution as the pair $(Z,Y)$, which satisfies the regression model in (\ref{3.1}) and let $U_n$ be defined as in (\ref{2.2}). Theorem \ref{Th1} establishes the limiting behavior of $U_n$ depending on the rate at which the bandwith $h$ tends to 0, which is ultimately related to the bias in the nonparametric estimation of the regression and density functions.

\renewcommand{\labelenumi}{(\arabic{enumi})}
\begin{Theorem}
	\label{Th1}
	Given the regression model in (\ref{3.1}), assume that (A1)-(A3) and (T1)-(T5) hold. Then:
	\begin{enumerate}
		\item $U_n$ is a consistent estimator of $\text{KCMD}(f(X)(Y-m(X))|Z)$. In particular,
		$$U_n = \text{KCMD}(f(X)(Y-m(X))|Z) + O(h^\nu) + O_\mathbb{P}\left(\frac{1}{\sqrt{n}}\right).$$
		\item  Under $\mathcal{H}_0$ and assuming $nh^{2\nu}\rightarrow \infty$,
		$$\frac{\sqrt{n}}{h^\nu}(U_n - \mu_{Bn}) \xrightarrow{\mathcal{D}} N(0,4\sigma_B^2),$$
		where $\xrightarrow{\mathcal{D}}$ means convergence in distribution and, given $(Z,Y)$ and $(Z',Y')$ i.i.d., $\mu_{Bn}$ and $\sigma_B^2$ are defined as
		\begin{align*}
			\mu_{Bn} &= \text{KCMD}(L(X)|Z) = O(h^{2\nu}),\\
			\sigma_B^2 &= \text{Var}\left(\mathbb{E}\left[\left(c(Z,Z') - b(X,Z')\right)\left\langle f(X)(Y-m(X)), B(X') \right\rangle|Z,Y\right]\right),
		\end{align*}
		where $L(x)= \mathbb{E}[K_h(X-X')(m(X)-m(X'))|X=x]$ and $B(x) = \lim_{h\rightarrow 0}L(x)/h^\nu$ verifies
		$$B(x) = \left(\int  K^u(t) t^\nu dt\right) \left( \sum\limits_{j=1}^{\nu -1}\frac{1}{j!(\nu-j)!}\sum\limits_{l=1}^d\frac{\partial^j m(x)}{(\partial x^l)^j}\frac{\partial^{\nu-j}f(x)}{(\partial x^l)^{\nu - j}} + \frac{f(x)}{\nu!} \sum\limits_{l=1}^d\frac{\partial^j m(x)}{(\partial x^l)^\nu } \right).$$
		\item  Under $\mathcal{H}_0$, assuming $nh^{2\nu}\rightarrow M \ge0$, $M<\infty$ ($d< 2\nu$),
		$$nU_n \xrightarrow{\mathcal{D}} ||\sqrt{M}\mathbb{E}[c(Z,\cdot)\otimes B(X)]+N_{{\mathcal{HS}}}||_\mathcal{HS}^2 - \mathbb{E}\left[||N_{{\mathcal{HS}}}||_\mathcal{HS}^2\right],$$
		where $N_{{\mathcal{HS}}}$ is a Gaussian random variable taking values in ${\mathcal{HS}}$ with zero mean and the same covariance operator as $(c(Z,\cdot)-b(X, \cdot))\otimes f(X)\mathcal{E}$.
	\end{enumerate}	
\end{Theorem}

We can compare the results in Theorem \ref{Th1} with the asymptotic behavior of a similar estimator of $\text{KCMD}(f(X)(Y-m(X)))$ when $m$ and $f$ are completely known. This estimator is defined as 
\begin{equation}\label{3.5}\tag{3.5}
	\text{KCMD}_n = \frac{1}{n(n-1)}\sum\limits_{i\neq j}C_{ij}\langle f(X_i)(Y_i-m(X_i)), f(X_j)(Y_j-m(X_j))\rangle_\mathcal{Y},
\end{equation}
where, for all $i\neq j$, $C_{ij}$ are defined as
$$C_{ij} =  \frac{1}{(n-2)(n-3)}\sum\limits_{\substack{k\neq l\\k,l\notin\{i,j\} }}\left(c(Z_i,Z_j) - c(Z_i,Z_l) - c(Z_k,Z_j) + c(Z_k,Z_l)\right).$$
Note that $\text{KCMD}_n$ can be expressed as a fourth order U-estatistic and its theoretical properties can be inferred from the results in \cite{Lee2020} and \cite{Lai2021}.

On the one hand, the nonparametric estimation of $f$ and $m$ does not prevent the consistency in probability of $U_n$ as an estimator of $\text{KCMD}(f(X)(Y-m(X)))$. In absence of kernel smoothing, however, almost sure consistency holds for $\text{KCMD}_n$ by the strong law of large numbers for U-statistics (see e.g. Theorem 3.3.1 in \citeauthor{Borovskich1996}, \citeyear{Borovskich1996}). On the other hand, the limitting distribution under $\mathcal{H}_0$ is affected by the bandwidth parameter $h$. The usual bandwidths for the density and regression function estimation, with an asymptotic order of $n^{-1/(2\nu +d)}$, verify $nh^{2\nu}\rightarrow \infty$ and lead to a limit Gaussian distribution for $U_n$ according to Theorem \ref{Th1}-(2). However, the condition $nh^{2\nu}\rightarrow 0$ is required for $U_n$ to have the same asymptotic distribution as $\text{KCMD}_n$ under $\mathcal{H}_0$ which, denoting by $\{\lambda_j\}_{j=1}^\infty$ the eigenvalues of the covariance operator of $(c(Z,\cdot)-b(X, \cdot))\otimes f(X)\mathcal{E}$, coincides with the distribution of $\sum_{j=1}^\infty \lambda_j (N^2_j-1)$, where $\{N_j\}_{j=1}^\infty$ is a collection of independent standard Gaussian random variables.

We do not provide a distributional result for $U_n$ under $\mathcal{H}_1$ in Theorem \ref{Th1} although, since we are assuming that $c$ is a characteristic kernel and, therefore, $\text{KCMD}(f(X)(Y-m(X))|X)>0$ under the alternative hypothesis, the following corollary trivially holds from Theorem \ref{Th1}-(1).

\begin{cor}
	\label{Cor1}
	Let (A1)-(A3) and (T1)-(T5) hold for the regression model in (\ref{3.1}). Under $\mathcal{H}_1$,
	$$nU_n \xrightarrow{\mathbb{P}}\infty \quad \text{and} \quad  \frac{\sqrt{n}}{h^\nu}(U_n - \mu_{Bn}) \xrightarrow{\mathbb{P}}\infty,$$
	where $\mu_{Bn}$ is defined as in Theorem \ref{Th1} and $\xrightarrow{\mathbb{P}}$ means convergence in probability.
\end{cor}

The combination of Theorem \ref{Th1} and Corollary \ref{Cor1} provides theoretical justification for the use of either $nU_n$ or $\sqrt{n}(U_n - \mu_{Bn})$ as test statistics when $nh^{2\nu}\rightarrow M \ge 0$ and $nh^{2\nu}\rightarrow \infty$, respectively, rejecting the null hypothesis for abnormally high values, according to their asymptotic distribution under $\mathcal{H}_0$. The latter statistic is more problematic since it involves a bias term which is asymptotically non negligible and unknown. In fact, a pair of nonparametric estimators $\hat{\sigma}^2_B$ and $\hat{\mu}_{Bn}$ verifying
$$\frac{\hat{\sigma}^2_B}{\sigma^2_B}\xrightarrow{\mathbb{P}} 1 \quad \text{and} \quad \sqrt{\frac{n}{h^\nu}}\left(\hat{\mu}_{Bn}-\mu_{Bn}\right)\xrightarrow{\mathbb{P}}0,$$
where $\sigma^2_B$ and $\mu_{Bn}$ are defined as in Theorem \ref{Th1}-(2), are required for that test statistic to be used in practice. Note also that $\sigma^2_B$ can be zero if $B = 0$, leading to a degenerate limiting distribution. However, Theorem \ref{Th1}-(3) establishes a non trivial limit for $nU_n$ if we assume that $\mathbb{P}(\mathcal{E}\neq 0)>0$ and $\sigma(W)\nsubseteq\sigma(X)$ since, in that case,  the covariance operator of $(c(Z,\cdot)-b(X, \cdot))\otimes f(X)\mathcal{E}$ has one or more positive eigenvalues. 

The smoothing-based test statistics in \cite{Fan1996}, \cite{Lavergne2000}, \cite{Lavergne2015} and \cite{Zhu2018} have a limiting normal distribution, which arises from the theory of degenerated U-statistics with sample size-dependent kernels developed in \cite{Hall1984}. Such results might seem preferable to the non-pivotal distribution in Theorem \ref{Th1}-(3). However, as the authors of \cite{Lavergne2000} and \cite{Lavergne2015} note, the Gaussian approximation can be quite inaccurate for small sample sizes, so they recommend calibrating their tests through resampling. As we show in Section \ref{Sec4}, the test based on $nU_n$ when $nh^{2\nu}\rightarrow 0$ can also be calibrated via bootstrap.

\subsection{Local alternative analysis}
\label{Sec3.3}

The following theorem establishes the limiting behavior of $U_n$ under $\{\mathcal{H}_{1n}\}_{n\in \mathbb{N}}$ in (\ref{3.2}), in terms of the exponent $\gamma>0$ as well as the convergence rate of $h$ to zero.

\begin{Theorem}
	\label{Th2}
	Suppose (A1)-(A3), (T1)-(T5) and $\mathcal{H}_{1n}$ hold for each $n\in \mathbb{N}$ and some $\gamma>0$. Let $\mu_{Bn}$, $\sigma^2_B$, $B$ and $N_{\mathcal{HS}}$ be defined as in Theorem \ref{Th1}. Moreover, for a given $0<\tilde{M}<\infty$ define $\mu_1$ and $\mu_2(\tilde{M})$ as
	$$\mu_1= -2\mathbb{E}\left[c(Z,Z')\left\langle B(X), f(X')\tilde{g}(Z')\right\rangle_\mathcal{Y}\right] \quad \text{and} \quad \mu_2(\tilde{M}) = \mu_1 + \frac{1}{\sqrt{\raisebox{0.1ex}{\(\tilde{M}\)}}}\text{KCMD}(f(X)\tilde{g}(Z)|Z).$$
	Then:
	\begin{enumerate}
		\item Assuming $nh^{2\nu}\rightarrow \infty$, we have:
		\begin{itemize}
			\item If $\gamma > \frac{1}{2}$, $\frac{\sqrt{n}}{h^\nu}(U_n - \mu_{Bn}) \xrightarrow{\mathcal{D}} N(0,4\sigma^2_B)$.
			\item If $\gamma =\frac{1}{2}$, $\frac{\sqrt{n}}{h^\nu}(U_n - \mu_{Bn})\xrightarrow{\mathcal{D}} N(\mu_1,4\sigma^2_B)$.
			\item If $\gamma <\frac{1}{2}$ and $n^{\gamma} h^\nu\rightarrow{0}$,  $\frac{\sqrt{n}}{h^\nu}(U_n - \mu_{Bn}) \xrightarrow{\mathbb{P}} \infty$.
			\item If $\gamma <\frac{1}{2}$ and $n^{\gamma} h^\nu\rightarrow{\tilde{M}}$, $\frac{\sqrt{n}}{h^\nu}(U_n - \mu_{Bn})=\frac{\sqrt{n}}{n^\gamma}\mu_2(\tilde{M}) + o_\mathbb{P}\left(\frac{\sqrt{n}}{n^\gamma}\right)$.
			\item If $\gamma <\frac{1}{2}$ and $n^{\gamma} h^\nu\rightarrow{\infty}$, $\frac{\sqrt{n}}{h^\nu}(U_n - \mu_{Bn})=\frac{\sqrt{n}}{n^\gamma}\mu_1 + o_\mathbb{P}\left(\frac{\sqrt{n}}{n^\gamma}\right)$.
		\end{itemize}
		\item Assuming $nh^{2\nu}\rightarrow M$ with $0\le M<\infty$, it holds: 
		\begin{itemize}
			\item If $\gamma > \frac{1}{2}$, $nU_n \xrightarrow{\mathcal{D}} ||\sqrt{M}\mathbb{E}[c(Z,\cdot)\otimes B(X)]+N_{{\mathcal{HS}}}||_{{\mathcal{HS}}}^2 -\mathbb{E}\left[||N_{{\mathcal{HS}}}||_{{\mathcal{HS}}}^2\right]$.
			\item If $\gamma =\frac{1}{2}$, $nU_n \xrightarrow{\mathcal{D}} ||\mathbb{E}[c(Z,\cdot)\otimes (\sqrt{M}B(X)-f(X)\tilde{g}(Z))]+N_{{\mathcal{HS}}}||_{{\mathcal{HS}}}^2 -\mathbb{E}\left[||N_{{\mathcal{HS}}}||^2_{{\mathcal{HS}}}\right]$.
			\item If $\gamma <\frac{1}{2}$, $nU_n \xrightarrow{\mathbb{P}} \infty$.
		\end{itemize}
	\end{enumerate}
\end{Theorem}

For ease of exposition, let us assume that $\mu_{Bn}$ and $\sigma^2_B$ are either known or can be estimated at a sufficiently fast rate, which is a requirement to use $\sqrt{n}(U_n -\mu_{Bn})/{h^\nu}$ as a test statistic. Then, the test based on the condition $nh^{2\nu}\rightarrow{0}$ has a clear theoretical advantage over the other two possibilities ($nh^{2\nu}\rightarrow M>0$ and $nh^{2\nu}\rightarrow{\infty}$). First, if $\gamma>1/2$ the test statistics converge to the corresponding limit distributions under the null hypothesis derived in Theorem \ref{Th1} and, therefore, alternative hypothesis converging at the null at a rate faster than $n^{-1/2}$ cannot be detected in any case. In addition, the limiting distribution of $nU_n$ when $\gamma=1/2$ and $nh^{2\nu}\rightarrow 0$ stochastically dominates the distribution of said test statistic under $\mathcal{H}_0$. This is a consecuence of the noncentral chi-squared distribution being stochastically increasing in the noncentrality parameter, as well as the following almost surely equivalent representation:
$$||N_{{\mathcal{HS}}} - \mathbb{E}[c(Z,\cdot)\otimes f(X)\tilde{g}(Z)]||_{{\mathcal{HS}}}^2 = \sum\limits_{j=1}^\infty \left(\left\langle N_{{\mathcal{HS}}},\phi_j \right\rangle_{{\mathcal{HS}}}-\left\langle \mathbb{E}[c(Z,\cdot)\otimes f(X)\tilde{g}(Z)],\phi_j \right\rangle_{{\mathcal{HS}}}\right)^2,$$
where $\{\left\langle N_{{\mathcal{HS}}},\phi_j \right\rangle\}_{j=1}^\infty$ is a collection of independent Gaussian random variables with zero mean and variance $\lambda_j\ge0$ and $\{\lambda_j,\phi_j\}_{j=1}^\infty$ is the orthonormal eigensystem associated with the covariance operator of $N_{\mathcal{HS}}$. This property is not necessarily preserved if $nh^{2\nu}\rightarrow M >0$. In fact, if it holds that $$||\mathbb{E}[c(Z,\cdot)\otimes (\sqrt{M}B(X)-f(X)\tilde{g}(Z))]||_{{\mathcal{HS}}}^2\le ||\mathbb{E}[c(Z,\cdot)\otimes \sqrt{M}B(X)]||_{{\mathcal{HS}}}^2,$$
then, for a given significance level $\alpha \in (0,1)$, the probability under $\{\mathcal{H}_{1n}\}_{n\in \mathbb{N}}$ of the test statistic exceding the $(1-\alpha)$th quantile of the limiting distribution under the null hypothesis will converge to some value bounded from above by $\alpha$. The same type of issue affects $\sqrt{n}(U_n -\mu_{Bn})/{h^\nu}$ in the case of $nh^{2\nu}\rightarrow\infty$, since the limiting Gaussian distribution under the null is shifted by a constant $\mu_1$, which can take negative values, under $\{\mathcal{H}_{1n}\}_{n\in \mathbb{N}}$. Finally, for $\gamma <1/2$ the test based on $nU_n$ with $nh^{2\nu}\rightarrow M\ge 0$ has an asymptotic power of $1$ whereas, if $n^{2\gamma} h^{2\nu}$ does not converge to 0 and either $\mu_1<0$ or $\mu_2<0$, the power of the test based on $\sqrt{n}(U_n -\mu_{Bn})/{h^2}$ will tend to zero.

\begin{remark}
	If we redefine the test based on $\sqrt{n}(U_n -\mu_{Bn})/{h^\nu}$ so that we reject the null hyphotesis whenever the squared statistic $n(U_n -\mu_{Bn})^2/{h^{2\nu}}$ excedes the $(1-\alpha)$th quantile of the corresponding limiting (under $\mathcal{H}_0$) scaled chi-squared distribution, local alternatives converging at the rate of $n^{-\gamma}$, for $\gamma\le1/2$, can be detected with non-trivial power as long as $\mu_1$ or $\mu_2(\tilde{M})$, depending on the particular case, are not zero. We still suggest, however, to select the bandwidth parameter $h$ such that it verifies $nh^{2\nu}\rightarrow{0}$ since, in practice, it is very difficult to find an adequate nonparametric estimator for $\mu_{Bn}$. Furthermore, the local power properties of $U_n$ when $nh^{2\nu}\rightarrow{0}$ and those of $\text{KCMD}_n$, the latter being defined as in (\ref{3.5}), coincide (see e.g. \citeauthor{Lee2020}, \citeyear{Lee2020} and \citeauthor{Lai2021}, \citeyear{Lai2021}).
\end{remark}

The detection of local alternatives converging to the null at the rate of $n^{-1/2}$ with nontrivial power, which holds for the test based on $nU_n$ when $nh^{2\nu}\rightarrow0$ by Theorem \ref{Th3}-(3), was also proved in \cite{Chen1999} and in \cite{Delgado2001}. In contrast, the tests in \cite{Fan1996} and \cite{Lavergne2000} can only detect local alternatives converging to the null at the rate of $n^{-1/2}h^{-(d+\bar{d})/4}$, where $d$ is the dimension of $X$ and $\bar{d}$ the dimension of $W$ (asuming the latter is also a vector covariate) and the tests in \cite{Lavergne2015} and \cite{Zhu2018} detect local alternatives converging to the null at the rate of $n^{-1/2}h^{-d/4}$ and $n^{-1/2}h^{-d_1/4}$, respectively, where $1\le d_1\le d$ is the number of significant components of $X$ in the regression model of $Y$ on $X$. Therefore, all previous smoothing-based tests are clearly subject to the curse of dimensionality in terms of power whereas our proposal, as well as the ones in \cite{Chen1999} and \cite{Delgado2001}, are not, at least, asymptotically.

\section{Bootstrap calibration}
\label{Sec4}

The limiting distribution of $nU_n$ when the bandwidth verifies $nh^{2\nu}\rightarrow{0}$, which is our suggested choice according to the discussion in Section \ref{Sec3.2}, coincides with the distribution of a weigthed sum of independent and centered chi-squared random variables with one degree of freedom. Such distribution is nonpivotal and depends on the data distribution through the, possible infinite, collection of eigenvalues of the covariance operator of the $\mathcal{HS}$-valued random variable $(c(Z,\cdot)-b(X,\cdot))\otimes f(X)\mathcal{E}$. Rather than attempting to estimate the eigenvalues up to some natural number or using an asymptotic Gaussian or chi-square approximation (see, for example, \citeauthor{Lindsay2014}, \citeyear{Lindsay2014}), we suggest to calibrate the test through resampling. Let us introduce the following notation:
$$\hat{m}'_{h}(X_i) = \frac{1}{n}\sum\limits_{k=1}^n\frac{|K_h(X_i-X_k)|}{\hat{f}'_{h}(X_i)}Y_k  \quad \text{and} \quad \hat{f}'_{h}(X_i) = \frac{1}{n}\sum\limits_{k=1}^n |K_h(X_i-X_k)|.$$
Given the sample \((Z_1,Y_1), \ldots, (Z_n,Y_n)\), we propose the following bootstrap scheme:

\begin{enumerate}
	\item Let $\{\kappa_n\}_{n=1}^\infty$ and $\{\tilde{h}\}_{n=1}^\infty$ be two sequences of nonnegative numbers verifying $\tilde{h}\rightarrow0$, $n\tilde{h}^d\rightarrow\infty$, $\kappa_n \tilde{h}^\nu \rightarrow0$ and $\kappa_n^2/(n\tilde{h}^d)\rightarrow 0$. Compute 
	$\{\hat{m}_{h,\tilde{h}}(X_i)\}_{i=1}^n$ and $\{\tilde{\hat{\mathcal{E}}}_i\}_{i=1}^n$, where 
	$$\hat{m}_{h,\tilde{h}}(X_i) = \hat{m}_{\tilde{h}}(X_i)\ind{\hat{f}'_{h}(X_i)\le |\hat{f}_{\tilde{h}}(X_i)|\kappa_n} + \hat{m}'_{h}(X_i)\ind{\hat{f}'_{h}(X_i)>|\hat{f}_{\tilde{h}}(X_i)|\kappa_n},$$
	and $\tilde{\hat{\mathcal{E}}}_i = Y_i - \hat{m}_{h,\tilde{h}}(X_i).$
	\item Draw an i.i.d. sample of size $n$, $\{r_i\}_{i=1}^n$, from a univariate distribution with zero mean, unit variance and finite $(2+\delta)$th order moment. 
	\item Compute the bootstrap statistic $nU_n^{*}$ defined as
	$$nU_n^{*} = \frac{1}{n-1}\sum\limits_{i\neq j} r_ir_jD_{ij}\left\langle \tilde{\hat{\mathcal{E}}}_i , \tilde{\hat{\mathcal{E}}}_j \right\rangle_{\mathcal{Y}},$$
	where $D_{ij}$ verifies
	$$D_{ij} = \frac{1}{n^2}\sum\limits_{k,l=1}^n K_h(X_i-X_k)K_h(X_j-X_l)\left(c(Z_i,Z_j)-c(Z_i,Z_l) - c(Z_k,Z_j) + c(Z_k,Z_l)\right).$$
	\item Repeat steps (2) and (3) to obtain $n_B$ conditionally i.i.d. copies of $nU_n^{*}$.
	\item For a fixed significance level $\alpha \in (0,1)$, the $1-\alpha$ quantile of the conditional distribution of $nU_n^{*}$, $Q^*_n(1-\alpha)$, is approximated by the empirical quantile of the sample obtained in step (4), $\hat{Q}^*_{n}(1-\alpha)$.
	\item Reject $\mathcal{H}_0$ if $nU_n > \hat{Q}^*_{n}(1-\alpha)$.
\end{enumerate}

Since we are imposing $nh^{2\nu}\rightarrow 0$, the residuals $\hat{\mathcal{E}}_i = Y_i-\hat{m}_h(X_i)$ will usually have less variability than the corresponding errors for small sample sizes. The modified nonparametric residuals defined in step (1) are a tool to improve the finite sample performance of the test since, as we discuss in more detail in Section \ref{Sec6}, $\tilde{\hat{\mathcal{E}}}_i \approx Y_i - \hat{m}_{\tilde{h}}(X_i)$ for small sample sizes and we can select $\tilde{h}$ such that the corresponding residual distribution is close to that of the regression errors. Note that those coefficients only need to be computed once so it is possible to implement this bootstrap procedure quite efficiently.

The results concerning the consistency of the resampling scheme and its non-trivial power under fixed and local alternatives are summarised in the following theorems.

\begin{Theorem}
	\label{Th3}
	Given the regression model in (\ref{3.1}), assume that (A1)-(A3) and (T1)-(T5) hold. Furthermore, assume that $\mathbb{P}(\mathcal{E}\neq 0)>0$ and $\sigma(W)\nsubseteq\sigma(X)$. If $nh^{2\nu}\rightarrow 0$, we have, under the null hypothesis, that
	$$\sup\limits_{x\in \mathbb{R}}|\mathbb{P}^*(nU^{*}_n\le x) - \mathbb{P}(nU_n\le x)| \xrightarrow{\mathbb{P}}0,$$
	where $\mathbb{P}^*$ denotes the conditional probability measure induced by the sample $\{(Z_i,Y_i)\}_{i=1}^n$. Moreover, denoting the $(1-\alpha)$th quantile of the conditional distribution of $nU_n^{*}$ by $Q_{n}^*(1-\alpha)$, it follows:
	\begin{enumerate}
		\item Under $\mathcal{H}_0$, $\mathbb{P}(nU_n > Q_{n}^*(1-\alpha))\rightarrow \alpha$.
		\item Under $\mathcal{H}_1$, $\mathbb{P}(nU_n > Q_{n}^*(1-\alpha))\rightarrow 1$.
	\end{enumerate}
\end{Theorem}

\begin{Theorem}
	\label{Th4}
	Let $F$ denote the cumulative distribution function of $||N_{\mathcal{HS}}||_{\mathcal{HS}}^2 - \mathbb{E}\left[||N_{\mathcal{HS}}||_{\mathcal{HS}}^2 \right]$, where $N_{\mathcal{HS}}$ is defined as in Theorem \ref{Th1}-(3).  Suppose (A1)-(A3) and (T1)-(T5) hold,  $\mathbb{P}(\mathcal{E}\neq 0)>0$ and $\sigma(W)\nsubseteq\sigma(X)$. Then, under $\{\mathcal{H}_{1n}\}_{n\in\mathbb{N}}$, for any $\gamma>0$ and assuming $nh^{2\nu}\rightarrow 0$, we have
	$$\sup\limits_{x\in \mathbb{R}}|\mathbb{P}^*(nU_n^{*}\le x) - F(x)| \xrightarrow{\mathbb{P}}0,$$
	where $\mathbb{P}^*$ denotes the conditional probability measure induced by the sample $\{(Z_i,Y_i)\}_{i=1}^n$. Furthermore, denote by $\tilde{F}$ the cumulative distribution function of $||\mathbb{E}[c(Z,\cdot)\otimes f(X)\tilde{g}(Z)]+N_{{\mathcal{HS}}}||_{\mathcal{HS}}^2 - \mathbb{E}\left[||N_{\mathcal{HS}}||_{\mathcal{HS}}^2 \right]$ and by $Q_{n}^*(1-\alpha)$ and $Q(1-\alpha)$ the $(1-\alpha)$th quantiles of the conditional distribution of $nU_n^{*}$ and $F$, respectively. Then:
	\begin{enumerate}
		\item If $\gamma > \frac{1}{2}$:  $\mathbb{P}(nU_n >  Q_{n}^*(1-\alpha)) \rightarrow \alpha$.
		\item If $\gamma =\frac{1}{2}$: $\mathbb{P}(nU_n >  Q_{n}^*(1-\alpha)) \rightarrow 1 - \tilde{F}(Q(1-\alpha))$.
		\item If $\gamma <\frac{1}{2}$: $\mathbb{P}(nU_n >  Q_{n}^*(1-\alpha)) \rightarrow 1$.
	\end{enumerate}
\end{Theorem}

\section{Extensions}
\label{Sec5}
The methodology presented in Section \ref{Sec2} and the results in Sections \ref{Sec3} and \ref{Sec4} can be naturally adapted to fit other kinds of specification and goodness-of-fit problems with minimal modifications to the test statistic, the bootstrap scheme and the corresponding proofs. 

The test can be directly used in a slightly more general framework than testing for the significance of a group of covariates. Recall that the definition of the null hypothesis, the test statistic and the theoretical results do not involve $W$ but $Z$. This means that the proposed test can be applied in the following goodness-of-fit problem:
\begin{equation}\label{5.1}\tag{5.1}
\mathcal{H}'_0 : \ \ \mathbb{E}[Y|Z] = \mathbb{E}[Y|\mathcal{P}(Z)]  \ \text{a.s.} \ \ \ \text{vs.} \ \ \ \mathcal{H}'_1: \ \ \mathbb{P}(\mathbb{E}[Y|Z]\neq  \mathbb{E}[Y|\mathcal{P}(Z)]) >0,
\end{equation}
for some known measurable function $\mathcal{P}:\mathcal{Z} \rightarrow \mathbb{R}^d$ where, obviously, $X = \mathcal{P}(Z)$.

\begin{exmp}\label{Ex1.1}
	Consider the nonparametric regression model of $Y$ on $Z$ given by
	$$Y = g(Z) + \mathcal{E},$$
	where \(\mathcal{E}\) is the regression error, verifying $\mathbb{E}[\mathcal{E}|Z] = 0$. If $\mathcal{Z}$ is an infinite dimensional, separable Hilbert space, we can identify $\mathcal{Z}$ with the space of square summable sequences, $\ell^2(\mathbb{R})$, through the isometric isomorphism 
	$$z = \sum\limits_{j=1}^\infty \langle z,\phi_j\rangle \phi_j\in \mathcal{Z} \leadsto \{\langle z,\phi_j\rangle\}_{j\in \mathbb{N}}\in \ell^2(\mathbb{R}),$$
	where $\{\phi_j\}_{j\in \mathbb{N}}$ is some orthonormal basis of $\mathcal{Z}$. Therefore, denoting by $\mathcal{P}(Z) = (\langle Z,\phi_1\rangle,\ldots,\langle Z,\phi_d\rangle)$, the hypothesis test in (\ref{5.1}) becomes a dimension reduction technique that helps to determine whether the information contained in $Z$ which is relevant to explain the behavior of the conditional expectation of $Y$ is concentrated in a $d$-dimensional projection. This is particularly useful if smoothing techniques are used to estimate $g(z)$ at $z\in \mathcal{Z}$ since, as shown in \cite{Ferraty2007}, the mean squared consistency of kernel smoothers in normed spaces depend on the \emph{small ball probability} condition, $n\mathbb{P}(||Z_i-z||_\mathcal{Z}<h)\rightarrow \infty $, where $h\equiv h(n)>0$ denotes the (sample size dependent) bandwidth parameter. Under $\mathcal{H}_0$, $g=m$ almost surely and the small ball probability condition in $\mathbb{R}^d$ becomes $nh^d\rightarrow \infty$, usually leading to a considerable reduction in the effect of the curse of dimensionality in the regression function estimation problem. This regression modelling approach consisting of projecting an infinite dimensional covariate and then nonparametrically estimating the regression function of the response on the covariate projection is implicitly used, for instance, whenever the authors in \cite{Ferraty2006} consider a projection type semi-metric in the definition of the regression kernel smoother.
\end{exmp}

A natural extension to this idea is to consider a semiparametric specification of $\mathcal{P}$ as in functional single-index models (see \citeauthor{Chen2011}, \citeyear{Chen2011}) or partially linear models (see \citeauthor{AneirosPerez2006}, \citeyear{AneirosPerez2006}). Under some regularity conditions, such as the existence of a $\sqrt{n}$-consistent estimator for the parametric part of $\mathcal{P}$, the test calibrated via (a conviniently modified) bootstrap might still be consistent and have non trivial power under the alternative hypothesis.

In the following sections, we present the extension of our proposal to account for discrete and mixed variables in the regression model and we also discuss the Hilbert space-valued $\mathcal{X}$ case, introducing a smoothing-less significance testing procedure valid whenever $X$ and $W$ are independent.

\subsection{Significance test involving discrete or mixed variables}
\label{Sec5.1}

It is important to note that the case where either $Y$ or $W$ (or both) have a discrete component is fundamentally different from the case of discrete or mixed $X$. In the former scenario, the only relevant issue is the definition of the Hilbert spaces $\mathcal{Y}$ and $\mathcal{W}$ where we assume that $Y$ and $W$ take values, respectively. The key idea is to consider an appropriate embedding for discrete components into a Hilbert space with an adequate inner product structure. For example, count variables can be straighforwardly embedded into $\mathbb{R}$, whereas non-ordinal categorical variables can be one-hot encoded, univocally mapping each distinct value to an element of the basis of the corresponding Euclidean space. Once we can identify $Y$ and $W$ with some Hilbert space-valued random variables, we can apply all the previously established results. 

In the particular case of discrete response, the embedding itself may have a considerable impact in the model interpretation. For example, if we one-hot encode $Y\in\{y_1,\ldots, y_{d_Y}\}$, it is trivial to verify that $m(X) = (\mathbb{P}(Y = y_1|X), \ldots, \mathbb{P}(Y = y_{d_Y}|X))^\top$ and $g(Z) = (\mathbb{P}(Y = y_1|Z), \ldots, \mathbb{P}(Y = y_{d_Y}|Z))^\top$. Thus, testing $\mathcal{H}_0$ is equivalent to testing whether $\mathbb{P}(Y = y_j|Z)$ is almost surely equal to $\mathbb{P}(Y = y_j|X)$ for all $j=1,\ldots, d_Y$ and, ultimately, if $Y$ and $W$ are conditionally independent given $X$.
 
Let us now assume that $X = \left(X^C,X^D\right)^\top$, where $X^C$ and $X^D$ contain the $d_C$ continuous and $d_D$ discrete components of $X$. We need to alter the definition of the test statistic to accomodate for the additional discrete variables. As in the previous discussion, we implicitly assume that $X^D$ takes values in a space embedded into some Hilbert space. In this context, the test statistic $U_n$ can be defined as
$$U_n =  \frac{\sum\limits_{i\neq j\neq k\neq l} c(Z_i,Z_j) K_h\left(X_i^C - X_k^C\right) K_h\left(X_j^C - X_l^C\right) \ind{X_i^D = X_k^D}  \ind{X_j^D = X_l^D} \left\langle Y_i-Y_k,Y_j-Y_l\right\rangle_\mathcal{Y}}{n(n-1)(n-2)(n-3)},$$
for the mixed $X$ case and
$$U_n =  \frac{1}{n(n-1)(n-2)(n-3)}\sum\limits_{i\neq j\neq k\neq l} c(Z_i,Z_j) \ind{X_i^D = X_k^D} \ind{X_j^D = X_l^D}\left\langle Y_i-Y_k,Y_j-Y_l\right\rangle_\mathcal{Y},$$
for the fully discrete case.
Alternatively, we could follow \cite{Racine2004} and consider a second product kernel, smoothing over the discrete covariates, instead of the indicator functions. For the sake of simplicity, we are restricting the discussion to the latter.

If $d_C>0$, let $f\left(x^C,x^D\right)$ denote the joint density function of $X=\left(X^C,X^D\right)$ at $\left(x^C,x^D\right)$; that is,
$$f\left(x^C,x^D\right) = \frac{\partial^{d_C}}{\partial x^1 \cdots \partial x^{d_C}} \mathbb{P}\left(X^1\le x^1,\ldots, X^{d_C}\le x^{d_C}, X^D = x^D\right).$$
In addition, denote by $\text{supp}{\left(\mathbb{P}_{X^D}\right)}$ the support of $X^D$, by $\text{supp}{\left(\mathbb{P}_{X^C|X^D = x^D}\right)}$ the support of the conditional distribution of $X^C$ given $X^D = x^D$ and by $\partial \text{supp}{\left(\mathbb{P}_{X^C|X^D = x^D}\right)}$ the boundary of the latter. With this notation, let us consider the following modified versions of technical conditions (T2), (T3) and (T4):

\renewcommand{\labelenumi}{($\tilde{\text{T}}$\arabic{enumi})}
\begin{enumerate}
	\setcounter{enumi}{1}
	\item As a function of $x^C$, both $f{\left(x^C,x^D\right)}$ and $m{\left(x^C,x^D\right)}$ are $\nu+1$ times continuously Fréchet differentiable in the interior of $\text{supp}{\left(\mathbb{P}_{X^C|X^D = x^D}\right)}$ for all $x^D\in \text{supp}{\left(\mathbb{P}_{X^D}\right)}$. Furthermore, there is some $h_1>0$ such that, for $a\in \{m,f\}$,
	\begin{itemize}
		\item $\sum\limits_{k_1,\ldots, k_j = 1}^{d_C} \mathbb{E}\left[\left|\left|\dfrac{\partial^j a(X^C,X^D)}{\partial x^{k_1}\cdots\partial x^{k_j}}\right|\right|^4\right]<\infty$ for $j=1,\ldots,\nu$.
		\item $\sum\limits_{k_1,\ldots, k_{\nu + 1} = 1}^{d_C} \mathbb{E}\left[\sup\limits_{||x^C_0||\le h_1}\left|\left|\dfrac{\partial^{\nu + 1} a(X^C+x^C_0,X^D)}{\partial x^{k_1}\cdots\partial x^{k_{\nu + 1}}}\right|\right|^4\ind{(X^C +x^C_0,x^D)\in \text{supp}(\mathbb{P}_X)}\right]<\infty$.
	\end{itemize}
	\item $b\left(x^C,x^D,z'\right)$ is differentiable with respect to $x^C$ for all $z'\in \mathcal{Z}$ and all $x^D\in \text{supp}\left(\mathbb{P}_{X^D}\right)$. Moreover, there will be some $h_2>0$ such that:
	\begin{itemize}
		\item $\sum\limits_{k=1}^{d_C} \mathbb{E}\left[\sup\limits_{||x^C_0||\le h_2}\left|\dfrac{\partial b(X^C+x^C_0,X^D,Z)}{\partial x^{k}}\right|^2\ind{(X^C + x^C_0,x^D)\in \text{supp}(\mathbb{P}_X)}\right]<\infty$.
	\end{itemize} 
	\item For all $x^D\in \text{supp}\left(\mathbb{P}_{X^D}\right)$ such that $\partial \text{supp}\left(\mathbb{P}_{X^C|X^D = x^D}\right)$ is not empty, we have, as $\epsilon>0$ goes to zero, 
	\begin{itemize}
		\item $\sup\limits_{d\left(x^C,\partial \text{supp}\left(\mathbb{P}_{X^C|X^D = x^D}\right)\right)\le \epsilon} f(x^C,x^D) = O(\epsilon^{\nu})$.
	\end{itemize} 
	Moreover, there is some $h_3>0$ such that 
	\begin{itemize}
		\item $\mathbb{E}\left[\sup\limits_{||x^C_0||\le h_3} ||m(X^C + x^C_0,X^D)||_\mathcal{Y}^2\ind{d\left(x^C,\partial \text{supp}\left(\mathbb{P}_{X^C|X^D = X^D}\right)\right)\le h_3, \ (X^C + x^C_0,X^D)\in \text{supp}(\mathbb{P}_X)}\right] = O(1)$.
	\end{itemize} 
\end{enumerate}

Under the assumptions (A1)-(A3), the technical conditions (T1), ($\tilde{\text{T}}$2), ($\tilde{\text{T}}$3), ($\tilde{\text{T}}$4) and (T5) and imposing that $nh^{d_C}\rightarrow\infty$ and $nh^{2\nu}\rightarrow0$, where $\nu$ is the order of the univariate kernel $K^u$, Theorem \ref{Th1}-(3), with $M=0$, still holds. In the purely discrete case, where $X = X^D$, Theorem \ref{Th1}-(3) also holds under just (A3) and (T1), with the interpretation that the function $f$ appearing in the definition of the covariance operator of $N_{\mathcal{HS}}$ is the probability mass function of $X^D$. Similarly, Theorem \ref{Th2}-(2) also holds in the mixed and discrete.

The bootstrap scheme presented in Section \ref{Sec4} is still consistent and the bootstrap-calibrated test will be able to identify local alternatives converging to the null at the rate of $n^{-1/2}$ in the mixed case if we define the nonparametric fitted values $\{\hat{m}_{h,\tilde{h}}(X_i)\}_{i=1}^n$ as
$$\hat{m}_{h,\tilde{h}}(X_i) = \hat{m}_{\tilde{h}}(X_i)\ind{\hat{f}'_{h}(X_i)\le |\hat{f}_{\tilde{h}}(X_i)|\kappa_n} + \hat{m}'_{h}(X_i)\ind{\hat{f}'_{h}(X_i)>|\hat{f}_{\tilde{h}}(X_i)|\kappa_n}$$
where $\tilde{h}$ and $\kappa_n$ are chosen such that $\tilde{h}\rightarrow0$, $n\tilde{h}^{d_C}\rightarrow\infty$, $\kappa_n \tilde{h}^\nu \rightarrow 0$ and $\kappa_n^2/(n\tilde{h}^{d_C})\rightarrow 0$; $\hat{m}_{\tilde{h}}(X_i)$ and $ \hat{f}_{\tilde{h}}(X_i)$ are defined as
$$\hat{m}_{\tilde{h}}(X_i) = \frac{1}{n}\sum\limits_{k=1}^n\frac{K_{\tilde{h}}\left(X^C_i-X^C_k\right)\ind{X^D_i = X^D_k}}{\hat{f}_{\tilde{h}}(X_i)}Y_k  \quad \text{and} \quad \hat{f}_{\tilde{h}}(X_i) = \frac{1}{n}\sum\limits_{k=1}^n K_{\tilde{h}}\left(X^C_i-X^C_k\right)\ind{X^D_i = X^D_k},$$
and, similarly, $\hat{m}'_{h}(X_i)$ and $\hat{f}'_{h}(X_i)$ correspond to the formulas
$$\hat{m}'_{h}(X_i) = \frac{1}{n}\sum\limits_{k=1}^n\frac{|K_{h}\left(X^C_i-X^C_k\right)|\ind{X^D_i = X^D_k}}{\hat{f}'_{h}(X_i)}Y_k  \quad \text{and} \quad \hat{f}'_{h}(X_i) = \frac{1}{n}\sum\limits_{k=1}^n |K_h\left(X^C_i-X^C_k\right)|\ind{X^D_i = X^D_k}.$$
In the fully discrete case, there is no need for the bandwidths $h$ and $\tilde{h}$ nor for the sequence $\{\kappa_n\}_{n\in \mathbb{N}}$. Instead, the fitted values are just computed as 
$$\hat{m}(X_i) = \frac{1}{n}\sum\limits_{k=1}^n\frac{\ind{X^D_i = X^D_k}}{\frac{1}{n}\sum\limits_{l=1}^n\ind{X^D_i = X^D_l}}Y_k.$$

\begin{remark}
	The limitations on the dimension of the continuous part of $X$ also apply in the mixed case ($d_C <2\nu$). However, there is no theoretical restriction on the dimension of $X^D$. 
\end{remark}

\subsection{Hilbert space-valued $X$: the case of independent $X$ and $W$}
\label{Sec5.2}

The use of high order kernels to overcome the dimensional constraint on $X$ cannot be extended, straighforwardly, to the infinite dimensional case. On the one hand, there are no densities, in the usual sense of the term, on general separable Hilbert spaces. On the other hand, the results in \cite{Ferraty2007}  indicate that the bias-variance tradeoff for the Nadaraya-Watson estimator becomes substantially more difficult to hande outside of the Euclidean case. In particular, the condition $nh^d\rightarrow\infty$ is extended to Hilbert spaces by means of the small ball probability condition: $n\mathbb{P}\left(||X_i-x||_{\mathcal{X}}<h\right) \rightarrow{0}$ for all $x\in \text{supp}(\mathbb{P}_X)$. This assumption is usually incompatible with a bias of order $o(n)$, which is a requisite for the consistency of the bootstrap scheme proposed in Section \ref{Sec4} and also for $nU_n$ to have a limit distribution, under $\mathcal{H}_0$, which can be expressed as the centered squared norm of a Hilbert space-valued Gaussian random variable. It is possible, however, to greatly simplify testing methodology under independence between $X$ and $W$.

It is easy to verify that
\begin{align*}
	\text{KCMD}(Y-m(X)|Z) &= \left|\left| \mathbb{E}\left[c(Z,\cdot)\otimes (Y-m(X))\right]\right|\right|_\mathcal{HS}^2\\
	 &= \left|\left| \mathbb{E}\left[\left(c(Z,\cdot)-b(X,\cdot)\right)\otimes Y\right]\right|\right|^2_\mathcal{HS}\\
	&= \mathbb{E}\left[\left(c(Z,Z') - b(X,Z') - b(X',Z) + \mathbb{E}[c(Z,Z')|X,X']\right)\langle Y, Y'\rangle_\mathcal{Y} \right].
\end{align*}
Let us assume that the bounded characteristic kernel $c$ verifies $c(z,z') = c_\mathcal{X}(x,x')c_\mathcal{W}(w,w')$ for all $z,z'\in \mathcal{Z}$, where $c_\mathcal{X}$ and $c_\mathcal{W}$ are two bounded characteristic kernels defined on $\mathcal{X}\times \mathcal{X}$ and $\mathcal{W}\times \mathcal{W}$, respectively. An example would be the Gaussian kernel, since 
$$c(z,z')= \exp\left(-\frac{||z-z'||_\mathcal{Z}^2}{\sigma_c^2}\right) = \exp\left(-\frac{||x-x'||_\mathcal{X}^2}{\sigma_c^2}\right)\exp\left(-\frac{||w-w'||_\mathcal{W}^2}{\sigma_c^2}\right) = c_\mathcal{X}(x,x')c_\mathcal{W}(w,w').$$
Then, by the independence between $X$ and $W$, the following equality chain almost surely holds:
$$b(X,Z') = c_\mathcal{X}(X,X')\mathbb{E}[c_\mathcal{W}(W,W')|X,W'] = c_\mathcal{X}(X,X')\mathbb{E}[c_\mathcal{W}(W,W')|W'].$$
Analogously, $\mathbb{E}[c(Z,Z')|X,X'] = c_\mathcal{X}(X,X')\mathbb{E}[c_\mathcal{W}(W,W')]$ a.s. and, therefore, 
$$  \mathcal{H}_0\iff \mathbb{E}\left[\left(c_\mathcal{W}(W,W') - \mathbb{E}[c_\mathcal{W}(W,W')|W'] - \mathbb{E}[c_\mathcal{W}(W,W')|W] + \mathbb{E}[c_\mathcal{W}(W,W')]\right)c_\mathcal{X}(X,X)\langle Y, Y'\rangle_\mathcal{Y} \right] = 0.$$
Denote by $\mathcal{HS}_{\mathcal{X}} = \mathcal{C}_\mathcal{X}\otimes \mathcal{Y}$ where $\mathcal{C}_\mathcal{X}$ is the RKHS generated by $c_\mathcal{X}$, so that 
$$c_\mathcal{X}(X,X)\langle Y, Y'\rangle_\mathcal{Y} = \langle c_\mathcal{X}(X,\cdot)\otimes Y, c_\mathcal{X}(X',\cdot)\otimes Y'\rangle_{\mathcal{HS}_\mathcal{X}},$$
where $c_\mathcal{X}(X,\cdot)\otimes Y, c_\mathcal{X}(X',\cdot)\otimes Y'\in \mathcal{HS}_\mathcal{X}$ a.s. Moreover, let $\mathcal{HS}_{\mathcal{W}\mathcal{X}} = \mathcal{C}_\mathcal{W}\otimes \mathcal{HS}_\mathcal{X}$ where $\mathcal{C}_\mathcal{W}$ is the RKHS generated by $c_\mathcal{W}$. Then, by straightforward manipulations, we have
\begin{align*}
	&\mathbb{E}\left[\left(c_\mathcal{W}(W,W') - \mathbb{E}[c_\mathcal{W}(W,W')|W'] - \mathbb{E}[c_\mathcal{W}(W,W')|W] + \mathbb{E}[c_\mathcal{W}(W,W')]\right)c_\mathcal{X}(X,X')\langle Y, Y'\rangle_\mathcal{Y} \right]\\
	=\ &\left|\left| \mathbb{E}\left[\left(c_\mathcal{W}(W,\cdot)-\mathbb{E}[c_\mathcal{W}(W,\cdot)]\right)\otimes c_\mathcal{X}(X,\cdot)\otimes Y\right]\right|\right|^2_{\mathcal{HS}_{\mathcal{W}\mathcal{X}}}\\
	=\ &\left|\left| \mathbb{E}\left[c_\mathcal{W}(W,\cdot)\otimes \left(c_\mathcal{X}(X,\cdot)\otimes Y- \mathbb{E}\left[c_\mathcal{X}(X,\cdot)\otimes Y\right]\right)\right]\right|\right|^2_{\mathcal{HS}_{\mathcal{W}\mathcal{X}}}\\
	=\ &\mathbb{E}\left[c_\mathcal{W}(W,W')\left\langle c_\mathcal{X}(X,\cdot)\otimes Y - \mathbb{E}[c_\mathcal{X}(X,\cdot)\otimes Y], c_\mathcal{X}(X',\cdot)\otimes Y' - \mathbb{E}[c_\mathcal{X}(X,\cdot)\otimes Y]\right\rangle_{\mathcal{HS}_\mathcal{X}} \right],
\end{align*}
and, ultimately,
$$ \text{KCMD}(Y-m(X)|Z) = \text{KCMD}(c_\mathcal{X}(X,\cdot)\otimes Y - \mathbb{E}[c_\mathcal{X}(X,\cdot)\otimes Y]|W).$$

In this simplified scenario, which only holds due to the independence of $X$ and $W$ or, alternatively, under the hypothesis of conditional mean independence of $c_\mathcal{W}(W,w')$ on $X$ for all $w'$ in the support of $W$, there is no need for an estimator of $m$ and, in fact, we can use the U-statistic approach proposed by \cite{Lee2020} and \cite{Lai2021}, and the corresponding bootstrap calibration schemes, to test the conditional mean independence of $c_\mathcal{X}(X,\cdot)\otimes Y$ on $W$ since, to compute their statistic, it is only required to obtain all the pairwise inner products associated with the samples $\{W_i\}_{i=1}^n$ and $\{c_\mathcal{X}(X_i,\cdot)\otimes Y_i\}_{i=1}^n$. We refer the readers to the aforementioned papers for the theoretic and practical aspects of that test.

\section{Practical aspects}
\label{Sec6}

The test proposed and analysed from a theoretical point of view in Sections \ref{Sec2}-\ref{Sec4} needs, in practice, some considerations in order to have the desired level and adequate power, especially if the sample size is small or moderate. As with other specification tests depending on smoothing bandwidth parameters, there is no unique set of rules for the parameter selection problem that is optimal in terms of preserving a fixed level and maximizing the power against every possible alternative hypothesis. Therefore, we heuristically explore in this section the issue of parameter tunning and its impact on the test performance and we provide some suggestions that, apparently, yield satisfactory results in simulations. Since our recommendations hold in the mixed $X$ case discussed in Section \ref{Sec5.1}, we denote the dimensions of the continous and discrete parts of $X$ by $d_C$ and $d_D$, respectively, throughout this section.

\subsection{Bootstrap, bandwidth selection and smoothing kernel}\label{Sec6.2}

The bootstrap-calibrated test depends on two bandwidth parameters $h$ and $\tilde{h}$, as well as a sequence of hyperparameters $\{\kappa_n\}_{n=1}^\infty$. We could simply take $\tilde{h} =  h$ and $\kappa_n= 1$ and the scheme would be consistent. However, for small samples, there would be an over-rejection of the null hypothesis due to the under-smoothing of the regression function
when trying to approximate the distribution of the regression errors via the nonparametric residuals. Instead, a reasonable approach is to use a bandwith selector leading to a pilot bandwith $\tilde{h}$ such that $\tilde{h}\propto n^{-1/(2\nu+d_C)}$, where $\propto$ denotes proportionality and $\nu$ is the order of $K^u$,  which is the optimal rate in terms of mean square error for the aforementioned approximation. Unfortunately, it is not straightforward to prove that the resampling scheme proposed in Section \ref{Sec4}, or the corresponding modification in Section \ref{Sec5.1}, but based on the fitted values $\{\hat{m}_{\tilde{h}}(X_i)\}_{i=1}^n$ instead of the modified fitted values defined in those sections, $\{\hat{m}_{h,\tilde{h}}(X_i)\}_{i=1}^n$, is still consistent. Note that, in any case, $\hat{m}_{h,\tilde{h}}$ is defined as a degenerate convex combination between 
$\hat{m}_{\tilde{h}}$ and another estimator such that, if $\kappa_n$ is large enough, $\hat{m}_{h,\tilde{h}} =\hat{m}_{\tilde{h}}$. For example, we can take $\kappa_n \propto \log(n)$ so that $\kappa_n^2(1/(n\tilde{h}^{d_C})+ \tilde{h}^{2\nu}) = O(\log(n)^2n^{-2\nu/(2\nu+{d_C})}) = o(1)$.

As a selector of the pilot bandwidth $\tilde{h}$, since the idea is to correctly approximate the distribution of the regression errors, sample-based selectors for the Nadaraya-Watson estimator and its modification in the mixed $X$ case, provided they have the asymptotic order mentioned above, are reasonable candidates. For example, we can take
$\tilde{h} = M_{\tilde{h}}h_{\text{GCV}}$ where  $M_{\tilde{h}}>0$,
$$h_{\text{GCV}} = \argmin\limits_{h_0\in [h_l,h_u]} \text{GCV}(h_0),$$
$\text{GCV}$ is the generalised least squares leave-one-out cross validation function defined as
\begin{equation}\label{6.1}\tag{6.1}
	 \text{GCV}(h_0) = \frac{1}{n}\sum\limits_{i=1}^n \left|\left|\frac{Y_i - \hat{m}_{h_0}(X_i)}{1-\frac{1}{n}\sum\limits_{j=1}^n \frac{K(0)}{nh_0\hat{f}_{h_0}(X_j)}}\right|\right|^2,
\end{equation}
$h_l = M_ln^{-1/(2\nu + d_C)}$, $h_u = M_un^{-1/(2\nu + {d_C})}$ and $0<M_l<M_u<\infty$. In fact, due to the bootstrap scheme only focusing on mimicking the behavior of the asymptotically dominant terms in $nU_n$ under $\mathcal{H}_0$, we have found that a slight oversmoothing ($M_{\tilde{h}}>1$) is preferable, particularly for high signal-to-noise scenarios, since an oversmooth estimator of $m(X)$ increases the variability of the resampled residuals in the bootstrap scheme which leads to a heavier right tail in the conditional distribution of $nU^*_n$ and, ultimately, a more conservative test. The theoretical results in the previous sections rely on deterministic bandwidths and, thus, the justification for a sample-based bandwidth selection mechanism is not trivial. Nevertheless, we believe that most of the proofs can be adapted so that the results in Sections \ref{Sec2}-\ref{Sec4} also hold for random bandwidths selected from compact sets such as $[M_ln^{-1/(2\nu + d_C)},M_un^{-1/(2\nu + {d_C})}]$.

To select $h$, a straightforward approach is to take $h\propto h_{\text{GCV}}n^{1/(2\nu+{d_C}) - \eta}$ with $n^{1-2\nu\eta}\rightarrow 0$, $n^{1-d\eta}\rightarrow \infty$. In theory, any $\eta \in \left(1/{2\nu},1/{{d_C}}\right)$ leads to a consistent resampling scheme, although $\eta = 2/(2\nu+{d_C})$ balances the second order moments of some asymptotically negligible terms in $nU_n$ and, therefore, is our suggested choice. The selection of the proportionality constant $M_h$, so that $h = M_hh_{\text{GCV}}n^{-1/(2\nu+{d_C})}$, should be ideally based on the search for maximum power, maintaining the desired significance level $\alpha$. Unfortunately, we cannot provide a universal selection mechanism for this purpose and, to the best of our knowledge,  the problem of bandwidth selection for specification tests based on kernel smoothing remains unsolved in the literature. There are some proposals, such as those of \cite{Horowitz2001} and \cite{Guerre2005} which consider $\sup_{h\in G_n} nU_n(h)$, for some mesh $G_n$, as the test statistic, with the corresponding modification of the resampling procedure. The present paper will not address such procedures; instead, we heuristically derive some proposals in the next section.

As with usual estimation and inference procedures involving kernel smoothing, there are, in practice, no relevant differences in the test performance associated with the choice of the univariate kernel $K^u$, as long as we do not change the kernel order $\nu$ which, in general terms, can be chosen to be the lowest even natural number such that $d_C<2\nu$. For example, for ${d_C}<4$ we use the Epanechnikov kernel, defined as
$$K^{u,2}(x) = \frac{3}{4}(1-x^2)\ind{x^2 \le 1},$$
for  all $x\in [-1,1]$, whereas for $4\le {d_C}<8$ we can rely on the fourth order polynomial kernel (see e.g. \citeauthor{Fan1992}, \citeyear{Fan1992}) defined as 
$$K^{u,4}(x) = \frac{15}{32}(1-x^2)(3-7x^2)\ind{x^2 \le 1}.$$

The bootstrap multipliers $\{r_i\}_{i=1}^n$ must have zero-mean, unit variance and finite $(2+\delta)$th order moment. For example, the two-point distribution proposed in \cite{Mammen1993}, where
$$\mathbb{P}^*\left(r_i = \frac{1-\sqrt{5}}{2}\right) = \frac{5+\sqrt{5}}{10}, \quad \mathbb{P}^*\left(r_i = \frac{1+\sqrt{5}}{2}\right) = \frac{5-\sqrt{5}}{10},$$
can be used in the calibration of our test. 

\subsection{The kernel $c$, the parameter $\sigma^2_c$ and covariate scaling}
\label{Sec6.1}

It is to be expected that different choices of $c$ will lead to different power properties of the correspoding test. In fact, as we have already mentioned in section \ref{Sec1}, the kernel that maximices power against a given family of alternatives might be neither bounded nor characteristic. Moreover, many kernels, such as the Gaussian kernel, depend on one or more hyperparameters. A very interesting discussion on the choice of such hyperparameters for testing purposes against a target family of alternatives can be found in \cite{Lindsay2014}. See also the proposal on kernel selection and parameter tunning in \cite{He2023}. We focus on the rest of the paper on the Gaussian kernel, which is usually regarded as an all-around solid choice, and apply a quantile heuristic to $\sigma_c^2$; that is, to take $\sigma_c^2$ as some sample quantile of the distances between the in-sample values of $Z$.

When the Gaussian kernel is chosen as $c$, another relevant issue is the scale of the regression model covariates. If $||X-X'||_\mathcal{X}^2$ tends to be much larger than $||W-W'||_\mathcal{W}^2$ then $||Z-Z'||_\mathcal{Z}^2 \approx ||X-X'||_\mathcal{X}^2 $. This is problematic since, in such case, $nU_n$ is approximately equal to the analogous statistic used to test whether $\mathbb{E}[Y-m(X)|X] = 0$ a.s., which is obviously true. Thus, we cannot expect an adequate test calibration for small sample sizes, due to the theoretical results stated in the previous sections probably not holding if $Z = X$. If $||X-X'||_\mathcal{X}$ tends to be much smaller than $||W-W'||_\mathcal{W}$, then 
$nU_n$ would be close to the corresponding statistic used to test whether $\mathbb{E}[Y-m(X)|W]$ is almost surely zero. Even if the calibration is correct in this scenario, the test would lack power, for small sample sizes, against alternatives where $\mathbb{E}[g(Z)-m(X)|W] = 0$ but $\mathbb{P}(g(Z)-m(X)\neq 0)>0$. Similarly, the discrepancies in scale between the components of $X = (X^1,\cdots, X^d)$ should also be taken into consideration. If it is possible to decompose $W$ into a collection of covariates with different scales, $W^1,\ldots, W^{\tilde{d}}$ for a given $\tilde{d}\in\mathbb{N}$, such information should also be included in the test.

Denote $\text{med}^X_j = \text{med}\left(||X^j - X^{'j}||_\mathcal{X}^2\right)$ (the median of $||X^j - X^{'j}||_\mathcal{X}^2$) for all $j=1, \ldots, d$, given two i.i.d. copies $X$ and $X'$. Analogously, let $\text{med}^W_k = \text{med}\left(||W^k- W^{'k}||_\mathcal{W}^2\right)$ for all $k=1, \ldots, \tilde{d}$, where $W$ and $W'$ are i.i.d. According to our empirical findings, a more robust test against the aforementioned scale issues can be obtained through a two step scaling procedure. First, define $\tilde{X}$ and $\tilde{W}$ as
$ \tilde{X}^j = X^j/\sqrt{\text{med}^X_j}$ and $\tilde{W}^k = W^k/\sqrt{\text{med}^W_k}$, respectively. Since conditional expectations are preserved through scaling, we suggest to use $\tilde{X}$ to compute both the Nadaraya-Watson estimator of $m(X) = \mathbb{E}[Y|X] = \mathbb{E}[Y|\tilde{X}]$ and the corresponding Parzen-Rosenblatt estimator of the density function of $\tilde{X}$, thus eliminating the scale effect on the bandwidth selection procedure for $\tilde{h}$. Then, define $X^{Sc}$ and $W^{Sc}$ as follows:
$$ X^{Sc} = \frac{\tilde{X}}{\text{med}\left(\sum\limits_{j=1}^d||\tilde{X}^j - \tilde{X}^{'j}||_\mathcal{X}^2\right)^\frac{1}{2}} \quad \text{and} \quad W^{Sc} = \frac{\tilde{W}}{\text{med}\left(\sum\limits_{k=1}^{\tilde{d}}||\tilde{W}^k - \tilde{W}^{'k}||_\mathcal{W}^2\right)^\frac{1}{2}}.$$ 
Denoting $Z^{Sc} = (X^{Sc},W^{Sc})$ we have
\begin{align*}
	||Z^{Sc}-Z^{'Sc}||_\mathcal{Z}^2  =\frac{\sum\limits_{j=1}^d\frac{||X^j - X^{'j}||_\mathcal{X}^2}{\text{med}^X_j}}{\text{med}\left(\sum\limits_{j=1}^d\frac{||X^j - X^{'j}||_\mathcal{X}^2}{\text{med}^X_j}\right)} + \frac{\sum\limits_{j=1}^{\tilde{d}}\frac{||W^j - W^{'j}||_\mathcal{W}^2}{\text{med}^W_j}}{\text{med}\left(\sum\limits_{j=1}^{\tilde{d}}\frac{||W^j - W^{'j}||_\mathcal{W}^2}{\text{med}^W_j}\right)}.
\end{align*}
Thus, $||Z^{Sc}-Z^{'Sc}||_\mathcal{Z}^2$ is a sum of two terms, each of them with median equal to one. In addition, no component of $X^{Sc}$ or $W^{Sc}$ will have a dominant scale over the rest. Then, $\sigma_c^2$ should be computed, following the aforementioned quantile heuristic, based on $Z^{Sc}$. Instead of medians, expected values could be used, although it would be necessary to impose additional conditions on the second order moments of the covariates, to guarantee their existence. In the particular case of discrete components, the medians can be zero, so the expected values are, in fact, a better choice. In any case, the population medians or expectations are unknown so, in practice, the scaling is done on the basis of the usual empirical estimators. 

It is beyond the scope of this paper to prove that the inclusion of estimators for the unknown quantities used in the scaling of the covariates and in the computation of $\sigma_c^2$ is either asymptotically negligible or preserves the structure of the limiting distribution of $nU_n$ and its bootstrap counterpart under $\mathcal{H}_0$, but simulation results suggest that this may be the case.

\subsection{Test sensitivity to hyperparameters and suggested values}
\label{Sec6.3}

This section provides a concise illustration of the impact of both bandwidths $\tilde{h}$ and $h$ as well as the scale parameter $\sigma_c^2$ of the Gaussian kernel in the performance of the test. We rely on the following data generation processes:

\begin{itemize}
 	\item \textbf{DGP1}: $Y = q_X\left(0.7X^1 + 2\hspace{-0.5mm}\cos^2(X^2)\right) + q_W1.5\langle W, \Upsilon \rangle + \mathcal{E}$,
 	\item \textbf{DGP2}: $Y = q_X\left(0.7X^1 + 2\hspace{-0.5mm}\cos^2(X^2)\right) + q_W2\sin(\pi\langle W, \Upsilon \rangle^2) + \mathcal{E}$.
\end{itemize}

In both cases, $X^1, X^2$, $W$ and $\mathcal{E}$ are independent, $X^1, X^2, \mathcal{E} \sim N(0,1)$,
$W\sim BM$, where $BM$ denotes a standard Brownian motion in $[0,1]$ ($\mathcal{W} = L^2(0,1)$) and $\Upsilon(\cdot) = \sin(0.5\hspace{0.5mm}\cdot)\in L^2(0,1)$.
The parameter $q_X\in \{1,4\}$ controls the signal-to-noise ratio of the regression of $Y$ over $X$, measured as $\text{Var}(m(X))/\text{Var}(\mathcal{E})$ which is, approximately, 1 and 16, respectively, whereas $q_W$ indicates whether the model is under the null ($q_W = 0$) or alternative ($q_W= 1$) hypothesis.
The simulations are conducted using a sample sizes of $n=100$,  $n_B = 1000$ bootstrap resamples and $n_{MC}=2000$ Monte Carlo replications and the test is calibrated taking into consideration a significance level of $\alpha=0.05$. Empirical rejection rates are rounded down to three decimal places.

Instead of directly manipulating $\sigma_c^2$, following the quantile heuristic mentioned in Section \ref{Sec6.1}, we consider
$$\sigma_c^2 = \hat{Q}_{||\hat{Z}^{Sc} - \hat{Z}^{'Sc}||_\mathcal{Z}^2}(p),$$
where $p \in \{0.1,0.2,\ldots, 0.8\}$, $\hat{Z}^{Sc}$ denotes the plug-in estimator of $Z^{Sc}$ based on the sample median instead of the true median and $\hat{Q}_{||\hat{Z}^{Sc} - \hat{Z}^{'Sc}||_\mathcal{Z}^2}(p)$ denotes the empirical quantile of order $p$ of the distribution of $||\hat{Z}^{Sc} - \hat{Z}^{'Sc}||_\mathcal{Z}^2$. 

As the univariate kernel $K^u$, we use the Epanechnikov kernel $K^{u,2}$ defined in Section \ref{Sec6.2}. Regarding the bandwidths, we take $h = M_hh_{\text{GCV}}n^{-1/6}$ and
$\tilde{h} = M_{\tilde{h}}h_{\text{GCV}}$ where  
$$h_{\text{GCV}} = \argmin\limits_{h_0\in [h_l,h_u]} \text{GCV}(h_0),$$
$\text{GCV}$ is the generalised least squares leave-one-out cross validation function defined in (\ref{6.1}),
$h_l=0.5n^{-1/6}$, $h_u=5n^{-1/6}$ and $M_{\tilde{h}}, M_h \in \{0.5,0.75,0.9,1,1.1,1.25,1.5,2\}$.
We also take $\kappa_n= 500\log(n)$.

In first place, we fix $M_h=1$, $p=0.5$ and we let $M_{\tilde{h}}$ vary. The empirical rejection rates of the test under the null (common to both data generating processes) and the alternatives in \textbf{DGP1} and \textbf{DGP2} for the different values of $M_{\tilde{h}}$ and $q_X$ are displayed in Table \ref{tab:Mhtilde}. Under the null hypothesis, common to both data generating processes, the test becomes slightly more conservative as we increase the value of $M_{\tilde{h}}$. This is to be expected, according to the discussion in Section \ref{Sec6.2}. The power against the alternatives in both \textbf{DGP1} and \textbf{DGP2} also decreases as the tests becomes more conservative under the null. Remarkably, the test calibration becomes more sensitive to the bandwidth selection as the signal-to-noise ratio increases and moreover, the test loses power agains both alternatives in the high signal-to-noise scenario ($q_X=4$) in comparison with the low signal-to-noise scenario ($q_X=1$). As mentioned in the previous section, this is caused by the bootstrap scheme only replicating the behavior of the asymptotically dominant terms in $nU_n$, whereas the asymptotically negligible terms become more relevant, for moderate sample sizes, precisely when the scale of the noise is small with respect to the error in the local approximation of $m$ by a Taylor polynomial. We have found this behavior to be consistent across multiple scenarios and, as a general rule, we suggest to take $M_{\tilde{h}}=1.25$; that is, $\tilde{h}= 1.25h_{\text{GCV}}$, with the search of $h_{\text{GCV}}$ being restricted to the interval $[h_l,h_u]$, where  $h_l=0.25d_Cn^{-1/(2\nu + d_C)}$ and $h_u=(3+d_C)n^{-1/(2\nu + d_C)}$. Note that, with this criteria, the selected bandwidths are forced to increase with the dimension of the continous part of $X$, which is required to avoid selecting ``extremely low'' values of $h_{\text{GCV}}$ and thus, of $h$, in order to reduce the relevance of some asymptotically negligible terms in $nU_n$. At the same time, $h_u$ prevents $\tilde{h}$ from not decreasing with order $n^{-1/(2\nu +d)}$ as $n$ goes to infinity. Of course, this selection mechanism is based on the scaling produce described in Section \ref{Sec6.1} and must be adapted if alternative scaling methods (or no scaling at all) are considered.

\begin{table}[t]
	 
	\centering
	\renewcommand{\arraystretch}{1.2}
	\begin{tabular}{ccccccccccc}
		\toprule
		&   &   & \multicolumn{8}{c}{$M_{\tilde{h}}$}  \\
		\cmidrule(l){4-11} 
		DGP & $q_W$ & \multicolumn{1}{c}{$q_X$} & \multicolumn{1}{c}{2} & \multicolumn{1}{c}{1.5} & \multicolumn{1}{c}{1.25} & \multicolumn{1}{c}{1.1} & \multicolumn{1}{c}{1} & \multicolumn{1}{c}{0.9} & \multicolumn{1}{c}{0.75} & \multicolumn{1}{c}{0.5} \\
		\midrule
		
		\multirow{2}{*}{\textbf{DGP1}}
		& \multirow{2}{*}{0}
		& 1  & \textbf{\textcolor{blue}{0.039}} & 0.046 & 0.052 & 0.055 & 0.058 & \textbf{\textcolor{red}{0.063}} & \textbf{\textcolor{red}{0.069}} & \textbf{\textcolor{red}{0.085}} \\
		& & 4  & \textbf{\textcolor{blue}{0.013}} & \textbf{\textcolor{blue}{0.032}} & 0.050 & 0.058 & \textbf{\textcolor{red}{0.067}} & \textbf{\textcolor{red}{0.074}} & \textbf{\textcolor{red}{0.086}} & \textbf{\textcolor{red}{0.012}} \\
		\midrule
		
		\multirow{2}{*}{\textbf{DGP1}}
		& \multirow{2}{*}{1}
		& 1  & 0.971 & 0.977 & 0.979 & 0.981 & 0.982 & 0.983 & 0.984 & 0.987 \\
		& & 4  & 0.715 & 0.838 & 0.865 & 0.881 & 0.892 & 0.897 & 0.906 & 0.929 \\
		\midrule
		
		\multirow{2}{*}{\textbf{DGP2}}
		& \multirow{2}{*}{1}
		& 1  & 0.279 & 0.320 & 0.350 & 0.373 & 0.385 & 0.399 & 0.427 & 0.469 \\
		& & 4  & 0.073 & 0.161 & 0.219 & 0.250 & 0.272 & 0.291 & 0.324 & 0.410 \\
		\midrule
		\bottomrule
	\end{tabular} 
	\caption{Rejection rates for different values of $M_{\tilde{h}}$, taking $M_h=1$ and $p=0.5$, under the null and alternative hypothesis in \textbf{DGP1} and \textbf{DGP2}. The test is calibrated with a nominal level of $\alpha = 0.05$. Rejection rates significantly lower (resp. higher) than the nominal level $\alpha$ under $\mathcal{H}_0$, with $95\%$ confidence, are boldfaced in blue (resp. in red).}
	\label{tab:Mhtilde}
\end{table}

Secondly, we take $M_{\tilde{h}}=1.25$, $p=0.5$ and we study the effect of varying $M_h$. The results are displayed in Table \ref{tab:Mh}. As the bandwidth $h$ increases, the Nadaraya-Watson estimator bias may become big enough so that the test detects the lack of conditional mean independence of the Nadaraya-Watson residuals $Y-\hat{m}_{h}(X)$ on the vector covariate $X$. Thus, the empirical level of the test is higher for big values of $M_h$ when $q_X =4$, although it appears to be relatively stable when $q_X =1$. The power of the test also increases with $M_h$ under both alternatives. We suggest to take  $M_h=1.1$ if $d_C<4$ and $M_h=1$ if $d_C\ge4$, which, as we shall see in Section \ref{Sec7}, appears to yield satisfactory results for the considered simulated models with small and moderate sample sizes and dimension $d_C\le 7$. 

\begin{table}[t]
	\vspace{5mm}
	\centering
	\renewcommand{\arraystretch}{1.2}
	\begin{tabular}{ccccccccccc}
		\toprule
		&   &   & \multicolumn{8}{c}{$M_h$}  \\
		\cmidrule(l){4-11} 
		DGP & $q_W$ & \multicolumn{1}{c}{$q_X$} & \multicolumn{1}{c}{2} & \multicolumn{1}{c}{1.5} & \multicolumn{1}{c}{1.25} & \multicolumn{1}{c}{1.1} & \multicolumn{1}{c}{1} & \multicolumn{1}{c}{0.9} & \multicolumn{1}{c}{0.75} & \multicolumn{1}{c}{0.5} \\
		\midrule
		
		\multirow{2}{*}{\textbf{DGP1}}
		& \multirow{2}{*}{0}
		& 1  & 0.059 & 0.053 & 0.053 & 0.051 & 0.052 & 0.049 & 0.047 & 0.042 \\
		& & 4 & \textbf{\textcolor{red}{0.074}} & 0.057 & 0.051 & 0.049 & 0.050 & 0.046 & 0.043 & \textbf{\textcolor{blue}{0.034}} \\
		\midrule
		
		\multirow{2}{*}{\textbf{DGP1}}
		& \multirow{2}{*}{1}
		& 1  & 0.994 & 0.990 & 0.984 & 0.977 & 0.972 & 0.965 & 0.937 & 0.827 \\
		& & 4  & 0.978 & 0.952 & 0.929 & 0.894 & 0.868 & 0.831 & 0.729 & 0.433 \\
		\midrule
		
		\multirow{2}{*}{\textbf{DGP2}}
		& \multirow{2}{*}{1}
		& 1  & 0.426 & 0.388 & 0.366 & 0.358 & 0.346 & 0.326 & 0.303 & 0.219 \\
		& & 4  & 0.366 & 0.310 & 0.278 & 0.253 & 0.229 & 0.208 & 0.170 & 0.100 \\
		\midrule
		\bottomrule
	\end{tabular} 
	\caption{Rejection rates for different values of $M_h$, taking $M_{\tilde{h}}=1.25$ and $p=0.5$, under the null and alternative hypothesis in \textbf{DGP1} and \textbf{DGP2}. The test is calibrated with a nominal level of $\alpha = 0.05$. Rejection rates significantly lower (resp. higher) than the nominal level $\alpha$ under $\mathcal{H}_0$, with $95\%$ confidence, are boldfaced in blue (resp. in red).}
	\label{tab:Mh}
\end{table}
 
Finally, we fix $M_h =1.1$ and $M_{\tilde{h}}=1.25$ in \textbf{DGP1} and \textbf{DGP2}, letting $p$ vary and we display the results in Table \ref{tab:Msigma}. The calibration is accurate for $p\le 0.6$ although for higher values the test slightly over-rejects the null hypothesis for $q_X=4$. Interestingly, for the linear alternative in \textbf{DGP1} the power is maximized for the biggest values of $p$ (where $\sigma^2_c$ is relatively big) whereas for the nonlinear alternative in \textbf{DGP2}, it is preferable to take smaller values of $p$. In general, $p = 0.3$ appears to be a reasonable choice.

\begin{table}[t]
	 
	\centering
	\renewcommand{\arraystretch}{1.2}
	\begin{tabular}{ccrrrrrrrrr}
		\toprule
		&   &   & \multicolumn{8}{c}{$p$}  \\
		\cmidrule(l){4-11} 
		DGP & $q_W$ & \multicolumn{1}{c}{$q_X$} & \multicolumn{1}{c}{0.1} & \multicolumn{1}{c}{0.2} & \multicolumn{1}{c}{0.3} & \multicolumn{1}{c}{0.4} & \multicolumn{1}{c}{0.5} & \multicolumn{1}{c}{0.6} & \multicolumn{1}{c}{0.7} & \multicolumn{1}{c}{0.8} \\
		\midrule
		
		\multirow{2}{*}{\textbf{DGP1}}
		& \multirow{2}{*}{0}
		& 1  & 0.044 & 0.051 & 0.053 & 0.051 & 0.052 & 0.053 & 0.055 & 0.054 \\
		& & 4  & 0.049 & 0.053 & 0.054 & 0.059 & 0.058 & 0.057 & \textbf{\textcolor{red}{0.060}} & \textbf{\textcolor{red}{0.063}}  \\
		\midrule
		
		\multirow{2}{*}{\textbf{DGP1}}
		& \multirow{2}{*}{1}
		& 1  & 0.956 & 0.974 & 0.982 & 0.987 & 0.991 & 0.993 & 0.994 & 0.995 \\
		& & 4  & 0.820 & 0.895 & 0.924 & 0.938 & 0.949 & 0.955 & 0.961 & 0.967 \\
		\midrule
		
		\multirow{2}{*}{\textbf{DGP2}}
		& \multirow{2}{*}{1}
		& 1  & 0.777 & 0.687 & 0.590 & 0.483 & 0.397 & 0.320 & 0.250 & 0.186 \\
		& & 4  & 0.562 & 0.488 & 0.397 & 0.327 & 0.266 & 0.211 & 0.165 & 0.126 \\
		\midrule
		\bottomrule
	\end{tabular} 
	\caption{Rejection rates for different values of $p$, taking $M_{\tilde{h}}=1.25$ and $M_h=1.1$, under the null and alternative hypothesis in \textbf{DGP1} and \textbf{DGP2}. The test is calibrated with a nominal level of $\alpha = 0.05$. Rejection rates significantly higher than the nominal level $\alpha$ under $\mathcal{H}_0$, with $95\%$ confidence, are boldfaced in red.}
	\label{tab:Msigma}
\end{table}

\section{Simulation study}
\label{Sec7}

In this section we conduct simulations to illustrate the performance of the proposed test for small and moderate sample sizes: $n\in \{50,100,250\}$. We generate data according to multiple models including scalar and functional responses as well as scalar, discrete and functional covariates. We consider, $n_B = 1000$ bootstrap resamples and $n_{MC} = 2000$ Monte Carlo replications. The test is calibrated according to the significance levels $\alpha = 0.05$ and $\alpha = 0.1$ and the empirical rejection rates are rounded down to three decimal places.

Following the discussion in Section \ref{Sec6}, we take: $\tilde{h}=1.25h_{\text{GCV}}$, where $h_{\text{GCV}}$ is the generalised least squares leave-one-out cross validation bandwidth restricted to the interval $[h_l,h_u]$ with $h_l=0.25d_Cn^{-1/(2\nu + d_C)}$ and $h_u=(3+d_C)n^{-1/(2\nu + d_C)}$; $h = \left(1.1-0.1\ind{d_C\ge 4}\right)\tilde{h}n^{-1/(2\nu + d_C)}$; $\kappa_n= 500\log(n)$  and $\sigma_c^2 = \hat{Q}_{||\hat{Z}^{Sc} - \hat{Z}^{'Sc}||_\mathcal{Z}^2}(0.3)$ where $\hat{Z}^{Sc}$ stands for the scaled covariates defined as in Section \ref{Sec6.1}. We use the Gaussian kernel as the charactertistic kernel $c$ and the Epanechnikov kernel $K^{u,2}$ and the fourth order kernel $K^{u,4}$, both defined in Section \ref{Sec6.2}, as the univariate kernel $K^u$ for $d_C\le3$ and $4\le d_C<7$, respectively. The functional data is assumed to take values in $L^2(0,1)$ and it is represented in practice by direct evaluation on an equispaced mesh of the interval $[0,1]$ with a step size of $0.001$. The approximation of the inner products and norms in $L^2(0,1)$ is based on the trapezoidal rule. 

We denote $W\sim BB$ if $W$ is a standard Brownian bridge on $[0,1]$ and $W\sim OU$ if $W$ is an Ornstein-Uhlenbeck process with mean 0, also in $[0,1]$, with a mean-reversion coefficient of $4.5$, an infinitesimal volatility parameter of $3$ and initial value $W(0) = 0$. In the latter case, $W$ is a Gaussian process such that $\mathbb{E}[W(t)|W(0)] = 0$ for all $t\in [0,1]$ and $\text{Cov}(W(t),W(s)|W(0)) = e^{-4.5|t-s|} - e^{-4.5(t+s)}$. We also denote by $t_1$ the Student's $t$ distribution with 1 degree of freedom, by $\chi_1^2$ the chi-squared distribution with 1 degree of freedom and by $N_2(\mu,\Sigma)$ the bivariate Gaussian distribution with mean $\mu$ and covariance matrix $\Sigma$. We define the element $\Upsilon\in L^2(0,1)$, based on Example 1 in \cite{Lee2020}, so that $\Upsilon = \sum\limits_{l=1}^{100} \xi_l\upsilon_l/||\boldsymbol{\upsilon}||_{\mathbb{R}^{100}}$, with $\xi_1(t) = 1$ and $\xi_l (t) = \frac{1}{2}\cos((l-1)\pi t)$ for $l>1$ and for all $t \in [0,1]$, $\boldsymbol{\upsilon} = (\upsilon_1,\ldots, \upsilon_{100})^\top$, $\upsilon_1 = 0.3$ and $\upsilon_l = 4 (-1)^l/l^2$ for $l>1$.

\subsection{Scalar response}
\label{Sec7.1}

This section presents the analysis of the test in terms of level and power for six regression models with scalar response and a functional covariate. In all cases, the models correspond to the expression
$$Y = m(X) + q_W\tilde{g}(Z) + \mathcal{E},$$
where $q_W\in\{0,1\}$, $\mathbb{E}[\tilde{g}(Z)|X] = 0$ and $\mathbb{P}(\tilde{g}(Z) \neq 0)>0$. Therefore, $\mathcal{H}_0$ holds when $q_W= 0$ and $\mathcal{H}_1$ holds when $q_W= 1$. To facilitate the comparability of results for different models, the signal-to-noise ratios verify
$$\frac{\text{Var}(m(X))}{\text{Var}(\mathcal{E})}\approx 4  \quad \text{and} \quad \frac{\text{Var}(\tilde{g}(Z))}{\text{Var}(\mathcal{E})}\approx 0.5.$$
In particular, we consider the following data generating processes:
 
\begin{itemize}
	\item \textbf{S.1}: $Y = 2X + q_W4.3\langle W,\Upsilon\rangle_{L^2(0,1)} + \mathcal{E}$, where $X, W$ and $\mathcal{E}$ are independent, $\mathcal{E}\sim (\chi_1^2-1)/\sqrt{2}$, $W \sim BB$ and $X\sim N(0,1)$.
	\item  \textbf{S.2}: $Y = X^1+X^2+X^3+X^4+ q_W4.3\langle W,\Upsilon\rangle_{L^2(0,1)} + \mathcal{E}$, where $X, W$ and $\mathcal{E}$ are independent, \\$\mathcal{E}\sim (\chi_1^2-1)/\sqrt{2}$, $W \sim BB$ and $X^1, X^2, X^3, X^4\sim N(0,1)$.
	\item  \textbf{S.3}: $Y = 2\left(\sum\limits_{j=1}^7X^j\right)/\sqrt{7} + q_W4.3\langle W,\Upsilon\rangle_{L^2(0,1)} + \mathcal{E}$, where $X, W$ and $\mathcal{E}$ are independent, \\ $\mathcal{E}\sim (\chi_1^2-1)/\sqrt{2}$, $W \sim BB$ and $X^1, \ldots, X^7\sim N(0,1)$.
	\item  \textbf{S.4}: $Y = 3.35\left(||X||_{\mathbb{R}^3}^2-2/3\right) + q_W4.4\left(||W||_{L^2(0,1)}^2-1/3\right)+ \mathcal{E}$, where $Z$ and $\mathcal{E}$ are independent,\\ $\mathcal{E}\sim N(0,1)$, $Z\sim OU$, $X^1 = \left\langle Z, \ind{ \cdot \in [0,1]}\right\rangle_{L^2(0,1)}$, $X^2= \left\langle Z, \sqrt{2}\cos(2\pi \cdot)\right\rangle_{L^2(0,1)}$,  \\$X^3=\left\langle Z, \sqrt{2}\sin(2\pi  \cdot)\right\rangle_{L^2(0,1)}$ and $W = Z - X^1\ind{\cdot\in [0,1]} - X^2\sqrt{2}\cos(2\pi  \cdot) - X^3\sqrt{2}\sin(2\pi  \cdot)$.
	\item  \textbf{S.5}: $Y = 5.2e^{-|X|^2} + q_W3.8\left(e^{-||W||_{L^2(0,1)}^2} - 0.42\right)+ \mathcal{E}$, where $X, W$ and $\mathcal{E}$ are independent, $\mathcal{E}\sim N(0,1)$, $W \sim OU$ and $X \sim t_1$. 
	\item  \textbf{S.6}: $Y = 2(\sin(\pi X^1) + 2\cos(\pi X^2))+ q_W\sin\left(\pi \langle W,\Upsilon\rangle_{L^2(0,1)}\right) +\mathcal{E}$, where $X$ and $W$ are independent,\\ $\mathcal{E}\sim 1.7/(1+||X||^2)N(0,1)$, $W \sim OU$ and $X\sim N_2(0,\Sigma)$ verifying $[\Sigma]_{11} = 1$, $[\Sigma]_{22} = 1$ and \\ $[\Sigma]_{12} = [\Sigma]_{21} = 0.9$.
\end{itemize}

In models  \textbf{S.1} -  \textbf{S.6}, $X$ is an absolutely continuous random vector covariate, with dimensions $d=1$, $d=2$, $d=4$ and $d=7$ (in all cases, $d_C=d$). Models \textbf{S.1} - \textbf{S.3} are linear, both under $\mathcal{H}_0$ and under $\mathcal{H}_1$, and examine the impact of increasing the dimension of the covariates under the null. Im model \textbf{S.4}, $X$ and $W$ are strongly related and features quadratic regression functions. Models \textbf{S.5} and \textbf{S.6} also present non-linear regression functions, considering a heavy tailed $X$ in the former case and both a dependence structure between the components of $X$ and conditional heteroscedasticity on the error term in the latter case.

Table \ref{tab4} shows the rejection rates for the data generating processes \textbf{S.1} -  \textbf{S.6}. On the one hand, the test generally preserves the chosen significance levels,  particularly for $n= 250$. There is, however a slight over-rejection of the null hypothesis in \textbf{S.3} for $n\le 100$. On the other hand, the proportion of rejections under $\mathcal{H}_1$ clearly increases with the sample size. Moreover, the power of the test does decrease with the dimension of $X$, as the results in \textbf{S.2} and \textbf{S.3} suggest, although such power loss is relatively mild, highlighting the robustness of this test against the curse of dimensionality. This, of course, does not mean that our testing procedure can be used in arbitrary high-dimension low-sample size scenarios. The bootstrap scheme requires the nonparametric residuals to have a distribution ``close enough'' to that of the unknown model errors to properly calibrate the test, meaning that, in practice, there is still a non-evident restriction on the dimension of $X$ associated with $n$. Furthermore, the test becomes more sensitive to the selection of the bandwidths $h$ and $\tilde{h}$ as the dimension increases, which is quite undesirable. Finally, nonlinear alternatives seem to be more difficult to identify than linear alternatives, although the convergence of the empirical power to 1 remains quite fast for \textbf{S.4} - \textbf{S.6}.

\begin{table}[t]
	\centering
	\renewcommand{\arraystretch}{1.2}
	\setlength{\tabcolsep}{4.8pt}
	\begin{tabular}{ccrrrrrrrrrrrr}
		\toprule
		& \hspace{0.3mm} $q_W$ & \multicolumn{6}{c}{0} & \multicolumn{6}{c}{1} \\
		\cmidrule(l){2-2}	\cmidrule(l){3-8} \cmidrule(l){9-14}
		& \hspace{0.3mm} $\alpha$ & \multicolumn{3}{c}{0.05} & \multicolumn{3}{c}{0.1} & \multicolumn{3}{c}{0.05} & \multicolumn{3}{c}{0.1} \\
		\cmidrule(l){2-2}	\cmidrule(l){3-5} \cmidrule(l){6-8} \cmidrule(l){9-11} \cmidrule(l){12-14}
		DGP & \hspace{0.3mm} $n$ &\multicolumn{1}{c}{50} &\multicolumn{1}{c}{100} & \multicolumn{1}{c}{250} & \multicolumn{1}{c}{50} &\multicolumn{1}{c}{100} & \multicolumn{1}{c}{250} &\multicolumn{1}{c}{50} &\multicolumn{1}{c}{100} & \multicolumn{1}{c}{250} & \multicolumn{1}{c}{50} &\multicolumn{1}{c}{100} & \multicolumn{1}{c}{250} \\
		\midrule
		
		\multirow{1}{*}{\textbf{S.1}}
		& & 0.054 & 0.059  & 0.055 & 0.103 & 0.106 & 0.099 & 0.802 & 0.970 & 1.000 & 0.864 & 0.983 & 1.000\\
		\midrule
		
		\multirow{1}{*}{\textbf{S.2}}
		& & 0.049 & 0.052 & 0.051 & 0.100 & 0.108 & 0.109 & 0.560 & 0.867  & 0.992 & 0.679 & 0.922 & 0.995\\
		\midrule
		
		\multirow{1}{*}{\textbf{S.3}}
		& & \textbf{\textcolor{red}{0.066}} & 0.053 & 0.044 & \textbf{\textcolor{red}{0.127}} & \textbf{\textcolor{red}{0.121}} & 0.103 & 0.438 & 0.724 & 0.972 & 0.563 & 0.816 & 0.985\\
		\midrule
		
		\multirow{1}{*}{\textbf{S.4}}
		& & \textbf{\textcolor{blue}{0.028}} & \textbf{\textcolor{blue}{0.032}} & 0.050 & \textbf{\textcolor{blue}{0.074}} & \textbf{\textcolor{blue}{0.080}} & 0.103 & 0.138 & 0.311  & 0.811 & 0.268 & 0.465 & 0.906\\
		\midrule
		
		\multirow{1}{*}{\textbf{S.5}}
		& & 0.050 & 0.045 & 0.059 & 0.102 & 0.088 & 0.105 & 0.270 & 0.673 & 0.995 & 0.429 & 0.832 & 0.998\\
		\midrule
		
		\multirow{1}{*}{\textbf{S.6}}
		& & 0.042 & \textbf{\textcolor{blue}{0.040}}  & 0.050 & \textbf{\textcolor{blue}{0.086}} & 0.096 & 0.092 & 0.232 & 0.532  & 0.962 & 0.355 & 0.680 & 0.984\\
		\bottomrule
	\end{tabular}
	\caption{\label{tab4}Rejection rates for the data generating processes (DGP) \textbf{S.1} - \textbf{S.6} under the null ($q_W= 0$) and alternative ($q_W=1$) hypothesis. Three sample sizes ($n=50$, $n=100$ and $n=250$) and two significance levels ($\alpha= 0.05$ and $\alpha= 0.1$) are considered for each scenario. Rejection rates significantly lower (resp. higher) than the nominal level $\alpha$ under $\mathcal{H}_0$, with $95\%$ confidence, are boldfaced in blue (resp. in red).}
\end{table}

\subsection{Functional response}
\label{Sec7.2}

In this section, six different scenarios with functional response are considered to illustrate the empirical level and power of the test for small and moderate sample sizes. As in the scalar case, every model satisfies the equation
$$Y = m(X) + q_W\tilde{g}(Z) + \mathcal{E},$$
where $q_W\in\{0,1\}$ although, in this case, we only require $\mathbb{E}[\tilde{g}(Z)|X] = y$ and $\mathbb{P}(\tilde{g}(Z) \neq y)>0$ for some $y\in L^2(0,1)$ independent of $X$. With this definition, if $y\neq 0$, then $\mathbb{E}[Y|X] = m(X) + y$ under the alternative and, thus, the regression function of $Y$ on $X$ under $\mathcal{H}_0$ and $\mathcal{H}_1$ differs by a constant. However, since the Nadaraya-Watson estimator is a linear smoother, this is not relevant for the test performance discussion. In addition, we assume that
$$\frac{\int_0^1 \text{Var}(m(X))(t) dt}{\int_0^1 \text{Var}(\mathcal{E})(t) dt}\approx 4 \quad \text{y} \quad \frac{ \int_0^1 \text{Var}(\tilde{g}(Z))(t) dt}{\int_0^1 \text{Var}(\mathcal{E})(t)dt}\approx 0.5 ,$$
except in one particular model where $m(X) = 0$. 

The selected data generating processes are defined as follows:

\begin{itemize}
\item \textbf{F.1}: $Y(t) = 1.3e^{-4(t-0.3)^2}X + q_W0.55tW + \mathcal{E}$, where $Z$ and $\mathcal{E}$ are independent,  $\mathcal{E}\sim 2.45BB$, \\ $Z \sim -\log\left(N_2((1,1),\Sigma)^2\right)$ where  $[\Sigma]_{11} = 1$, $[\Sigma]_{22} = 1$ and  $[\Sigma]_{12} = [\Sigma]_{21} = 0.9$, $X = Z^1$ and $W = Z^2$.
\item \textbf{F.2}: $Y(t) = 25.5(t-0.5)^2X^1/(1+|X^2|^2) + q_W1.1X^1\sin(1.75\pi tW) + \mathcal{E}$, where $X, W$ and $\mathcal{E}$ are independent, $\mathcal{E}\sim 2.45BB$ and $W,X^1,X^2 \sim N(0,1)$. 
\item \textbf{F.3}: $Y(t) = 6.6t\sin(2|X|^2) + q_W1.75\int_0^1e^{(t^2+s^2)/2}W(s)ds + \mathcal{E}$, where $X, W$ and $\mathcal{E}$ are independent, $\mathcal{E}\sim 2.45BB$, $W \sim BB$ and $X\sim N(0,1)$. 
\item \textbf{F.4}: $Y(t) = q_W0.8\int_0^1e^{(t^2+s^2)/2}|W(s)|^2ds + \mathcal{E}$, where $X, W$ and $\mathcal{E}$ are independent, $\mathcal{E}, W \sim OU$ and $X^1,X^2,X^3\sim N(0,1)$. 
\item \textbf{F.5}: $Y(t) = 2.3(1-tX^1 - t^2|X^2|^2) +  q_W0.33\exp(W(t)) + \mathcal{E}$, where $X^1, X^2, W$ and $\mathcal{E}$ are independent, $\mathcal{E}, W \sim OU$ and $X^1, X^2 \sim N(0,1)$.
\item \textbf{F.6}: $Y(t) = 0.23e^{-4(t-0.3)^2}X^1X^2(0.95 + 0.3q_W(\ind{W=w_1}-\ind{W\neq w_1})) + \mathcal{E}$, where $X, W$ and $\mathcal{E}$ are independent, $\mathcal{E}\sim OU$, $W\in \{w_1,w_2,w_3,w_4\}$ with $\mathbb{P}(W = w_1)=0.5$, $\mathbb{P}(W = w_2)=0.25$,\\ $\mathbb{P}(W = w_3)=0.15$ and $\mathbb{P}(W = w_4)=0.1$ , $X^1\sim \text{Bernoulli}(0.3)$ and $\log(X^2)\sim N(3,0.5)$. 
\end{itemize}

\textbf{F.1} and \textbf{F.2} feature a functional response and scalar covariates, whereas in \textbf{F.3} - \textbf{F.5} we have included both a functional response and a functional covariate and in \textbf{F.6} there are discrete variables, both in $X$ and $W$. The null hypothesis in \textbf{F.1} is based on Example 3 in \cite{Lee2020}, involves a linear alternative and we have included a dependence structure between $X$ and $W$. \textbf{F.2} presents a different regression model under the null hyphotesis and a nonlinear interaction effect between $X^1$ and $W$ under the alternative. The alternative hypotheses in \textbf{F.3} and \textbf{F.4}, linear and quadratic respectively, are based on Example 7 of \cite{Lai2021}. Moreover, in $\textbf{F.4}$ the regression function of $Y$ on $X$ is constantly zero. In \textbf{F.5} a nonlinear concurrent alternative is introduced. The null in \textbf{F.6} is a modification of that in \textbf{F.1} and \textbf{F.2} and the alternative defines an interaction effect between $X^1$, $X^2$ and $W$. In the latter scenario, $X^1$ in straightforwardly embedded into $\mathbb{R}$ and $W$ is one-hot-encoded, as suggested in Section \ref{Sec5.2}.

Table \ref{tab6} shows the rejection rates for models \textbf{F.1} - \textbf{F.6}. As in the scalar response case, the test generally preserves the significance level under the null and the power increases with the sample size in all scenarios. However, this increase is remarkably slower in \textbf{F.2} and \textbf{F.6} which correspond to alternatives hypothesis with interaction effects between some components of $X$ and $W$.

\begin{table}[t]
\centering
\renewcommand{\arraystretch}{1.2}
\setlength{\tabcolsep}{4.8pt}
\begin{tabular}{ccrrrrrrrrrrrr}
	\toprule
	& \hspace{0.3mm} $q_W$ & \multicolumn{6}{c}{0} & \multicolumn{6}{c}{1} \\
	\cmidrule(l){2-2}	\cmidrule(l){3-8} \cmidrule(l){9-14}
	& \hspace{0.3mm} $\alpha$ & \multicolumn{3}{c}{0.05} & \multicolumn{3}{c}{0.1} & \multicolumn{3}{c}{0.05} & \multicolumn{3}{c}{0.1} \\
	\cmidrule(l){2-2}	\cmidrule(l){3-5} \cmidrule(l){6-8} \cmidrule(l){9-11} \cmidrule(l){12-14}
	DGP & \hspace{0.3mm} $n$ &\multicolumn{1}{c}{50} &\multicolumn{1}{c}{100} & \multicolumn{1}{c}{250} & \multicolumn{1}{c}{50} &\multicolumn{1}{c}{100} & \multicolumn{1}{c}{250} &\multicolumn{1}{c}{50} &\multicolumn{1}{c}{100} & \multicolumn{1}{c}{250} & \multicolumn{1}{c}{50} &\multicolumn{1}{c}{100} & \multicolumn{1}{c}{250} \\
	\midrule
	
	\multirow{1}{*}{\textbf{F.1}}
	& & 0.059 & 0.050  & 0.050 & \textbf{\textcolor{red}{0.123}} & 0.100 & 0.093 & 0.347 & 0.702  & 0.992 & 0.498 & 0.822 & 0.997\\
	\midrule
	
	\multirow{1}{*}{\textbf{F.2}}
	& & 0.057 & 0.051  & 0.054 & 0.108 & 0.101 & 0.107 & 0.105 & 0.139  & 0.393 & 0.183 & 0.257 & 0.596\\
	\midrule
	
	\multirow{1}{*}{\textbf{F.3}}
	& & 0.055 & 0.055 & 0.059 & 0.106 & 0.103 & \textbf{\textcolor{red}{0.114}} & 0.931 & 1.000 & 1.000 & 0.963 & 1.000 & 1.000\\
	\midrule
	
	\multirow{1}{*}{\textbf{F.4}}
	& & 0.053 & 0.048 & 0.046 & 0.111 & 0.095 & 0.099 & 0.610 & 0.962  & 1.000 & 0.770 & 0.990 & 1.000\\
	\midrule
	
	\multirow{1}{*}{\textbf{F.5}}
	& & 0.049 & \textbf{\textcolor{blue}{0.038}} & 0.049 & 0.100 & 0.093 & 0.100 & 0.675 & 0.971 & 1.000 & 0.793 & 0.987 & 1.000\\
	\midrule
	
	\multirow{1}{*}{\textbf{F.6}}
	& & 0.059 & 0.053  & 0.040 & 0.109 & 0.103 & 0.096 & 0.239 & 0.380  & 0.782 & 0.325 & 0.489 & 0.867\\
	\bottomrule
\end{tabular}
\caption{\label{tab6}Rejection rates for the data generating processes (DGP) \textbf{F.1} - \textbf{F.6} under the null ($q_W= 0$) and alternative ($q_W=1$) hypothesis. Three sample sizes ($n=50$, $n=100$ and $n=250$) and two significance levels ($\alpha= 0.05$ and $\alpha= 0.1$) are considered for each scenario. Rejection rates significantly lower (resp. higher) than the nominal level $\alpha$ under $\mathcal{H}_0$, with $95\%$ confidence, are boldfaced in blue (resp. in red).}
\end{table}

\section{Real data}
\label{Sec8}

We apply our test to a real dataset featuring a functional response and a functional covariate: the Hawaii Ocean Time-series data. This dataset is available from the Hawaii Ocean Time-series program's official website (http://hahana.soest.hawaii.edu/hot/hot-dogs/cextraction.html) and includes repeated measurements of the oceanic water's chemical, physical, and biological profiles, taken in relation to hydrostatic pressure (and, therefore, depth) at the ALOHA station in Hawaii. For this analysis, we focus on oxygen concentration (\SI{}{\micro\mol}/kg) and chloropigment concentration (\SI{}{\micro\g}/l), measured every 2 decibars of pressure from 2 to 200 decibars (approximately corresponding to depths between 1.98 and 198.6 meters). We use measurements collected from January 1, 2010, to December 31, 2022.

\begin{figure}[t]
\centering
\includegraphics[width=\textwidth]{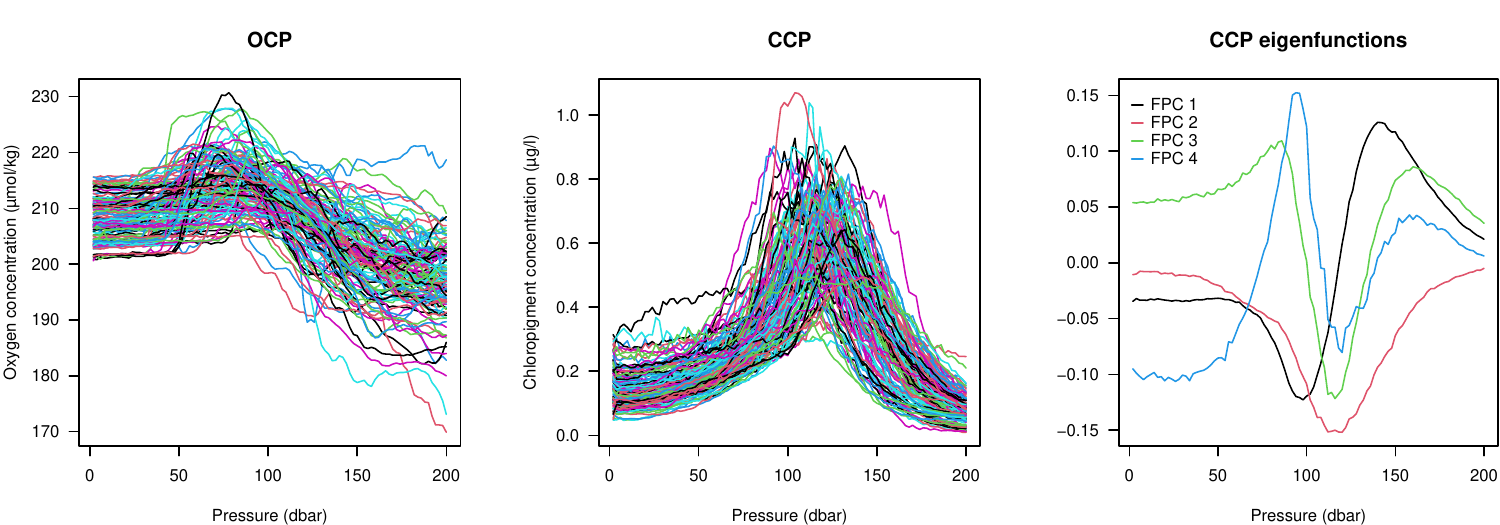}
\caption{From left to right: oxygen concentration profiles (OCP), chloropigment concentration profiles (CCP) and the first four estimated functional principal component (FPC) eigenfunctions of the CCP.}
\label{fig1}
\end{figure}

A high concentration of chloropigments indicates the presence of photosynthetic organisms, which influence oxygen concentration levels as both a source (through photosynthesis) and a sink (through respiration and bacterial decomposition of organic matter). Our primary objective is to explain the conditional mean behavior of the oceanic oxygen concentration profile (OCP) concerning the chloropigment concentration profile (CCP), both shown in Figure \ref{fig1} (left and center). After removing some curves with measurement errors (negative values), we have a sample size of $n=118$ and we assume that both functional variables take values in the $L^2(0,1)$ space. The independence hypothesis is clearly not fulfilled in this context, although the time span between profile measurements (generally over a month) is likely wide enough to assume a weak dependence structure, under which we believe Theorems \ref{Th1} and \ref{Th3} still hold.

\begin{figure}[t]
\centering
\includegraphics[width=11.5cm]{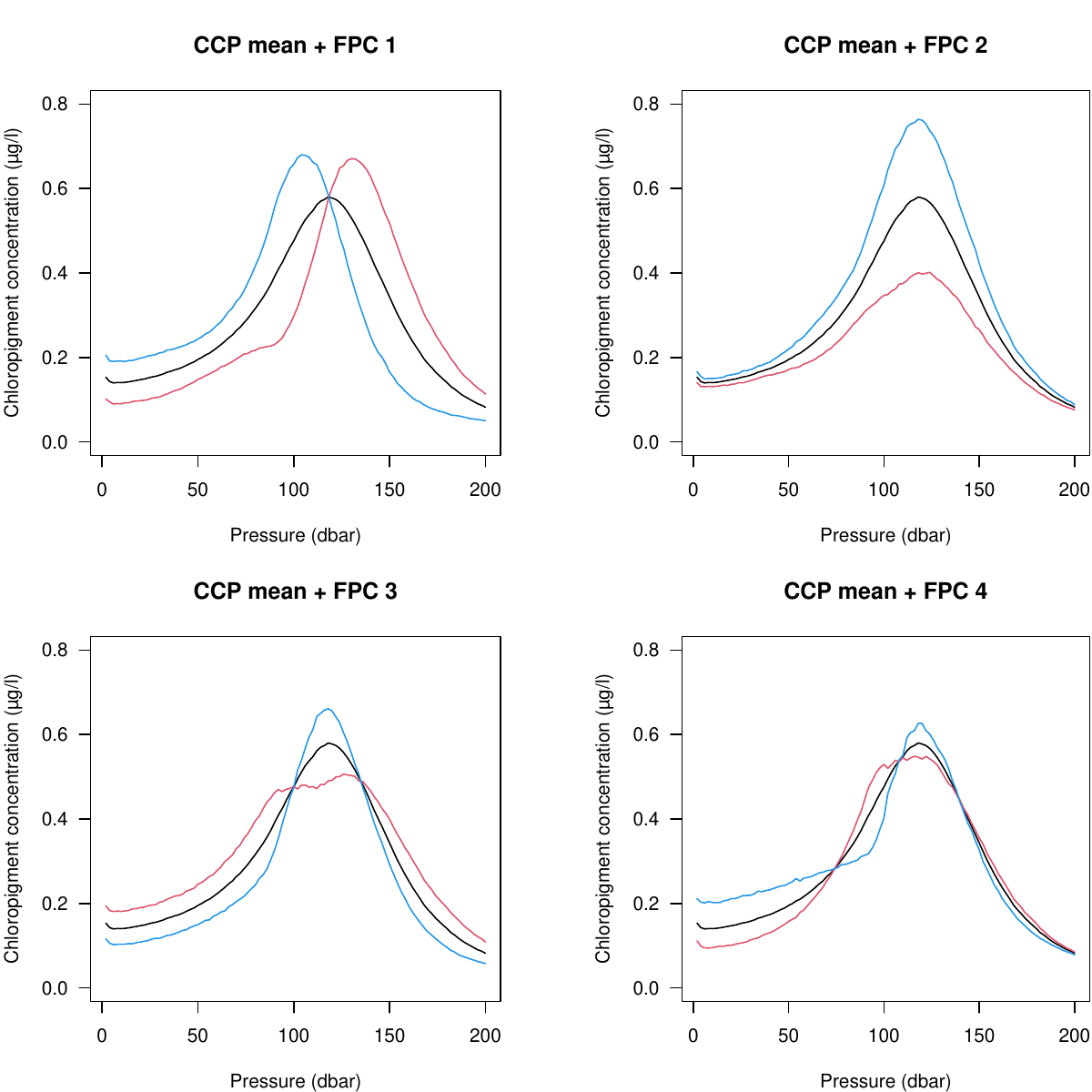}
\caption{Empirical mean curve (black) and perturbations due to the addition of the 0.05 (blue) and 0.95 (red) quantiles of the first (top left), second (top right), third (bottom left) and fourth (bottom right) FPC scores of the CCP, multiplied by the corresponding eigenfuncions.}
\label{fig2}
\end{figure}

The first question we should answer in this context is whether there is statistical evidence of an effect of the CCP in the average behavior of the OCP. To this end, we apply the bootstrap-calibrated KCMD test of
\cite{Lai2021}, based on the characterization of the conditional mean independence between the OCP and the CCP through $\text{KCMD}(\text{OCP}|\text{CCP})$, where the $\text{KCMD}$ is defined as in Section \ref{Sec1}. Since we obtain a $p$-value of 0, with a total of $n_B=5000$ bootstrap resamples, we can reject the null hyphotesis of conditional mean independence and establish a relation between this two functional variables through a regression model. In particular, we try to fit a simplified nonparametric regression model where, instead of using the entire covariate curve, we take into account the scores of some functional principal components (FPCs). The first four estimated FPCs of the CCP explain over 90$\%$ of the total variability. These FPC eigenfunctions are represented in Figure \ref{fig1} (right). Note that the estimated FPC scores also suffer from estimation-induced dependence, an issue that requires further theoretical justification and will be addressed in a future paper. However, note that the same procedure can be applied to any alternative functional basis. Particularly, some relevant profile characteristics may be captured by projecting of the CCP onto an ad hoc set of linearly independent directions, proposed by subject-matter experts, which can be further extended to a basis of the $L^2(0,1)$ space. The functional principal component basis was selected for its interpretability in this case. To illustrate the interpretation, we have plotted in Figure \ref{fig2} the CCP mean curve along with the addition of the FPC eigenfunctions multiplied by the 0.05 and 0.95 level quantiles of the corresponding scores. For instance, the first FPC determines the position of the maximum chloropigment concentration over the curve domain whereas the second FPC is a measurement of the scale of this variable. The third and fourth FPCs influence the concentration increase and decrease steepness.

Although the conditional mean information is not necessarily contained in any of the first FPCs, given the interpretation mentioned earlier, we find it reasonable to test if, given the aforementioned FPC scores, the ``CCP remainder'' is superfluous on the regression model of the OCP on the CCP. That is, denoting by $Y$ the OCP, by $X$ the first four FPC scores of the CCP and letting $W$ be defined as
$$W = \text{CCP} - \sum\limits_{j=1}^4\langle \text{CCP},\phi_j \rangle\phi_j ,$$
where $\phi_j$ denotes $j$th CCP FPC eigenfunction, 
we want to test the null hypothesis of $\mathbb{E}[Y-m(X)|(X,W)] = 0$ a.s. Note that, since both the Gaussian characteristic kernel and the Nadaraya-Watson estimators are invariant with respect to shifts in the distribution of $X$ and $W$, the test results are not affected by centering the covariates; all relevant information is contained in the in-sample distances. Then, applying our test proposal, taking into account the suggestions in Section \ref{Sec6}, with $n_B=5000$ bootstrap replicates, we obtain a p-value of $0.357$ and, thus, we do not find evidence against the nonparametric simplified model that only considers the first four FPC scores of the CCP as covariates.

\begin{remark}\label{remark4}
	Since we have performed our test conditionally on the rejection of the previous null of conditional mean independence of the OCP on the CCP it is formally correct to control type I errors independently for each test. Indeed, given two null hypothesis $\mathcal{H}^1_0$ and $\mathcal{H}^2_0$ such that $\mathcal{H}^2_0$ is only tested if $\mathcal{H}^1_0$  is rejected, we have
	$$\mathbb{P}\left(\text{Reject any true } \mathcal{H}^j_0, j\in\{1,2\}| \mathcal{H}^1_0 \text{ is true}\right) = \mathbb{P}\left(\text{Reject } \mathcal{H}^1_0| \mathcal{H}^1_0 \text{ is true}\right), $$
	and, moreover, assuming that the calibration of the test of $\mathcal{H}^2_0$ is independent of $\mathcal{H}^1_0$,
	$$\mathbb{P}\left(\text{Reject } \mathcal{H}^2_0 | \mathcal{H}^1_0 \text{ is false}, \mathcal{H}^2_0 \text{ is true}\right) = \mathbb{P}\left(\text{Reject } \mathcal{H}^2_0 | \mathcal{H}^2_0 \text{ is true}\right).$$
\end{remark}

A natural question would then be: can we further simplify the model? If we were to eliminate any of the first four FPCs, which one should we drop drop from the model? We apply a backward elimination process based on our test to try to answer those questions. In particular, we start by applying our test in the following four scenarios, where $Y$ is still the OCP:
\begin{itemize}
	\item $X = ( \langle \text{CCP},\phi_2 \rangle,  \langle \text{CCP},\phi_3 \rangle,  \langle \text{CCP},\phi_4 \rangle )$ and $W = \langle \text{CCP},\phi_1 \rangle$ $\leadsto$ $p$-value = $0$.
	\item $X = ( \langle \text{CCP},\phi_1 \rangle,  \langle \text{CCP},\phi_3 \rangle,  \langle \text{CCP},\phi_4 \rangle )$ and $W = \langle \text{CCP},\phi_2 \rangle$ $\leadsto$ $p$-value = $0.714$.
	\item $X = ( \langle \text{CCP},\phi_1 \rangle,  \langle \text{CCP},\phi_2 \rangle,  \langle \text{CCP},\phi_4 \rangle )$ and $W = \langle \text{CCP},\phi_3 \rangle$ $\leadsto$ $p$-value = $0.309$.
	\item $X = ( \langle \text{CCP},\phi_1 \rangle,  \langle \text{CCP},\phi_2 \rangle,  \langle \text{CCP},\phi_3 \rangle )$ and $W = \langle \text{CCP},\phi_4 \rangle$ $\leadsto$ $p$-value = $0.092$.
\end{itemize}
Considering a significance level of $\alpha=0.01$ (see Remark \ref{remark5}) we cannot reject the null hypothesis of covariate redundancy for the second, third or fourth scores. We eliminate the second score since it obtained the largest p-value. Similarly, considering only two FPCs under the null hyphotesis, we obtained the following results:
\begin{itemize}
	\item $X = (\langle \text{CCP},\phi_3 \rangle,  \langle \text{CCP},\phi_4 \rangle)$ and $W = \langle \text{CCP},\phi_1 \rangle$ $\leadsto$ $p$-value = $0$.
	\item $X = ( \langle \text{CCP},\phi_1 \rangle,  \langle \text{CCP},\phi_4 \rangle)$ and $W = \langle \text{CCP},\phi_3 \rangle$ $\leadsto$ $p$-value = $0.092$.
	\item $X = ( \langle \text{CCP},\phi_1 \rangle, \langle \text{CCP},\phi_3 \rangle)$ and $W = \langle \text{CCP},\phi_4 \rangle$ $\leadsto$ $p$-value = $0.036$.
\end{itemize}
According to these results, there is no evidence against removing the third FPC from the nonparametric regression model. Finally, the $p$-values obtained from the analogous model simplification tests where $X$ is either $\langle \text{CCP},\phi_1 \rangle$ or $\langle \text{CCP},\phi_4 \rangle$ and $W$ the corresponding remaining score are $0$ and $0.009$, respectively, implying that an appropiate regression model should consider both $\langle \text{CCP},\phi_1 \rangle$ and $\langle \text{CCP},\phi_4 \rangle$.

\begin{remark}\label{remark5}
	We have performed an iterative backwards elimination procedure where, in each step, the covariate with the maximum p-value is eliminated as long as such maximum exceedes $\alpha= 0.01$.
	The rationale for that criteria is as follows:
	\renewcommand{\labelenumi}{(\arabic{enumi})}
	\begin{enumerate}
		\item There were a total of four steps after the initial test of conditional mean independence of the OCP on the CCP, which should be calibrated independently (see Remark \ref{remark4}). The second, third and fourth steps are only performed conditionally on the non-rejection of some null hypothesis of the previous step. A potential fifth step could have been necessary if the last significance tests for $\langle \text{CCP},\phi_1 \rangle$ or $\langle \text{CCP},\phi_4 \rangle$ yielded p-values over $\alpha = 0.01$, although we would not have tested for the conditional mean independence of the OCP on a single CCP FPC score.
		\item The probability of the maximum of a collection of p-values exceeding $\alpha$ is bounded by the probability of any of those p-values exceeding $\alpha$ which, under the null, converges to $\alpha$. Therefore, in each step, the probability of the variable selection mechanism not eliminating any of the remainding non-relevant covariates is bounded by $\alpha$; that is, this stepwise procedure aims to control the probability of selecting non-relevant covariates as significant. This is equivalent to establishing a test with level $\alpha$ for the iterative null hypothesis defined as:
		there is, at least, one more superfluous covariate in the model. The level bound by $\alpha$ is, in fact, asymptotically tight if there is exactly one redundant variable remaining in the model at the start of some step.
		\item The rejection of the combined null hypothesis in one step is incompatible with rejecting the null hypothesis in a previous step. The type I error associated with the multi-step procedure is (asymptotically) bounded by $k\alpha$ where $k$ is the maximum number of steps. In this real data application, $k=5$.
	\end{enumerate}
\end{remark}

\begin{figure}[t]
	\centering
	\includegraphics[width=11.5cm]{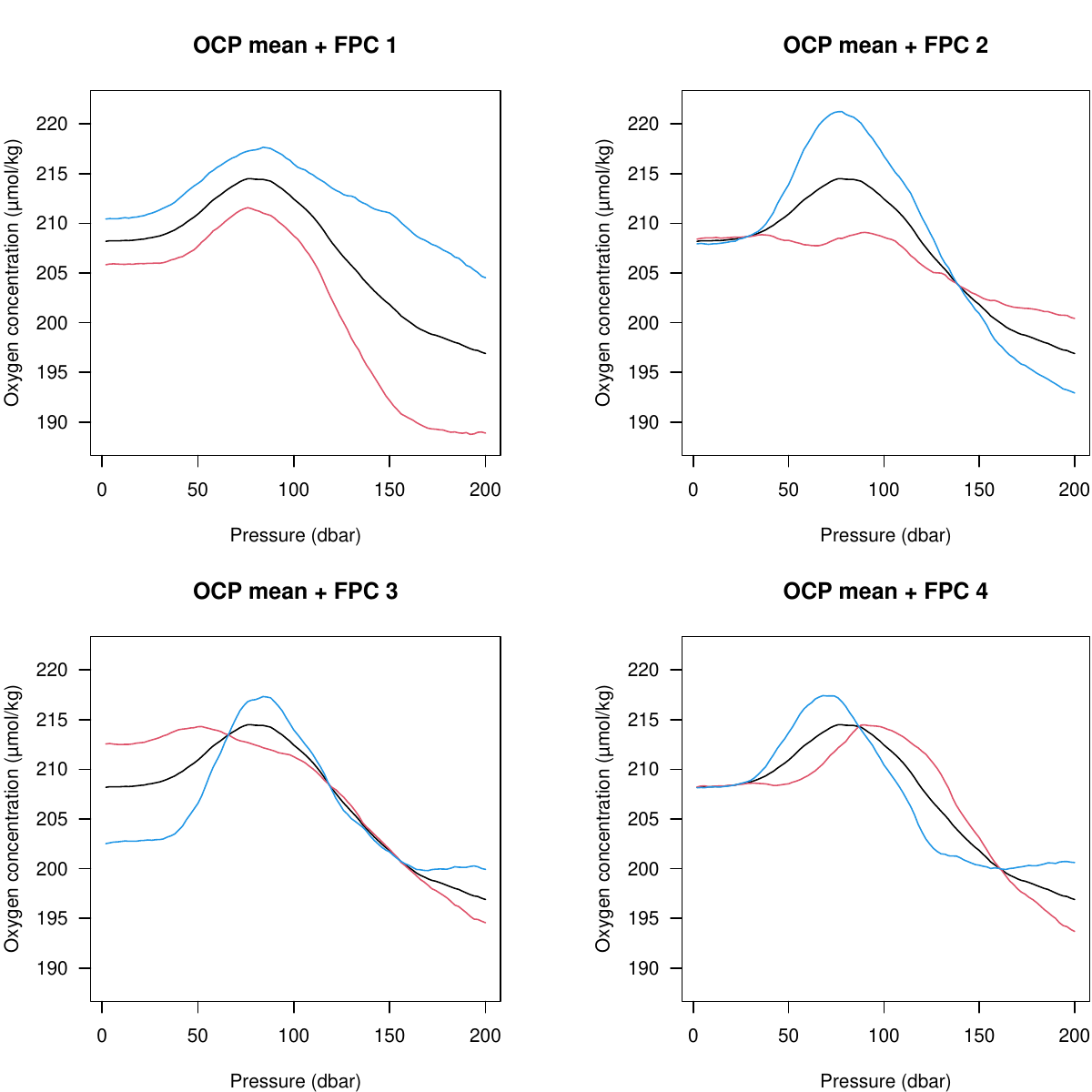}
	\caption{Empirical mean curve (black) and perturbations due to the addition of the 0.05 (blue) and 0.95 (red) quantiles of the first (top left), second (top right), third (bottom left) and fourth (bottom right) FPC scores of the OCP, multiplied by the corresponding eigenfuncions.}
	\label{fig3}
\end{figure}

\begin{figure}[t]
	\centering
	\includegraphics[width=13.5cm]{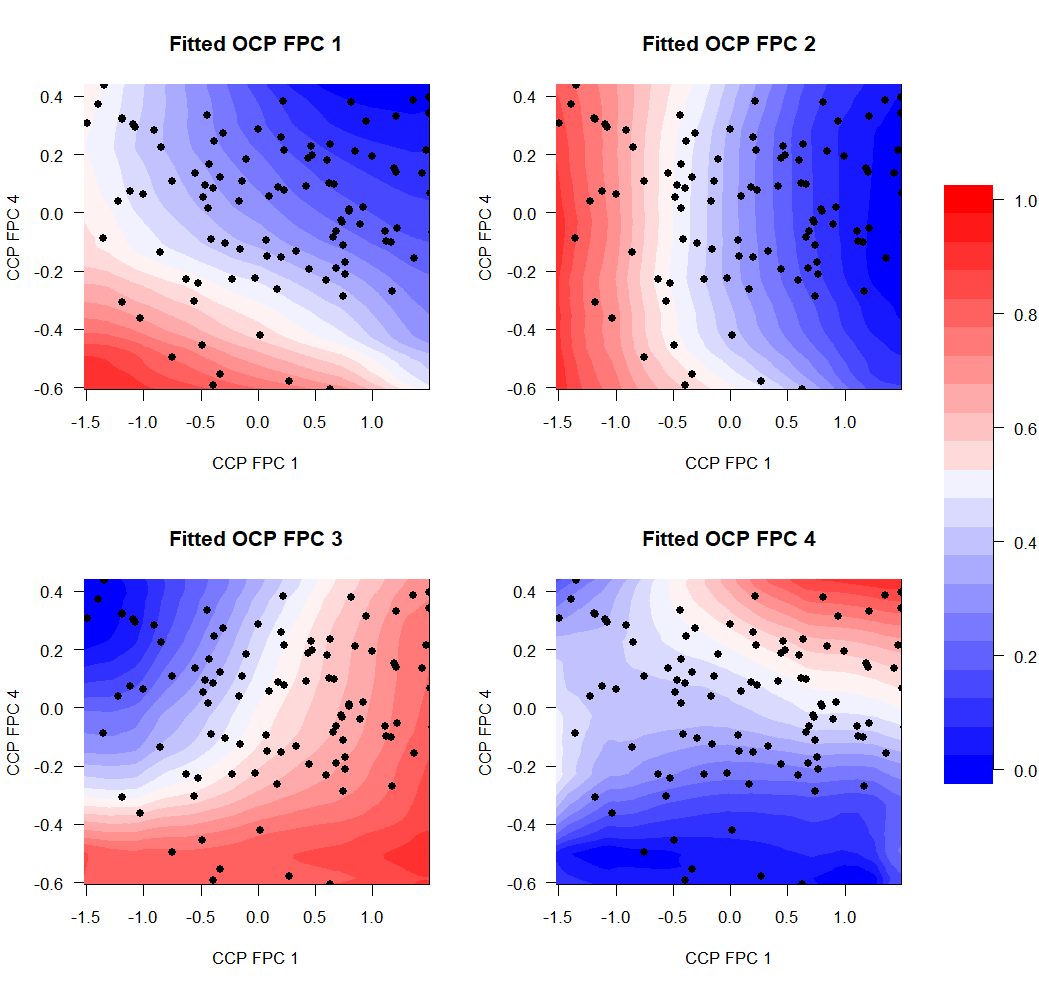}
	\caption{Heatmaps of the fitted values from the regression of the OCP on the first and fourth CCP FPC scores, projected onto the first four OCP FPC eigenfunctions and scaled to the interval $[0,1]$.}
	\label{fig4}
\end{figure}

To gain more insight on the modelled effect of the chloropigment concentration, through the first and fourth FPC scores, on the oxygen concentration profile, we provide additional plots in Figures \ref{fig3} and \ref{fig4}. The representation of a nonparametric estimator in a function-on-scalar regression context is not an easy task. Thus, we will be restricting our analysis to the first four functional principal components of the OCP, which also explain over 90$\%$ of the OCP variability. By analogy with Figure \ref{fig2}, the OCP mean curve and the perturbations corresponding to the 0.05 and 0.95 level quantiles are represented in Figure \ref{fig3}. The first OCP FPC can be interpreted as a scale parameter whereas the second an third FPCs are related to the presence or absence of a maximum in the oxigen concentration curves. The fourth OCP FPC modifies the location at which the maximum concentration is achieved.

In Figure \ref{fig4} we are displaying the heatmaps of the OCP fitted values, projected onto the corresponding eigenfunctions of the first four OCP FPCs, based on the Nadaraya-Watson estimator associated with the first and fourth FPC scores of the CCP. The projected fitted values are then scaled to the $[0,1]$ interval, for ease of comparison and display. Observe that computing the functional Nadaraya-Wason estimator and then projecting it onto some direction is equivalent to computing the scalar Nadaraya-Wason estimator based on the projected response, since said nonparametric estimator is a linear smoother. According to the plots in Figure \ref{fig4}, there may be a linear relationship between the chosen CCP scores and the first three OCP scores. Furthermore, given the FPCs interpretation implied by the plots in Figures \ref{fig1} and \ref{fig2}, along with the selected signs for said FPCs, we infer the following:
\begin{itemize}
	\item The oxygen concentration increases when the chloropigment concentration is relatively low at low pressure and it achieves its maximum value at high pressures.
	\item The oxygen concentration profile tends to exhibit a pronounced maximum between 50 and 100 dbars of pressure when the CCP achieves its maximum value at high pressures. There is little to no relationship between the OCP maximum and the CCP shape at lower pressures.
	\item When the CCP peaks at low pressures but the chloropigment concentration is relatively low around the maximum, the oxygen concentration tends to be low at low pressures.
	\item A relatively high chloropigment concentration at low pressures may be associated with the OCP achieving its maximum value at lower pressures.
\end{itemize}
Although we are not experts in the analysis of oceanic profiles, we find the implied relation between the CCP maximum value location, in terms of pressure, and the OCP behavior quite intriguing. The fitted values suggest that the deeper the maximum concentration of chloropigment is reached, the higher the expected oxygen concentration will be observed, even in the shallower ocean layers. This phenomenon may be partially explained by the aforementioned oxygen consumption processes as well as near-surface waters being oxygen-saturated by default. However, it could also be due to a confounding effect, as we are not accounting for the seasonality of the life cycle of photosintetic organisms, the atmospheric oxygen diffusion in water, as well as the change in oxygen solubility with temperature and salinity. 

\section{Conclusions and further research}
\label{Sec9}

The proposed kernel-based covariate significance test is, according to the theoretical and numerical results provided in this paper, a solid proposal designed to avoid the curse of dimensionality. It improves the theoretical performance of the family of smoothing-based significance tests since it is able to detect local alternatives in Pitman's sense converging to the null at the rate of $n^{-1/2}$. Furthermore, the test can deal with vector and Hilbert-valued response and covariates, except for functional $X$. In the latter case, when $X$ and $W$ are independent, a simpler test, which does not require any form of nonparametric estimation of $m$, was derived in Section \ref{Sec5.2}. A general procedure, based on the approximation of $m(X)$ through the conditional expectation of $Y$ on a growing dimension sequence of projections of $X$, is currently being investigated to adapt the testing methodology to the case of Hilbert-valued $X$ when there is no independence between $X$ and $W$.

Some natural extensions to the test introduced in this paper are goodness-of-fit tests for semiparametric models, such as single-index \citep{Chen2011} or partially linear models \citep{AneirosPerez2006} as well as completely parametric models. In \cite{Lee2020} and \cite{Lai2021}, the possibility of using the FMDD and KCMD to test for the correct specification of functional linear models is suggested in the data applications therein. Moreover, a goodness-of-fit test for parametric regression models with vector data, based on the MDD of \cite{Shao2014}, was proposed in \cite{Su2017} and later generalized to kernel-based tests by \cite{Escanciano2024}, where Neyman-orthogonal kernels are introduced as a very useful tool to eliminate the effect of parameter estimation in the test statistic limit distribution.  However, a rigorous adaptation of these ideas to the functional context is still missing. 

Although we have assumed sample independence in this paper, some preliminar results indicate that the proposed test, with some minor modifications on the bootstrap scheme, is still consistent for dependent data under commonly considered dependence structures such as mixing conditions and near epoch dependence (see e.g. \cite{Ibragimov1962}).

It is also possible to develop a test for conditional independence of $Y$ and $W$ given $X$ for scalar (or vector-valued) $Y$ using our testing methodology. Note that $Y\in \mathbb{R}$ is independent of $W$ given $X$ if and only if $\mathbb{P}(Y\le y|(X,W)) = \mathbb{P}(Y\le y|X)$ a.s. for all $y\in \mathbb{R}$. Therefore, denoting by $\tilde{Y}(y) = \ind{Y\le y}$, we can equivalently test for conditional mean independence of $\tilde{Y}(y)$ on $W$ given $X$, for all $y$, using the analogously defined test statistic $nU_n(y)$. Furthermore, we can consider the empirical process $\{nU_n(y): y\in \mathbb{R}\}_{n\in \mathbb{N}}$ and compute, for instance, $\sup_{y\in \mathbb{R}}n|U_n(y)|$ as a test statistic of conditional independence between $Y$ and $W$ given $X$. Under some regularity conditions, we believe that the corresponding modification of the bootstrap scheme presented in Section \ref{Sec4} would be consistent and the test would preserve the theoretical properties introduced in Section \ref{Sec3}.

The main focus of this contribution is to provide a covariate significance test that admits functional data. Due to the lack of competitors with complete theoretical support, the simulation study in Section \ref{Sec7} is purely descriptive. We believe that an additional study comparing the finite sample performance of our proposal and some of the tests referenced in Section \ref{Sec1} in the non-functional context along with the development of a characteristic kernel selection procedure, as well as the corresponding parameter tuning mechanism,  are interesting topics for another paper.

\newpage

\bibliography{biblio}
\bibliographystyle{apalike}

\end{document}


\maketitle

In this document we provide proofs for the theoretical results in the main document. We  state and prove some intermediate results required in the proofs of Theorems 1-4 and we briefly discuss the differentiability assumptions in technical conditions (T2)-(T3). We also provide a sufficient condition to ensure that the latter assumption holds.

Explicit references to the spaces where inner products and norms are computed are omitted for ease of notation.

\section{Fréchet differentiability assumptions}

In techical conditions (T2) and (T3) we require the regression function $m$, the density function $f$ and also the real valued function $b$ defined as $b(x,z') = \mathbb{E}[c(Z,z')|X=x]$ 
for all $z'\in \mathcal{Z}$ and all $x\in \mathbb{R}^d$, to verify some differentiability and moment assumptions. This requirements are used in a technical lemma (Lemma \ref{Lm1}) to bound the moments of the terms and remainders of Taylor expansions of these functions. 

By virtue of Taylor's theorem (see, for example, Theorem 5.6.3 in \citeauthor{Cartan1971}, \citeyear{Cartan1971}), if $\mathcal{S}$ is a Banach space and $a:\mathbb{R}\rightarrow \mathcal{S}$ is $\nu$ times differentiable in an open neighborhood of $x\in \mathbb{R}^d$, $O_x$, and the segment joining $x$ and $x_0$ is contained in $O_x$ for some $x_0\in \mathbb{R}^d$, we can write 
$$a(x + x_0) = a(x) + \sum\limits_{j=1}^\nu\frac{1}{j!}D^{j}a(x)(x_0,\ldots,x_0) + \mathcal{R}^\nu_{a}(x,x_0),$$
where $D^ja(x):\mathbb{R}^{jd}\rightarrow \mathcal{S}$ is a multilinear symmetric map for all $1\le j\le \nu$ and 
$\mathcal{R}^\nu_{a}(x,x_0)\in \mathcal{S}$ is the Taylor expansion remainder verifying $||\mathcal{R}^\nu_{a}(x,x_0)|| = o(||x_0||^\nu)$. If, in addition, $l$ is $\nu+1$ times continuously differentiable in $O_x$, 
$$||\mathcal{R}^\nu_{a}(x,x_0)||\le \frac{1}{(\nu+1)!}\sup_{t\in[0,1]}\left|\left| D^{\nu+1}a(x+tx_0)(x_0,\ldots,x_0)\right|\right|,$$ 
where, for all $t\in[0,1]$, $D^{\nu+1}a(x+tx_0):\mathbb{R}^{(\nu+1)d}\rightarrow \mathcal{S}$ is also a multilinear symmetric map.

We can relate the collection of multilinear operators $\{D^ja(x)\}_{j=1}^\nu$ and $D^{\nu+1}a(x+tx_0)$ to the notion of (Gateaux) partial derivatives. Let $\{e_k\}_{k=1}^d$ denote the canonical basis of $\mathbb{R}^d$ so that $x_0 = \sum_{k=1}^d x_0^k e_k$. It follows from the definition of Fréchet differentiability that
$$ D^1a(x)(e_{k}) = \lim\limits_{t \rightarrow 0} \frac{1}{t}(a(x+te_{k}) - a(x)) =  \frac{\partial a(x)}{\partial x^{k}} \in \mathcal{S}.$$
Similarly, for all $1<j\le \nu+1$ and all $1\le k_1,\ldots, k_j\le d$, we have
$$ D^ja(x)(e_{k_1},\ldots, e_{k_j}) =  \lim\limits_{t \rightarrow 0} \frac{1}{t}\left(\frac{\partial^{j-1} a(x+te_{k_j})}{\partial x^{k_{j-1}}\cdots \partial x^{k_1}} - \frac{\partial^{j-1} a(x)}{\partial x^{k_{j-1}}\cdots \partial x^{k_1}}\right) = \frac{\partial^ja(x)}{\partial x^{k_j}\cdots \partial x^{k_1}},$$
which is also an element of $\mathcal{S}$. Moreover, the order of partial derivatives is exchangable, since $D^ja(x)$ is symmetric for all $j$. Then, by the definition of multilinear maps,
\begin{equation*}
	D^{j}a(x)(x_0,\ldots,x_0) = \sum\limits_{k_1,\ldots,k_j=1}^d x_0^{k_1}\cdots x_0^{k_j}\frac{\partial^j a(x)}{\partial x^{k_1}\cdots \partial x^{k_j}}.
\end{equation*}
Furthermore, by the Cauchy-Schwarz inequality, the remainder of the Taylor expansion of order $\nu$ verifies
\begin{align*}
	||\mathcal{R}_{a}^\nu(x,x_0)|| &\le \frac{1}{(\nu+1)!}\sup_{t\in[0,1]}\left|\left|\sum\limits_{k_1,\ldots,k_{\nu + 1}=1}^d x_0^{k_1}\cdots x_0^{k_{\nu + 1}}\frac{\partial^{\nu + 1} a(x +tx_0)}{\partial x^{k_1}\cdots \partial x^{k_{\nu + 1}}}\right|\right|\\
	&\le \frac{\left|\left|x_0\right|\right|^{\nu+1}}{(\nu+1)!}\left(\sum\limits_{k_1,\ldots,k_{\nu + 1}=1}^d \sup_{t\in[0,1]}\left|\left|\frac{\partial^{\nu + 1}a(x +tx_0)}{\partial x^{k_1}\cdots \partial x^{k_{\nu + 1}}}\right|\right|^2\right)^\frac{1}{2}, \label{Sm.1}\tag{Sm.1}
\end{align*}	
which, by the continuity of the partial derivatives of order $\nu +1$ of $a$, is a finite bound.

If we assume that $f(x)$, $m(x)$ and $b(x,z')$ are $\nu+1$ times continuously Fréchet differentiable with respect to $x$, the corresponding Taylor expansions of order $\nu$ will hold in a convex neighborhood of $x$. Moreover, bounding the moments of a Taylor expansion is reduced to bounding the moments of the partial derivatives up to order $\nu + 1$. 

\section{An insight into condition (T3)}

Recall the definition of technical condition (T3):

\renewcommand{\labelenumi}{(T\arabic{enumi})}
\begin{enumerate}\setcounter{enumi}{2}
	\item For every  $z'\in \mathcal{Z}$, $b(x,z')=\mathbb{E}[c(Z,z')|X=x]$ is continuously differentiable with respect to $x\in \mathbb{R}^d$. Moreover, there will be some $h_2>0$ such that
	\begin{itemize}
		\item $\sum\limits_{k=1}^d \mathbb{E}\left[\sup\limits_{||x_0||\le h_2}\left|\dfrac{\partial b(X+x_0,Z)}{\partial x^{k}}\right|^2\ind{X+ x_0 \in \text{supp}(\mathbb{P}_X)}\right]<\infty$.
	\end{itemize} 
\end{enumerate}
\renewcommand{\labelenumi}{(\arabic{enumi})}
	
It is by no means straighforward to verify whether (T3) holds since $b(x,z')$ and its derivative with respect to $x$ depend on the joint distribution of $Z = (X,W)$. However, the following proposition states that, when $c$ is the Gaussian kernel, (T3) holds for a reasonably wide family of data generating processes. 

\begin{prop}
	\label{Pr1}
	Let $c(z,z') = \exp\left(-||z-z'||^2/\sigma_c^2\right)$ for some $\sigma_c^2>0$ and suppose that the following regression model of $W$ on $X$ holds:
	\begin{equation}\label{Pr1.1}\tag{Pr1.1}
		W = m_W(X) + \Sigma_W(X)(\mathcal{E}_W),
	\end{equation}
	where $\Sigma_W(x):\mathcal{S}\rightarrow\mathcal{W}$ is a bounded linear operator for all $x\in\mathbb{R}^d$, $\mathcal{E}_W\in \mathcal{S}$ is independent of $X$, where $\mathcal{S}$ is some normed space and $\mathbb{E}\left[||\mathcal{E}_W||\right]<\infty$. Assume that $m_W$ and $\Sigma_W$ are continuously Fréchet differentiable (the latter with respect to the operator norm) with respect to $x\in \mathbb{R}^d$ and also that there is some $h_2>0$ such that
	\begin{equation}\label{Pr1.2}\tag{Pr1.2}
		\sum\limits_{k=1}^d \mathbb{E}\left[\sup\limits_{||x_0||\le h_2}\left(\left|\left|\frac{\partial m_W(X_i + x_0)}{\partial x^k}\right|\right|^2+\left|\left|\frac{\partial \Sigma_W(X_i + x_0)}{\partial x^k}\right|\right|^2\right)\ind{X+ x_0 \in \text{supp}(\mathbb{P}_X)}\right]<\infty.
	\end{equation}
	Then, (T3) holds. 
\end{prop}

\begin{proof}[\bfseries\textup{Proof of Proposition 1}]
	Given the regression model in (\ref{Pr1.1}) it trivially holds that
	$$b(x,z') = \exp\left(-\frac{||x-x'||^2}{\sigma_c^2}\right)\mathbb{E}\left[\exp\left(-\frac{||m_W(x) + \Sigma_W(x)(\mathcal{E}_W)-w'||^2}{\sigma_c^2}\right)\right],$$
	for any $z' = (x',w')$. Denote $v(x,w')=\exp\left(-||m_W(x) + \Sigma_W(x)(\mathcal{E}_W)-w'||^2/\sigma_c^2\right)$. We have, almost surely,
	$$\frac{\partial v(x,w')}{\partial x^k}  = \left\langle \frac{\partial m_W(x)}{\partial x^k} + \frac{\partial \Sigma_W(x)}{\partial x^k}(\mathcal{E}_W), -\frac{2}{\sigma_c^2}\left(m_W(x) + \Sigma_W(x)(\mathcal{E}_W)-w'\right)v(x,w')\right\rangle,$$
	where $\partial/\partial x^k\Sigma_W(x): \mathcal{S}\rightarrow\mathcal{W}$ is also a bounded linear operator. Chooose any $\tilde{x}\in \mathbb{R}^d$ such that $x$ belongs to the closed ball centered in $\tilde{x}$ for some radius $R>0$, $Cb(\tilde{x},R)$. Since $l(y) = 2y/\sigma_c^2e^{-y^2/\sigma_c^2}$ is bounded in $\mathbb{R}$ and the partial derivatives of both $m_W$ and $\Sigma_W$ are continuous, it follows that, for all $x_0\in Cb(\tilde{x},R)$,
	 $$\left|\frac{\partial v (x_0,w')}{\partial x^k} \right| \le ||l||_\infty \left|\left|\frac{\partial m_W(x_0) }{\partial x^k}+ \frac{\partial \Sigma_W(x_0)}{\partial x^k}(\mathcal{E}_W) \right|\right| \le ||l||_\infty (M_{\tilde{x},R,1} + M_{\tilde{x},R,2} ||\mathcal{E}_W||),$$
	 for some positive constants $M_{\tilde{x},R,1}$ and $M_{\tilde{x},R,2}$, independent of $x_0$. We assume $\mathbb{E}[||\mathcal{E}_W||]<\infty$ and, thus, by virtue of the dominated convergence theorem (in the neighborhood of $\tilde{x}$) we can exchange the expectation and the differential operators so that $\partial/\partial x^k\mathbb{E}\left[v(x,w')\right] = \mathbb{E}\left[\partial/\partial x^k v(x,w')\right]$.
	 Therefore,        
	 \begin{align}
	 	\frac{\partial b(x,z')}{\partial x^k} = &-\frac{2(x^k-x^{\prime k})}{\sigma_c^2}\exp\left(-\frac{||x-x'||^2}{\sigma_c^2}\right)\mathbb{E}\left[v(x,w')\right] \nonumber + \exp\left(-\frac{||x-x'||^2}{\sigma_c^2}\right)\mathbb{E}\left[\frac{\partial v(x,w')}{\partial x^k}\right], \nonumber
	 \end{align}
	and the norm of $\partial/\partial x^kb(x,z')$ verifies
	$$ \left|\frac{\partial b(x,z')}{\partial x^k}\right|  \le ||l||_\infty\left(1+\left|\left|\frac{\partial m_W(x)}{\partial x^k}\right|\right| + \left|\left|\frac{\partial \Sigma_W(x)}{\partial x^k}\right|\right|\mathbb{E}\left[||\mathcal{E}_W|| \right]\right).$$
	Then, (T3) trivially follows form (\ref{Pr1.2}).
\end{proof}

The regression model in (\ref{Pr1.1}) is not identifiable, since we are not imposing any structure on the probability law of $\mathcal{E}_W$ besides the existence of first order moment. However, this is not specially concerning in this context as we are only interested in providing a sufficient condition for (T3) to hold.

\section{Notation and technical lemmas}

Let  $\psi_n:(\mathcal{Z}\times \mathcal{Y})^2\rightarrow \mathcal{HS}$ be defined as 
$$\psi_n((z,y),(z',y')) = \frac{1}{2}(c(z,\cdot)-c(z',\cdot))\otimes K_h(x-x')(y-y'),$$
for all $(z,y),(z',y')$ in the support of $(Z,Y)$. Then, the plug-in estimator of $\text{KCMD}(f(X)(Y-m(X))|Z)$, $V_n$, verifies
\begin{align*}
	V_n &= \frac{1}{n^4}\sum\limits_{i,j,k,l}c(Z_i,Z_j)\langle K_h(X_i-X_k)(Y_i-Y_k), K_h(X_j-X_l)(Y_k-Y_l)\rangle\\
	&= \frac{(n-1)^2}{n^2}\left|\left| \frac{1}{{n\choose 2}}\sum\limits_{i<k} \psi_n((Z_i,Y_i),(Z_k,Y_k))  \right|\right|^2, \label{Sm.2}\tag{Sm.2}
\end{align*}
and its asymptotically unbiased counterpart, $U_n$, is given by the following expression:
\begin{align*}
	U_n &= \frac{1}{n(n-1)(n-2)(n-3)}\sum\limits_{i\neq j\neq k\neq l}c(Z_i,Z_j)\langle K_h(X_i-X_k)(Y_i-Y_k), K_h(X_j-X_l)(Y_k-Y_l)\rangle\\
	&=\frac{1}{n(n-1)(n-2)(n-3)}\sum\limits_{i\neq j\neq k\neq l} \langle \psi_n((Z_i,Y_i),(Z_k,Y_k)), \psi_n((Z_j,Y_j),(Z_l,Y_l)) \rangle.\label{Sm.3}\tag{Sm.3}
\end{align*}

The proof of Theorems 1 and 2 is based, on the one hand, on the asymptotic relation between $V_n$ and $U_n$ defined as in (\ref{Sm.2}) and (\ref{Sm.3}), respectively and, on the other hand, on the Hoeffding decomposition of the second order U-statistic
$$\frac{1}{{n\choose 2}}\sum\limits_{i< k} \psi_n((Z_i,Y_i),(Z_k,Y_k)).$$ 
Let $\psi_n((Z_i,Y_i),(Z_k,Y_k))$ be denoted by $\psi_n(i,k)$. Moreover, let $\psi^1_n(i) = \mathbb{E}[\psi_n(i,k)|Z_i,Y_i]-\mathbb{E}[\psi_n(i,k)]$ and $\psi^2_n(i,k) = \psi_n(i,k)- \psi^1_n(i) - \psi^1_n(k) + \mathbb{E}[\psi_n(i,k)]$. By the Hoeffding decomposition we have
\begin{equation}\label{Sm.4}\tag{Sm.4}
	\frac{1}{{n\choose 2}}\sum\limits_{i< k} \psi_n(i,k) =  \mathbb{E}[\psi_n(i,k)] + \frac{2}{n}\sum\limits_{i=1}^n \psi^1_n(i)  + \frac{1}{{n \choose 2}}\sum\limits_{i<k}^n \psi^2_n(i,k),
\end{equation}
where $i\neq k$ in $\mathbb{E}[\psi_n(i,k)]$.
The asymptotic behavior of the U-statistic is then characterized by the behavior of every term in the Hoeffding decomposition in (\ref{Sm.4}). These terms depend on the distribution of $\{(Z_i,Y_i)\}_{i=1}^n$ which is different under $\mathcal{H}_0$, $\mathcal{H}_1$ and $\{\mathcal{H}_{1n}\}_{n\in \mathbb{N}}$.
We treat jointly the three types of hypothesis by considering the following regression model
\begin{equation}\label{Sm.5}\tag{Sm.5}
	Y = m(X) + \frac{\tilde{g}(Z)}{n^\gamma} + \mathcal{E},
\end{equation}
where $\mathbb{E}[\mathcal{E}|Z] = 0$ a.s., $\mathbb{E}[\tilde{g}(Z)|X] = 0$ a.s. and $\gamma\ge 0$. Thus, if the model in (\ref{Sm.5}) holds, we have, for each $n\in \mathbb{N}$:
\begin{itemize}
	\item $\mathcal{H}_0\hspace{1.7mm} \iff \tilde{g}(Z) = 0$ a.s.
	\item $\mathcal{H}_1\hspace{1.7mm} \iff \mathbb{P}\left(\tilde{g}(Z) \neq 0\right)>0$ and $\gamma = 0$.
	\item $\mathcal{H}_{1n} \iff \mathbb{P}\left(\tilde{g}(Z) \neq 0\right)>0$ and $\gamma > 0$.
\end{itemize}

Let us introduce the notation of $o_{L^p}$ and $O_{L^p}$ which will recur throughout the remainder of the document.

\begin{definition}\label{Def3}
	Given a sequence of non-zero real numbers $\{M_n\}_{n=1}^\infty$, we say that a sequence of random variables $\{\eta_n\}_{n=1}^\infty$ taking values in some normed space $\mathcal{S}$ is $o_{L^p}(M_n)$ if
	$$\mathbb{E}\left[\left|\left|\frac{\eta_n}{M_n}\right|\right|^p\right] = o(1).$$
	We also say that $\{\eta_n\}_{n=1}^\infty$ is $O_{L^p}(M_n)$ if 
	$$\mathbb{E}\left[\left|\left|\frac{\eta_n}{M_n}\right|\right|^p\right] = O(1).$$
\end{definition}

The notation in Definition \ref{Def3} is convenient since the proofs of many intermediate results required to prove Theorems 1-4 rely on bounding the first or second order moments of some asymptotically negligible terms. Note that for $p\ge 1$, $o_{L^p}(M_n) \implies o_{\mathbb{P}}(M_n)$ and $O_{L^p}(M_n) \implies O_{\mathbb{P}}(M_n)$, where $O_{\mathbb{P}}(1)$ and $o_{\mathbb{P}}(1)$ denote bound in probability and convergence to zero in probability, respectively.

Now, we state and prove two technical lemmas that are essential in the proofs of Theorems 1 and 2.

\begin{lemma}
	\label{Lm1}
	Assume that (A1)-(A3) hold. Let $B$ be defined as
	$$B(x) = \left(\int  K^u(t) t^\nu dt\right) \left( \sum\limits_{j=1}^{\nu -1}\frac{1}{j!(\nu-j)!}\sum\limits_{l=1}^d\frac{\partial^j m(x)}{(\partial x^l)^j}\frac{\partial^{\nu-j}f(x)}{(\partial x^l)^{\nu - j}} + \frac{f(x)}{\nu!} \sum\limits_{l=1}^d\frac{\partial^j m(x)}{(\partial x^l)^\nu } \right),$$  
	for all $x\in\text{supp}(\mathbb{P}_X)$. Given $(Z_i,Y_i)$ and $(Z_k,Y_k)$ i.i.d. verifying the regression model in (\ref{Sm.5}), it follows:
	\renewcommand{\labelenumi}{(\arabic{enumi})}
	\begin{enumerate}
		\item $\mathbb{E}[K_h(X_i-X_k)(Y_k-Y_i)|X_i,Y_i] = f(X_i)(m(X_i) - Y_i) + O_{L^1}(h^{\nu})$ under (T1), (T2), (T4) and (T5).
		\item $\mathbb{E}[K_h(X_i - X_k)(Y_k - Y_i) | X_i] = h^\nu B(X_i) +o_{L^2}(h^{\nu}) = O_{L^2}(h^{\nu})$ under (T1), (T2), (T4) and (T5).
		\item $\mathbb{E}\left[ |K_h(X_i-X_k)| \ |X_i\right] \le 2^d||f||_\infty||K||_\infty<\infty$ under (T5).
		\item $\mathbb{E}[||K_h(X_i-X_k)(Y_i-Y_k)||^2]\le 4||K||_\infty\mathbb{E}[|K_h(X_i-X_k)| \ ||Y_i||^2]/h^d = O\left(1/{h^d}\right)$  under (T1) and (T5).
		\item $\mathbb{E}[K_h(X_i - X_k)|X_i] = f(X_i) + O_{L^4}(h^\nu)$ under (T2) and (T4).
		\item $\mathbb{E}\left[|K_h(X_i-X_k)| \ ||m(X_i)-m(X_k)||\ |X_i\right] = O_{L^2}(h)$ under (T1), (T2), (T4) and (T5).
		\item $\mathbb{E}[K_h(X_i - X_k)(b(X_k,\cdot)-b(X_i,\cdot))|X_i] =  o_{L^4}(1)$ under (T3)-(T5).
		\item $\mathbb{E}\left[(c(Z_i,\cdot)-b(X_k,\cdot))\otimes K_h(X_i-X_k)(Y_i-m(X_k))|Z_i,Y_i\right] = (c(Z_i,\cdot)-b(X_i,\cdot))\otimes f(X_i)\mathcal{E}_i  + o_{L^{2}}(1)$  under 
		(T1)-(T5) and $\mathcal{H}_0$.
	\end{enumerate}
\end{lemma}

\begin{proof}[\bfseries\textup{Proof of Lemma \ref{Lm1}}]
	
	We will prove (1)-(8) sequentially, assuming that $\partial \text{supp}(\mathbb{P}_X)$ is not empty. The proof when $\text{supp}(\mathbb{P}_X) = \mathbb{R}^d$ is simpler and, thus, is omitted. 
	
	By the i.i.d. hypothesis and the definition of conditional expectation, we have
	\begin{align*}
		\mathbb{E}[K_h(X_i-X_k)(Y_k-Y_i)|X_i,Y_i]  &= \mathbb{E}\left[K_h(X_i-X_k)m(X_k)|X_i\right] - Y_i\mathbb{E}\left[K_h(X_i-X_k)|X_i\right]\\
		&= \mathbb{E}\left[K_h(X_i-X_k)(m(X_k)-m(X_i))|X_i\right]\\
		&\quad + (m(X_i)-Y_i)\mathbb{E}\left[K_h(X_i-X_k)|X_i\right]\\
		&= (m(X_i)-Y_i)f(X_i) + \mathbb{E}\left[K_h(X_i-X_k)(m(X_k)-m(X_i))|X_i\right] \\
		&\quad + (m(X_i)-Y_i)\left(\mathbb{E}\left[K_h(X_i-X_k)|X_i\right]-f(X_i)\right). \tag{Lm1.1}\label{Lm1.1}
	\end{align*}	
	Since $X_i$ is absolutely continuous, it belongs to the interior of $\text{supp}(\mathbb{P}_X)$ almost surely. The support of $K$ is $[-1,1]^d$ by (A1) and, thus, for all $x$ in the support of $K$, $||xh||\le \sqrt{d}h$. Let us first assume that $n$ is large enough so that $\sqrt{d}h<d(X_i,\partial \text{supp}(\mathbb{P}_X))$, where $d(X_i,\partial \text{supp}(\mathbb{P}_X))$ denotes the distance from $X_i$ to the boundary of its support. Then, the open ball centered in $X_i$ with radius $\sqrt{d}h$ is contained in the interior of $\text{supp}(\mathbb{P}_X)$. By the smoothness assumptions on $f$, Taylor's theorem and the properties of kernel $K$, it follows
	\begin{align*}
		\mathbb{E}\left[K_h(X_i-X_k)|X_i\right] - f(X_i) &= \int K(x)f(X_i+xh)dx - f(X_i) = \int K(x)(f(X_i+xh)-f(X_i))dx\\
		&=  \int K(x)\left(\sum\limits_{j=1}^\nu\frac{h^{j}}{j!} \sum\limits_{k_1,\ldots,k_j=1}^d \frac{\partial^jf(X_i)}{\partial x^{k_1}\cdots \partial x^{k_j}}x^{k_1}\cdots x^{k_j} + \mathcal{R}_{f}^\nu(X_i,xh)\right)dx\\
		&= \sum\limits_{j=1}^\nu\frac{h^{j}}{j!} \sum\limits_{k_1,\ldots,k_j=1}^d  \frac{\partial^jf(X_i)}{\partial x^{k_1}\cdots \partial x^{k_j}}\int K(x) x^{k_1}\cdots x^{k_j}dx + \int\mathcal{R}_{f}^\nu(X_i,xh)dx\\
		&= \frac{h^{\nu}}{\nu!} \sum\limits_{k_1,\ldots, k_\nu=1}^d \frac{\partial^\nu f(X_i) }{\partial x^{k_1}\cdots \partial x^{k_\nu}}\int K(x) x^{k_1}\cdots x^{k_\nu}dx + \int K(x)\mathcal{R}_{f}^\nu(X_i,xh)dx\\
		&= \frac{h^{\nu}}{\nu!} \sum\limits_{k=1}^d \frac{\partial^\nu f(X_i) }{(\partial x^{k})^\nu}\left(\int K^u(t)t^\nu dt\right)+ \int K(x)\mathcal{R}_{f}^\nu(X_i,xh)dx, \tag{Lm1.2}\label{Lm1.2}
	\end{align*}	
	since $\int K^u(t)t^idt = \ind{i=0}$ for all $0\le i\le \nu-1$ by (A1) and, therefore,
	$$ \int K(x) x^{k_1}\cdots x^{k_j}dx = \int  \prod_{l=1}^d  K^u(x^{l}) x^{k_1}\cdots x^{k_j}dx^l = \ind{k_1 = \ldots = k_j, \ j= \nu} \left(\int  K^u(t) t^\nu dt\right) .$$
	Similarly, it follows from the smoothness assumptions on $m$ that
	\begin{align*}
		&\mathbb{E}\left[K_h(X_i-X_k)(m(X_k)-m(X_i))|X_i\right]\\
		= &f(X_i)\int K(x)(m(X_i + xh)-m(X_i))dx + \int K(x)(m(X_i + xh)-m(X_i))(f(X_i+xh)-f(X_i))dx\\
		= &f(X_i)\left(\frac{h^{\nu}}{\nu!} \sum\limits_{k=1}^d \frac{\partial^\nu m(X_i) }{(\partial x^{k})^\nu}\left(\int K^u(t)t^\nu dt\right) + \int K(x)\mathcal{R}_{m}^\nu(X_i,xh)dx\right)\\
		&\ + \int K(x)\left(\sum\limits_{j=1}^\nu\frac{h^{j}}{j!} \sum\limits_{k_1,\ldots,k_j=1}^d \frac{\partial^jm(X_i)}{\partial x^{k_1}\cdots \partial x^{k_j}}x^{k_1}\cdots x^{k_j} + \mathcal{R}_{m}^\nu(X_i,xh)\right)(f(X_i+xh)-f(X_i))dx\\
		= &f(X_i)\left(\frac{h^{\nu}}{\nu!} \sum\limits_{k=1}^d \frac{\partial^\nu m(X_i) }{(\partial x^{k})^\nu}\left(\int K^u(t)t^\nu dt\right) + \int K(x)\mathcal{R}_{m}^\nu(X_i,xh)dx\right) \tag{Lm1.3}\label{Lm1.3}\\
		&\ + \int K(x)\mathcal{R}_{m}^\nu(X_i,xh)\left(\sum\limits_{j=1}^\nu\frac{h^{j}}{j!} \sum\limits_{k_1,\ldots,k_j=1}^d \frac{\partial^jf(X_i)}{\partial x^{k_1}\cdots \partial x^{k_j}}x^{k_1}\cdots x^{k_j} + \mathcal{R}_{f}^\nu(X_i,xh)\right)dx\\
		&\ + \int K(x)\left(\sum\limits_{j=1}^\nu\frac{h^{j}}{j!} \sum\limits_{k_1,\ldots,k_j=1}^d \frac{\partial^jm(X_i)}{\partial x^{k_1}\cdots \partial x^{k_j}}x^{k_1}\cdots x^{k_j}\right)\left(\sum\limits_{j=1}^\nu\frac{h^{j}}{j!} \sum\limits_{k_1,\ldots,k_j=1}^d \frac{\partial^jf(X_i)}{\partial x^{k_1}\cdots \partial x^{k_j}}x^{k_1}\cdots x^{k_j}\right)dx\\
		&\ + \int K(x)\left(\sum\limits_{j=1}^\nu\frac{h^{j}}{j!} \sum\limits_{k_1,\ldots,k_j=1}^d \frac{\partial^jm(X_i)}{\partial x^{k_1}\cdots \partial x^{k_j}}x^{k_1}\cdots x^{k_j}\right)\mathcal{R}_{f}^\nu(X_i,xh)dx. 
	\end{align*}
	Straightforward algebraic manipulations yield
	\begin{align*}
		&\int K(x)\left(\sum\limits_{j=1}^\nu\frac{h^{j}}{j!} \sum\limits_{k_1,\ldots,k_j=1}^d \frac{\partial^jm(X_i)}{\partial x^{k_1}\cdots \partial x^{k_j}}x^{k_1}\cdots x^{k_j}\right)\left(\sum\limits_{j=1}^\nu\frac{h^{j}}{j!} \sum\limits_{k_1,\ldots,k_j=1}^d \frac{\partial^jf(X_i)}{\partial x^{k_1}\cdots \partial x^{k_j}}x^{k_1}\cdots x^{k_j}\right)dx\\
		= &\sum\limits_{j=1}^\nu\sum_{\substack{l=1 \\ j+l<\nu}}^\nu\sum\limits_{k_1,\ldots, k_j=1}^{d}\sum\limits_{q_1,\ldots, q_l=1}^{d} \frac{h^{j+l}}{j!l!}\frac{\partial^j m(X_i)}{\partial x^{k_1} \cdots \partial x^{k_j}}\frac{\partial^l f(X_i)}{\partial x^{q_1} \cdots \partial x^{q_l}}\int K(x)  x^{k_1}\cdots x^{k_j}x^{q_1}\cdots x^{q_l} dx \tag{Lm1.4}\label{Lm1.4}\\
		&\ + \sum\limits_{j=1}^\nu\sum_{\substack{l=1 \\ j+l>\nu}}^\nu\sum\limits_{k_1,\ldots, k_j=1}^{d}\sum\limits_{q_1,\ldots, q_l=1}^{d} \frac{h^{j+l}}{j!l!}\frac{\partial^j m(X_i)}{\partial x^{k_1} \cdots \partial x^{k_j}}\frac{\partial^l f(X_i)}{\partial x^{q_1} \cdots \partial x^{q_l}}\int K(x)  x^{k_1}\cdots x^{k_j}x^{q_1}\cdots x^{q_l} dx\\
		&\ + \sum\limits_{j=1}^{\nu -1} \sum\limits_{k_1,\ldots, k_j=1}^{d}\sum\limits_{q_1,\ldots, q_{\nu-j}=1}^{d} \frac{h^{\nu}}{j!(\nu-j)!}\frac{\partial^j m(X_i)}{\partial x^{k_1} \cdots \partial x^{k_j}}\frac{\partial^{\nu - j} f(X_i)}{\partial x^{q_1} \cdots \partial x^{q_{\nu-j}}}\int K(x)  x^{k_1}\cdots x^{k_j}x^{q_1}\cdots x^{q_{\nu-j}} dx. 
	\end{align*}
	Using, once again, that $\int K^u(t)t^idt = 0$ for all $1\le i\le \nu-1$, we have
	\begin{equation}\tag{Lm1.5}\label{Lm1.5}
		\sum\limits_{j=1}^\nu\sum_{\substack{l=1 \\ j+l<\nu}}^\nu\sum\limits_{k_1,\ldots, k_j=1}^{d}\sum\limits_{q_1,\ldots, q_l=1}^{d} \frac{h^{j+l}}{j!l!}\frac{\partial^j m(X_i)}{\partial x^{k_1} \cdots \partial x^{k_j}}\frac{\partial^l f(X_i)}{\partial x^{q_1} \cdots \partial x^{q_l}}\int K(x)  x^{k_1}\cdots x^{k_j}x^{q_1}\cdots x^{q_l} dx = 0, 
	\end{equation}
	and, moreover, 
	\begin{align*}
		&\sum\limits_{j=1}^{\nu -1} \sum\limits_{k_1,\ldots, k_j=1}^{d}\sum\limits_{q_1,\ldots, q_{\nu-j}=1}^{d} \frac{h^{\nu}}{j!(\nu-j)!}\frac{\partial^j m(X_i)}{\partial x^{k_1} \cdots \partial x^{k_j}}\frac{\partial^{\nu - j} f(X_i)}{\partial x^{q_1} \cdots \partial x^{q_{\nu-j}}}\int K(x)  x^{k_1}\cdots x^{k_j}x^{q_1}\cdots x^{q_{\nu-j}} dx\\
		= & \sum\limits_{j=1}^{\nu -1}\frac{h^{\nu}}{j!(\nu-j)!} \sum\limits_{p=1}^{d} \frac{\partial^j m(X_i)}{(\partial x^{p})^j}\frac{\partial^{\nu - j} f(X_i)}{(\partial x^{p})^{\nu - j}}\int K(x)  (x^p)^\nu dx\\
		= & \sum\limits_{j=1}^{\nu -1}\frac{h^{\nu}}{j!(\nu-j)!} \sum\limits_{p=1}^{d} \frac{\partial^j m(X_i)}{(\partial x^{p})^j}\frac{\partial^{\nu - j} f(X_i)}{(\partial x^{p})^{\nu - j}}\left(\int K^u\left(t\right)  (t)^\nu dt\right). \tag{Lm1.6}\label{Lm1.6}
	\end{align*}
	Let $D(X_i,Y_i)= \mathbb{E}[K_h(X_i-X_k)(Y_k-Y_i)|X_i,Y_i] - (m(X_i)-Y_i)f(X_i)$. By combination of (\ref{Lm1.2}), (\ref{Lm1.3}), (\ref{Lm1.4}), (\ref{Lm1.5}) and (\ref{Lm1.6}), it follows that
	\begin{align*}
		D(X_i,Y_i) &= (m(X_i)-Y_i)\left( \frac{h^{\nu}}{\nu!} \sum\limits_{k=1}^d \frac{\partial^\nu f(X_i) }{(\partial x^{k})^\nu}\left(\int K^u(t)t^\nu dt\right)+ \int K(x)\mathcal{R}_{f}^\nu(X_i,xh)dx\right) \tag{Lm1.7}\label{Lm1.7}\\
		&\quad  + f(X_i)\left(\frac{h^{\nu}}{\nu!} \sum\limits_{k=1}^d \frac{\partial^\nu m(X_i) }{(\partial x^{k})^\nu}\left(\int K^u(t)t^\nu dt\right) + \int K(x)\mathcal{R}_{m}^\nu(X_i,xh)dx\right)\\
		&\quad +  \int K(x)\mathcal{R}_{m}^\nu(X_i,xh)\left(\sum\limits_{j=1}^\nu\frac{h^{j}}{j!} \sum\limits_{k_1,\ldots,k_j=1}^d \frac{\partial^jf(X_i)}{\partial x^{k_1}\cdots \partial x^{k_j}}x^{k_1}\cdots x^{k_j} + \mathcal{R}_{f}^\nu(X_i,xh)\right)dx\\
		&\quad  +\sum\limits_{j=1}^\nu\sum_{\substack{l=1 \\ j+l>\nu}}^\nu\sum\limits_{k_1,\ldots, k_j=1}^{d}\sum\limits_{q_1,\ldots, q_l=1}^{d} \frac{h^{j+l}}{j!l!}\frac{\partial^j m(X_i)}{\partial x^{k_1} \cdots \partial x^{k_j}}\frac{\partial^l f(X_i)}{\partial x^{q_1} \cdots \partial x^{q_l}}\int K(x)  x^{k_1}\cdots x^{k_j}x^{q_1}\cdots x^{q_l} dx\\
		&\quad  + \sum\limits_{j=1}^{\nu -1}\frac{h^{\nu}}{j!(\nu-j)!} \sum\limits_{p=1}^{d} \frac{\partial^j m(X_i)}{(\partial x^{p})^j}\frac{\partial^{\nu - j} f(X_i)}{(\partial x^{p})^{\nu - j}}\left(\int K^u\left(t\right)  (t)^\nu dt\right)\\
		&\quad + \int K(x)\left(\sum\limits_{j=1}^\nu\frac{h^{j}}{j!} \sum\limits_{k_1,\ldots,k_j=1}^d \frac{\partial^jm(X_i)}{\partial x^{k_1}\cdots \partial x^{k_j}}x^{k_1}\cdots x^{k_j}\right)\mathcal{R}_{f}^\nu(X_i,xh)dx. 
	\end{align*}
	Let us show that the expectation of the norm of every term on the right hand side of (\ref{Lm1.7}) is $O(h^\nu)$. When no Taylor remainder is involved, the result is immediate by the Cauchy-Schwarz inequality, (T1), (T2) and (T5). Let us illustrate how to proceed in a particular case involving $R^{\nu}_f$. Since the support of the univariate kernel $K^u$ is $[-1,1]$, it follows that
	\begin{align*}
		&\mathbb{E}\left[\left|\left|\int K(x)\left(\sum\limits_{j=1}^\nu\frac{h^{j}}{j!} \sum\limits_{k_1,\ldots,k_j=1}^d \frac{\partial^jm(X_i)}{\partial x^{k_1}\cdots \partial x^{k_j}}x^{k_1}\cdots x^{k_j}\right)\mathcal{R}_{f}^\nu(X_i,xh)dx  	\right|\right|\right]\\
		\le &\sum\limits_{j=1}^\nu\frac{h^{j}}{j!}\sum\limits_{k_1,\ldots,k_j=1}^d\mathbb{E}\left[\left|\left| \frac{\partial^jm(X_i)}{\partial x^{k_1}\cdots \partial x^{k_j}}\right|\right|\int |K(x)|    |x^{k_1}\cdots x^{k_j}| \ |\mathcal{R}_{f}^\nu(X_i,xh)| dx \right]\\
		\le &\sum\limits_{j=1}^\nu\frac{h^{j}}{j!}\sum\limits_{k_1,\ldots,k_j=1}^d\mathbb{E}\left[\left|\left| \frac{\partial^jm(X_i)}{\partial x^{k_1}\cdots \partial x^{k_j}}\right|\right|  \sup_{||x_0||\le \sqrt{d}h}|\mathcal{R}_{f}^\nu(X_i,x_0)|  \right]\int |K(x)|    |x^{k_1}\cdots x^{k_j}| dx\\
		\le &\mathbb{E}\left[\sup_{||x_0|| \le \sqrt{d}h}|\mathcal{R}_{f}^\nu(X_i,x_0)|^2  \right]^\frac{1}{2} \left(\int |K(x)| dx\right) \sum\limits_{j=1}^\nu\frac{h^{j}}{j!}\sum\limits_{k_1,\ldots,k_j=1}^d\mathbb{E}\left[\left|\left| \frac{\partial^jm(X_i)}{\partial x^{k_1}\cdots \partial x^{k_j}}\right|\right|^2 \right]^\frac{1}{2}. \tag{Lm1.8}\label{Lm1.8}
	\end{align*}
	By (\ref{Sm.1}), for every $x_0$, the Taylor expansion remainder $\mathcal{R}_{f}^\nu(X_i,x_0)$ verifies the bound
	\begin{align*}
		|R^{\nu}_f(X_i,x_0)| &\le   \frac{||x_0||^{\nu+1}}{(\nu + 1)!}\left(\sum\limits_{k_1,\ldots,k_{\nu+1}=1}^d \sup_{t\in[0,1]}\left| \frac{\partial^{\nu+1}f(X_i+tx_0)}{\partial x^{k_{1}}\cdots \partial x^{k_{\nu+1}}}\right|^2\right)^\frac{1}{2},
	\end{align*}	
	Therefore, for $n$ large enough so that $\sqrt{d}h\le h_1$, where $h_1$ is defined as in condition (T2), it holds
	\begin{align*}
		\mathbb{E}\left[\sup_{||x_0|| \le \sqrt{d}h}|\mathcal{R}_{f}^\nu(X_i,x_0)|^2  \right]^\frac{1}{2} &\le  \frac{(\sqrt{d}h)^{\nu + 1}}{(\nu + 1)!}  \mathbb{E}\left[\sup_{||x_0|| \le \sqrt{d}h}\sum\limits_{k_1,\ldots,k_{\nu+1}=1}^d \sup_{t\in[0,1]}  \left| \frac{\partial^{\nu+1}f(X_i+tx_0)}{\partial x^{k_{1}}\cdots \partial x^{k_{\nu+1}}}\right|^2\right]^\frac{1}{2} \\
		&\le \frac{(\sqrt{d}h)^{\nu + 1}}{(\nu + 1)!}  \left(\sum\limits_{k_1,\ldots,k_{\nu+1}=1}^d  \mathbb{E}\left[ \sup_{||x'_0|| \le \sqrt{d}h}\left| \frac{\partial^{\nu+1}f(X_i+x'_0)}{\partial x^{k_{1}}\cdots \partial x^{k_{\nu+1}}}\right|^2\right]\right)^\frac{1}{2}\\
		&\le   \sum\limits_{k_1,\ldots,k_{\nu+1}=1}^d  \frac{(\sqrt{d}h)^{\nu + 1}}{(\nu + 1)!}\mathbb{E}\left[ \sup_{||x'_0|| \le h_1}\left| \frac{\partial^{\nu+1}f(X_i+x'_0)}{\partial x^{k_{1}}\cdots \partial x^{k_{\nu+1}}}\right|^2 \ind{X_i+ x'_0 \in \text{supp}(\mathbb{P}_X)}\right]^\frac{1}{2}\\
		&= O(h^{\nu + 1}),
	\end{align*}	
	since $||tx_0||\le \sqrt{d}h\le h_1$. Thus, by (\ref{Lm1.8}) and (T2),
	$$\mathbb{E}\left[\left|\left|\int K(x)\left(\sum\limits_{j=1}^\nu\frac{h^{j}}{j!} \sum\limits_{k_1,\ldots,k_j=1}^d \frac{\partial^jm(X_i)}{\partial x^{k_1}\cdots \partial x^{k_j}}x^{k_1}\cdots x^{k_j}\right)\mathcal{R}_{f}^\nu(X_i,xh)dx  	\right|\right|\right] = O(h^{\nu +1})O(h) = O(h^{\nu +2}).$$
	The rest of the terms can be addressed in a similar manner, so $D(X_i,Y_i)\ind{\sqrt{d}h<d(X_i,\partial \text{supp}(\mathbb{P}_X))} = O_{L^1}(h^\nu)$. To prove (1), by (\ref{Lm1.1}), we also need to show $D(X_i,Y_i)\ind{\sqrt{d}h\ge d(X_i,\partial \text{supp}(\mathbb{P}_X))} = O_{L^1}(h^\nu)$. It follows from previous arguments, condition (T4), the Cauchy-Schwarz inequality and Fubuni's theorem, that
	\begin{align*}
		&\mathbb{E}\left[\left|\left|(m(X_i)-Y_i)\left(\mathbb{E}\left[K_h(X_i-X_k)|X_i\right] - f(X_i)\right)\right|\right|\ind{\sqrt{d}h\ge d(X_i,\partial \text{supp}(\mathbb{P}_X))}\right]\\
		\le\ &\mathbb{E}\left[\left|\left|(m(X_i)-Y_i)\right|\right|^2\right]^\frac{1}{2}\mathbb{E}\left[ \left|\int K(x)(f(X_i+xh)-f(X_i))dx\right|^2\ind{\sqrt{d}h\ge d(X_i,\partial \text{supp}(\mathbb{P}_X))}\right]^\frac{1}{2} \\
		\le\ &\mathbb{E}\left[\left|\left|(m(X_i)-Y_i)\right|\right|^2\right]^\frac{1}{2}\int |K(x)| dx \mathbb{E}\left[ \sup_{x_0\in [-1,1]^d} |f(X_i+x_0h) - f(X_i)|^2 \ind{\sqrt{d}h\ge d(X_i,\partial \text{supp}(\mathbb{P}_X))}\right]^\frac{1}{2} \\
		\le\ &\mathbb{E}\left[\left|\left|(m(X_i)-Y_i)\right|\right|^2\right]^\frac{1}{2}\int |K(x)| dx\ 2\mathbb{E}\left[ \sup_{||x'_0||\le \sqrt{d}h} |f(X_i+x'_0)|^2 \ind{\sqrt{d}h\ge d(X_i,\partial \text{supp}(\mathbb{P}_X))}\right]^\frac{1}{2} \\
		=\ &O(h^{\nu}).
	\end{align*}
	Similarly, if $n$ is large enough so that $\sqrt{d}h\le h_3$ where $h_3$ is defined as in (T4), we have
	\begin{align*}
		&\mathbb{E}\left[\left|\left|\mathbb{E}\left[K_h(X_i-X_k)(m(X_k)-m(X_i))|X_i\right]\right|\right|\ind{\sqrt{d}h\ge d(X_i,\partial \text{supp}(\mathbb{P}_X))}\right]\\
		=\ &\mathbb{E}\left[\left|\left|\int K(x) \left(m(X_i+xh) - m(X_i)\right) f(X_i+xh)dx \right|\right|\ind{\sqrt{d}h\ge d(X_i,\partial \text{supp}(\mathbb{P}_X))}\right]\\
		\le\ & \int |K(x)| dx \  \mathbb{E}\left[ \sup_{||x_0||\le h_3} ||m(X_i+x_0) - m(X_i)|| \ind{h_3\ge d(X_i,\partial \text{supp}(\mathbb{P}_X)), \ X_i+ x_0 \in \text{supp}(\mathbb{P}_X)}\right] \times \\
		&\quad \sup\limits_{d(x,\partial \text{supp}(\mathbb{P}_X))\le 2\sqrt{d}h} f(x)\\
		=\ &O(h^\nu),
	\end{align*}
	where we have implicitly used that, if $d(X_i,\partial \text{supp}(\mathbb{P}_X))\le \sqrt{d}h$, then $d(X_i + x_0,\partial \text{supp}(\mathbb{P}_X))\le 2\sqrt{d}h$ for all $x_0$ such that $||x_0||\le \sqrt{d}h$.
	Therefore, $D(X_i,Y_i)\ind{\sqrt{d}h\ge d(X_i,\partial \text{supp}(\mathbb{P}_X))} = O_{L^1}(h^\nu)$ and (1) holds.
	
	We only give a sketch proof of (2) due to the similarity with the proof of (1). Note that
	$$\mathbb{E}[K_h(X_i - X_k)(Y_k - Y_i) | X_i] - h^\nu B(X_i) = \mathbb{E}[K_h(X_i - X_k)(m(X_k) - m(X_i)) | X_i] - h^\nu B(X_i).$$
	Trivially, since $h=o(1)$, $\left|\left|\mathbb{E}[K_h(X_i - X_k)(m(X_k) - m(X_i)) | X_i]/h^\nu -  B(X_i)\right|\right|\ind{\sqrt{d}h\ge d(X_i,\partial \text{supp}(\mathbb{P}_X))}\rightarrow 0$ a.s. Moreover, by (T1), (T2) and (T4), and for $n$ large enough, there will be some constant $M_0>0$ such that
	\begin{align*}
		&\left|\left|\frac{\mathbb{E}[K_h(X_i - X_k)(m(X_k) - m(X_i)) | X_i]}{h^\nu} -  B(X_i)\right|\right|^2\ind{\sqrt{d}h\ge d(X_i,\partial \text{supp}(\mathbb{P}_X))}\\
		\le\ &  2\left|\int |K(x)| dx\right|^2 \sup_{||x_0||\le h_3} ||m(X_i+x_0) - m(X_i)||^2 \ind{h_3\ge d(X_i,\partial \text{supp}(\mathbb{P}_X)), \ X_i+ x_0 \in \text{supp}(\mathbb{P}_X)}\times\\
		&\quad \sup\limits_{d(x,\partial \text{supp}(\mathbb{P}_X))\le 2\sqrt{d}h} \left|\frac{f(x)}{h^\nu}\right|^2 + 2||B(X_i)||^2\\
		\le\ &  M_0 \sup_{||x_0||\le h_3} ||m(X_i+x_0) - m(X_i)||^2 \ind{h_3\ge d(X_i,\partial \text{supp}(\mathbb{P}_X)),\ X_i+ x_0 \in \text{supp}(\mathbb{P}_X)} + 2||B(X_i)||^2\\
		=\ &O_{L^1}(1).
	\end{align*}
	Thus, it follows from the dominated convergence theorem that
	$$\left|\left|\mathbb{E}[K_h(X_i - X_k)(Y_k - Y_i) | X_i]-  h^\nu B(X_i)\right|\right|\ind{\sqrt{d}h\ge d(X_i,\partial \text{supp}(\mathbb{P}_X))} = o_{L^2}(h^\nu).$$ 
	If $\sqrt{d}h<d(X_i,\partial \text{supp}(\mathbb{P}_X))$ we have, by (\ref{Lm1.3}), (\ref{Lm1.4}), (\ref{Lm1.5}), (\ref{Lm1.6}) and (T2),
	\begin{align*}
		&\mathbb{E}[K_h(X_i - X_k)(m(X_k) - m(X_i)) | X_i] - h^\nu B(X_i)\\
		=\ &f(X_i)\int K(x)\mathcal{R}_{m}^\nu(X_i,xh)dx\\
		&\quad + \int K(x)\mathcal{R}_{m}^\nu(X_i,xh)\left(\sum\limits_{j=1}^\nu\frac{h^{j}}{j!} \sum\limits_{k_1,\ldots,k_j=1}^d \frac{\partial^jf(X_i)}{\partial x^{k_1}\cdots \partial x^{k_j}}x^{k_1}\cdots x^{k_j} + \mathcal{R}_{f}^\nu(X_i,xh)\right)dx\\
		&\quad  +\sum\limits_{j=1}^\nu\sum_{\substack{l=1 \\ j+l>\nu}}^\nu\sum\limits_{k_1,\ldots, k_j=1}^{d}\sum\limits_{q_1,\ldots, q_l=1}^{d} \frac{h^{j+l}}{j!l!}\frac{\partial^j m(X_i)}{\partial x^{k_1} \cdots \partial x^{k_j}}\frac{\partial^l f(X_i)}{\partial x^{q_1} \cdots \partial x^{q_l}}\int K(x)  x^{k_1}\cdots x^{k_j}x^{q_1}\cdots x^{q_l} dx\\
		&\quad + \int K(x)\left(\sum\limits_{j=1}^\nu\frac{h^{j}}{j!} \sum\limits_{k_1,\ldots,k_j=1}^d \frac{\partial^jm(X_i)}{\partial x^{k_1}\cdots \partial x^{k_j}}x^{k_1}\cdots x^{k_j}\right)\mathcal{R}_{f}^\nu(X_i,xh)dx\\
		=\  &O_{L^2}(h^{\nu + 1}).
	\end{align*}
	Thus,
	$$\left|\left|\mathbb{E}[K_h(X_i - X_k)(Y_k - Y_i) | X_i]-  h^\nu B(X_i)\right|\right|\ind{\sqrt{d}h< d(X_i,\partial \text{supp}(\mathbb{P}_X))} = o_{L^2}(h^{\nu}).$$ 
	Moreover, $B(X) = O_{L^2}(1)$ by (T2) and (T5), so (5) holds.
	
	The proof of (3) is quite simple. By (A1) and (T5),  $||f||_\infty||K||_\infty <\infty$ and, since the support of $K$ is $[-1,1]^d$,
	$$\mathbb{E}[|K_h(X_i-X_k)|\ |X_i] = \int |K(x)| f(X_i + hx) dx \le ||f||_\infty \int |K(x)| dx \le 2^d||f||_\infty||K||_\infty.$$
	
	To prove (4) note that, by the i.i.d. hypothesis and the symmetry of $K^u$ around $0$,
	\begin{align*}
		\mathbb{E}[|K_h(X_i-X_k)|^2||Y_i - Y_k||^2] &\le \frac{4||K||_\infty}{h^d}\mathbb{E}[|K_h(X_i-X_k)| \ ||Y_i||^2]\\
		&= \frac{4||K||_\infty}{h^d}\mathbb{E}\left[\mathbb{E}\left[|K_h(X_i-X_k)| \ | X_i,Y_i\right]||Y_i||^2\right] \\
		&= \frac{4||K||_\infty}{h^d}\mathbb{E}\left[\mathbb{E}\left[|K_h(X_i-X_k)| \ | X_i\right]||Y_i||^2\right]\\
		&\le \frac{2^{d+2}||f||_\infty ||K||^2_\infty}{h^d}\sup_{n\in \mathbb{N}}\mathbb{E}\left[||Y_i||^2\right] = O\left(\frac{1}{h^d}\right).
	\end{align*}
	
	Now we prove (5). On the one hand, if $\sqrt{d}h<d(X_i,\partial \text{supp}(\mathbb{P}_X))$ we have, by (\ref{Lm1.2}),
	$$\mathbb{E}\left[K_h(X_i-X_k)-f(X_i)|X_i\right] =\frac{h^{\nu}}{\nu!} \sum\limits_{k=1}^d \frac{\partial^\nu f(X_i) }{(\partial x^{k})^\nu}\left(\int K^u(t)t^\nu dt\right)+ \int K(x)\mathcal{R}_{f}^\nu(X_i,xh)dx.$$
	The support of $K$ is $[-1,1]^d$ and, since $f$ is $\nu+1$ times continuously differentiable, it follows that
	\begin{align*}
		&\mathbb{E}\left[|\mathbb{E}\left[K_h(X_i-X_k)-f(X_i)|X_i\right]|^4\ind{\sqrt{d}h<d(X_i,\partial \text{supp}(\mathbb{P}_X))}\right] \\
		\le\ &\frac{2^3h^{4\nu}}{|\nu!|^4}\left(\int K^u(t)t^\nu dt\right)^4\mathbb{E}\left[\left|\left|\sum\limits_{k=1}^d \frac{\partial^\nu f(X_i) }{(\partial x^{k})^\nu}\right|\right|^4\right]\\
		&\quad + \frac{2^3(\sqrt{d}h)^{4(\nu+1)}}{|(\nu+1)!|^4}\mathbb{E}\left[\left(\int |K(x)| \left(\sum\limits_{k_1,\ldots,k_{\nu+1}=1}^d \sup_{t\in[0,1]}\left|\frac{\partial^{\nu+1}f(X_i+txh)}{\partial x^{k_{1}}\cdots \partial x^{k_{\nu+1}}}\right|^2\right)^\frac{1}{2} dx\right)^4\right]\\
		\le\ &\frac{2^3h^{4\nu}}{|\nu!|^4}\left(\int K^u(t)t^\nu dt\right)^4d^3\sum\limits_{k=1}^d \mathbb{E}\left[\left|\left|\frac{\partial^\nu f(X_i) }{(\partial x^{k})^\nu}\right|\right|^4\right]\\
		&\quad + \frac{2^3(\sqrt{d}h)^{4(\nu+1)}}{|(\nu+1)!|^4}\left(\int |K(x)|   dx\right)^4 d^{\nu+1}\sum\limits_{k_1,\ldots,k_{\nu+1}=1}^d\mathbb{E}\left[\sup_{||x_0|| \le h_1} \left|\frac{\partial^{\nu+1}f(X_i+x_0)}{\partial x^{k_{1}}\cdots \partial x^{k_{\nu+1}}}\right|^4\ind{X_i + x_0 \in \text{supp}(\mathbb{P}_X)}\right],
	\end{align*}
	where the last inequality holds for $n$ large enough so that $||t xh||\le ||\sqrt{d}h||\le h_1$. Thus,  by (T2), we have
	$$\left|\mathbb{E}\left[K_h(X_i-X_k)-f(X_i)|X_i\right]\right|^4\ind{\sqrt{d}h<d(X_i,\partial \text{supp}(\mathbb{P}_X))}= O_{L^1}(h^{4\nu}).$$
	On the other hand, by (T4),
	\begin{align*}
		&\left|\mathbb{E}\left[K_h(X_i-X_k)-f(X_i)|X_i\right]\right|^4\ind{\sqrt{d}h\ge d(X_i,\partial \text{supp}(\mathbb{P}_X))}\\
		=\ &\left|\int K(x)(f(X_i+xh)-f(X_i))dx\right|^4\ind{\sqrt{d}h\ge d(X_i,\partial \text{supp}(\mathbb{P}_X))}\\
		\le\ &\left|\int |K(x)| dx \right|^4 \sup_{||x_0||\le \sqrt{d}h}\left|f(X_i+x_0)-f(X_i)\right|^4\ind{\sqrt{d}h\ge d(X_i,\partial \text{supp}(\mathbb{P}_X))}\\
		\le\ &\left|\int |K(x)| dx \right|^42^4 \sup_{||x_0||\le \sqrt{d}h}\left|f(X_i+x_0)\right|^4\ind{\sqrt{d}h\ge d(X_i,\partial \text{supp}(\mathbb{P}_X))}\\
		=\ &O_{L^1}(h^{4\nu}),
	\end{align*}
	so (5) holds.
	
	We omit the proof of (6) as it is analogous to those of (1) and (5).
	
	The proof of (7) and (8) is as follows. Under $\mathcal{H}_0$,
	\begin{align*}
		&\mathbb{E}\left[(c(Z_i,\cdot)-b(X_k,\cdot))\otimes K_h(X_i-X_k)(Y_i-m(X_k))|Z_i,Y_i\right]\\
		=\ &\mathbb{E}\left[(c(Z_i,\cdot)-b(X_k,\cdot)) K_h(X_i-X_k)|Z_i\right]\otimes\mathcal{E}_i\\
		&+ \mathbb{E}\left[(c(Z_i,\cdot)-b(X_k,\cdot))\otimes K_h(X_i-X_k)(m(X_i)-m(X_k))|Z_i\right]\\
		=\ &\left(c(Z_i,\cdot)\mathbb{E}\left[K_h(X_i-X_k)|X_i\right] - \mathbb{E}\left[b(X_k,\cdot)K_h(X_i-X_k)|X_i\right]\right)\otimes \mathcal{E}_i\\
		&+ \mathbb{E}\left[(c(Z_i,\cdot)-b(X_k,\cdot))\otimes K_h(X_i-X_k)(m(X_i)-m(X_k))|Z_i\right].\\	
		=\ &(c(Z_i,\cdot)-b(X_i,\cdot))\mathbb{E}\left[K_h(X_i-X_k)|X_i\right]\otimes \mathcal{E}_i - \mathbb{E}\left[(b(X_k,\cdot)-b(X_i,\cdot))K_h(X_i-X_k)|X_i\right]\otimes \mathcal{E}_i \tag{Lm1.9}\label{Lm1.9}\\
		&+ \mathbb{E}\left[(c(Z_i,\cdot)-b(X_k,\cdot))\otimes K_h(X_i-X_k)(m(X_i)-m(X_k))|Z_i\right].				 
	\end{align*}
	By (5) and the boundedness of $c$, we have
	\begin{align*}
		(c(Z_i,\cdot)-b(X_i,\cdot))\mathbb{E}\left[K_h(X_i-X_k)|X_i\right]\otimes \mathcal{E}_i &= (c(Z_i,\cdot)-b(X_i,\cdot))(f(X_i)+O_{L^4}(h))\otimes \mathcal{E}_i\\
		&= (c(Z_i,\cdot)-b(X_i,\cdot))f(X_i)\otimes \mathcal{E}_i+O_{L^4}(h)\otimes \mathcal{E}_i, \tag{Lm1.10}\label{Lm1.10}			 
	\end{align*}
	since $||b(X_i,\cdot)||^2 = \mathbb{E}[\langle c(Z_i,\cdot), b(X_i,\cdot)\rangle|X_i] = \mathbb{E}[b(X_i,Z_i)|X_i]\le 1$. Furthermore, using the Cauchy-Schwarz inequality twice (based on $|K_h(X_i-X_k)| = \sqrt{|K_h(X_i-X_k)|}^2$) and also (3), we have
	$$\mathbb{E}\left[||\mathbb{E}\left[(b(X_k,\cdot)-b(X_i,\cdot))K_h(X_i-X_k)|X_i\right]||^4\right] \le 2^{3d}||f||_\infty^3||K||_\infty^3 \mathbb{E}\left[||b(X_k,\cdot)-b(X_i,\cdot)||^4 |K_h(X_i-X_k)|\right].$$
	Since $\langle b(X_i,\cdot),b(X_k,\cdot) \rangle = \mathbb{E}[b(X_i,Z_k)|X_i,X_k] = \mathbb{E}[b(X_k,Z_i)|X_i,X_k]$,
	it follows, by a symmetry argument, that
	\begin{align*}
		&\mathbb{E}\left[||b(X_k,\cdot)-b(X_i,\cdot)||^4|K_h(X_i-X_k)|\right]\\
		\le \ &\mathbb{E}\left[|\mathbb{E}[b(X_i,Z_i)-b(X_k,Z_i) + b(X_k,Z_k) - b(X_i,Z_k)|X_i,X_k]|^2|K_h(X_i-X_k)|\right]\\
		\le \ &\mathbb{E}\left[\mathbb{E}[|b(X_i,Z_i)-b(X_k,Z_i) + b(X_k,Z_k) - b(X_i,Z_k)|^2|K_h(X_i-X_k)| \ |X_i,X_k]\right]\\
		\le \ &4\mathbb{E}\left[|b(X_i,Z_i)-b(X_k,Z_i)|^2|K_h(X_i-X_k)|\right]\\
		= \ &4\mathbb{E}\left[\int |b(X_i,Z_i)-b(X_i+xh,Z_i)|^2|K(x)|f(X_i+xh)dx\right]\\
		\le \ &4||f||_\infty\mathbb{E}\left[\int |b(X_i,Z_i)-b(X_i+xh,Z_i)|^2|K(x)|dx \ind{\sqrt{d}h<d(X_i,\partial \text{supp}(\mathbb{P}_X))}\right]\\
		& + 16\int |K(x)| dx\  \sup\limits_{d(x,\partial \text{supp}(\mathbb{P}_X))\le 2\sqrt{d}h} f(x),
	\end{align*}
	where, in the last inequality, we have used that $d(X_i+xh, \partial \text{supp}(\mathbb{P}_X))\ind{\sqrt{d}h\ge d(X_i,\partial \text{supp}(\mathbb{P}_X))}\le 2\sqrt{d}h$.
	By the differentiability assumptions on $b(x,z')$ with respect to the first argument we have, for every $x\in [-1,1]^d$,
	$$|b(X_i,Z_i)-b(X_i+xh,Z_i)|^2\ind{\sqrt{d}h<d(X_i,\partial \text{supp}(\mathbb{P}_X))} \le  dh^2\sum\limits_{j=1}^d \sup\limits_{t\in [0,1]}\left| \frac{\partial b(X_i+txh,Z_i)}{\partial x^j} \right|^2.$$
	Therefore, since $K$ has support in $[-1,1]^d$ and assuming $n$ is large enough so that $||t xh||\le  \sqrt{d}h\le h_2$,
	\begin{align*}
		&\mathbb{E}\left[||b(X_k,\cdot)-b(X_i,\cdot)||^4|K_h(X_i-X_k)|\right]\\
		\le \ &4||f||_\infty dh^2 \int |K(x)| dx \sum\limits_{j=1}^d\mathbb{E}\left[ \sup\limits_{||x_0||\le h_2}\left| \sum\limits_{j=1}^d \frac{\partial b(X_i+x_0,Z_i)}{\partial x^j} \right|^2\right] + 16\int |K(x)| dx\  \sup\limits_{d(x,\partial \text{supp}(\mathbb{P}_X))\le 2\sqrt{d}h} f(x),
	\end{align*}
	and, ultimately,
	\begin{equation}\tag{Lm1.11}\label{Lm1.11}	
		\mathbb{E}\left[||\mathbb{E}\left[(b(X_k,\cdot)-b(X_i,\cdot))K_h(X_i-X_k) |X_i\right]||^4\right] =O(h^2) + O(h^\nu) = O(h^2),
	\end{equation}	
	which is equivalent to $\mathbb{E}\left[(b(X_k,\cdot) - b(X_i,\cdot))K_h(X_i-X_k)|X_i\right] =  O_{L^4}(\sqrt{h})$, so (7) holds. In addition, the following bound is inmediate, 
	\begin{align*}
		&||\mathbb{E}\left[(c(Z_i,\cdot)-b(X_k,\cdot))\otimes K_h(X_i-X_k)(m(X_i)-m(X_k))|Z_i,Y_i\right]||\\
		\le \ &\mathbb{E}\left[||c(Z_i,\cdot)-b(X_k,\cdot)|| \ |K_h(X_i-X_k)| \ ||m(X_i)-m(X_k)||\ |Z_i,Y_i\right] \\
		\le \ &2\mathbb{E}\left[|K_h(X_i-X_k)| \ ||m(X_i)-m(X_k)||\ |X_i\right]\\
		=\ &O_{L^2}(h), 	 \tag{Lm1.12}\label{Lm1.12}		 
	\end{align*}
	where the last equality follows from (6). Therefore, by combination of (\ref{Lm1.9}), (\ref{Lm1.10}), (\ref{Lm1.11}) and (\ref{Lm1.12}), we have
	$$\mathbb{E}\left[(c(Z_i,\cdot)-b(X_k,\cdot))\otimes K_h(X_i-X_k)(Y_i-m(X_k))|Z_i,Y_i\right] = (c(Z_i,\cdot)-b(X_i,\cdot))\otimes f(X_i)\mathcal{E}_i + o_{L^2}(1),$$
	since $o_{L^4}(1)\otimes\mathcal{E}_i = o_{L^2}(1)$ by the Cauchy-Schwarz inequality.
\end{proof}

\begin{lemma}
	\label{Lm2}
	Assume that (A1)-(A3) and (T1)-(T5) hold for the model in (\ref{Sm.5}) and let $\{\eta_n(Z,Y)\}_{n=1}^\infty$ be any sequence of measurable functions of $(Z,Y)$, taking values in some normed space, such that $\eta_n(Z,Y) = O_{L^2}(1)$. Let $\hat{f}'_{h}(x), \hat{f}_{h}(x)$ and $\hat{m}_{h}(x)$ be defined as 
	$$ \hat{f}'_{h}(x) = \frac{1}{n}\sum\limits_{k=1}^n |K_h(x-X_k)|, \quad \hat{f}_{h}(x) = \frac{1}{n}\sum\limits_{k=1}^n K_h(x-X_k), \quad \hat{m}_{h}(x) = \frac{1}{n\hat{f}_{h}(x)}\sum\limits_{k=1}^n K_h(x-X_k) Y_k.$$
	Then:
	\begin{enumerate}
		\item $(\hat{f}'_{h}(X_i)-f(X_i))\eta_n(Z_i,Y_i) = O_{L^2}(1)$, $\hat{f}'_{h}(X_i)\eta_n(Z_i,Y_i) = O_{L^2}(1)$ and $\hat{f}_{h}(X_i)\eta_n(Z_i,Y_i) = O_{L^2}(1)$.
		\item $(\hat{f}_h(X_i)-f(X_i))\mathcal{E}_i = O_{L^2}\left(\sqrt{\frac{1}{nh^{d}} + h^{2\nu}}\right)$.
		\item $\hat{f}_h(X_i) (m(X_i) - \hat{m}_h(X_i)) = O_{L^2}\left(\sqrt{\frac{1}{nh^{d}} + h^{2\nu}}\right)$.
		
		\item $\left(\frac{1}{n}\sum\limits_{k=1}^n K_h(X_i-X_k)(c(Z_i,\cdot)-c(Z_k,\cdot)) -\hat{f}_h(X_i)(c(i,\cdot)-b(i,\cdot))\right)\otimes\mathcal{E}_i = o_{L^2}(1)$
	\end{enumerate}
\end{lemma}

\begin{proof}[\bfseries\textup{Proof of Lemma \ref{Lm2}}]
	
	Let us start by decomposing and bounding the squared norm of $\hat{f}'_{h}(X_i)-f(X_i)$:
	\begin{align*}
		\left| \hat{f}'_{h}(X_i)-f(X_i) \right|^2 &= \frac{1}{n^2}\sum\limits_{k=1}^n\sum\limits_{l=1}^n(|K_h(X_i-X_k)| - f(X_i))(|K_h(X_i-X_l)|- f(X_i))\\
		&= \frac{1}{n^2}\sum\limits_{k=1}^n(|K_h(X_i-X_k)| - f(X_i))^2\\
		&\quad + \frac{1}{n^2} \sum_{\substack{k,l=1 \\ k\neq l}}^{n}(|K_h(X_i-X_k)| - f(X_i))(|K_h(X_i-X_l)|- f(X_i))\\
		&= \frac{1}{n^2}\left(\frac{K(0)}{h^d} - f(X_i)\right)^2 + \frac{1}{n^2} \sum_{\substack{k=1 \\ k\neq i}}^{n}(|K_h(X_i-X_k)| - f(X_i))^2\\
		&\quad + \frac{1}{n^2}\sum_{\substack{k,l=1 \\ k,l\neq i \\ k\neq l}}^{n}(|K_h(X_i-X_k)| - f(X_i))(|K_h(X_i-X_l)|- f(X_i))\\
		&\quad + \frac{2}{n^2}\sum_{\substack{k=1 \\ k\neq i}}^{n}(|K_h(X_i-X_k)| - f(X_i))\left(\frac{K(0)}{h^d} - f(X_i)\right)\\
		&\le \frac{2||K||_\infty^2}{n^2h^{2d}} + \frac{2||f||_\infty^2}{n^2} + \frac{3}{n}\left(\frac{||K||_\infty}{h^d}+||f||_\infty\right)\frac{1}{n} \sum_{\substack{k=1 \\ k\neq i}}^{n}| \ |K_h(X_i-X_k)| - f(X_i)| \label{Lm2.1}\tag{Lm2.1}\\
		&\quad + \frac{1}{n^2}\sum_{\substack{k,l=1 \\ k,l\neq i \\ k\neq l}}^{n}(|K_h(X_i-X_k)| - f(X_i))(|K_h(X_i-X_l)|- f(X_i)).
	\end{align*}
	By Lemma \ref{Lm1}-(3), $\mathbb{E}[|\ |K_h(X_i-X_l)|- f(X_i)||X_i] \le ||f||_\infty(1+2^d||K||_\infty)$ and, thus, it follows that
	\begin{align*}
		\mathbb{E}\left[||\eta_n(Z_i,Y_i)||^2| \ |K_h(X_i-X_k)| - f(X_i)|\right] &= \mathbb{E}\left[||\eta_n(Z_i,Y_i)||^2\mathbb{E}\left[| \ |K_h(X_i-X_k)| - f(X_i)|\ |X_i,Y_i\right]\right]\\
		&= \mathbb{E}\left[||\eta_n(Z_i,Y_i)||^2\mathbb{E}\left[|\ |K_h(X_i-X_k)| - f(X_i)|\ |X_i\right]\right]\\
		&\le ||f||_\infty(1+2^d||K||_\infty)\mathbb{E}\left[||\eta_n(Z_i,Y_i)||^2\right],
	\end{align*}
	for all $k\neq i$. Similarly, given three different indices $k,l,i$, we have
	\begin{align*}
		&\mathbb{E}\left[||\eta_n(Z_i,Y_i)||^2(|K_h(X_i-X_k)| - f(X_i))(|K_h(X_i-X_l)| - f(X_i))\right]\\  
		= \ &\mathbb{E}\left[||\eta_n(Z_i,Y_i)||^2\mathbb{E}\left[|K_h(X_i-X_k)| - f(X_i) | X_i,Y_i\right]^2\right] \\
		= \ &\mathbb{E}\left[||\eta_n(Z_i,Y_i)||^2\mathbb{E}\left[|K_h(X_i-X_k)| - f(X_i) | X_i\right]^2\right] \\
		\le \ &||f||^2_\infty(1+2^d||K||_\infty)^2\mathbb{E}\left[||\eta_n(Z_i,Y_i)||^2\right].
	\end{align*}
	Therefore, taking into account the inequality in (\ref{Lm2.1}) and also that $\mathbb{E}\left[||\eta_n(Z_i,Y_i)||^2\right]=O(1)$, it holds that
	$$\mathbb{E}\left[\left| \hat{f}'_{h}(X_i)-f(X_i)\right|^2||\eta_n(Z_i,Y_i)||^2\right] = O\left(\frac{1}{n^2h^{2d}}\right) + O\left(\frac{1}{n^2}\right) +  O\left(\frac{1}{nh^d}+\frac{1}{n}\right) + O(1) = O(1).$$
	Moreover, since $||f||_\infty<\infty$ by (T5),
	$$\mathbb{E}\left[\left| \hat{f}'_{h}(X_i)-f(X_i)\right|^2||\eta_n(Z_i,Y_i)||^2\right] =O(1) \implies \mathbb{E}\left[\left| \hat{f}'_{h}(X_i)\right|^2||\eta_n(Z_i,Y_i)||^2\right] =O(1).$$
	We omit the proof of $\hat{f}_{h}(X_i)\eta_n(Z_i,Y_i) = O_{L^2}(1)$ since it is completely analogous.
	
	The proof of (2) is based on bounding $\left| \hat{f}_{h}(X_i)-f(X_i) \right|^2$ in the following manner:
	\begin{align*}
		\left| \hat{f}_{h}(X_i)-f(X_i) \right|^2 &\le \frac{2||K||_\infty^2}{n^2h^{2d}} + \frac{2||f||_\infty^2}{n^2} + \frac{3}{n}\left(\frac{||K||_\infty}{h^d}+||f||_\infty\right)\frac{1}{n} \sum_{\substack{k=1 \\ k\neq i}}^{n}| K_h(X_i-X_k) - f(X_i)|\\
		&\quad + \frac{1}{n^2}\sum_{\substack{k,l=1 \\ k,l\neq i \\ k\neq l}}^{n}(K_h(X_i-X_k)- f(X_i))(K_h(X_i-X_l)- f(X_i)).
	\end{align*}
	By Lemma \ref{Lm1}-(5), we have
	\begin{align*}
		\mathbb{E}\left[||\mathcal{E}_i||^2\mathbb{E}\left[K_h(X_i-X_k) - f(X_i) |X_i\right]^2\right] &\le \mathbb{E}\left[||\mathcal{E}_i||^4\right]^\frac{1}{2}\mathbb{E}\left[\mathbb{E}\left[K_h(X_i-X_k) - f(X_i) |X_i\right]^4\right]^\frac{1}{2} = O(h^{2\nu}).
	\end{align*}
	Therefore, by previous arguments, the following asymptotic bound is verified:
	\begin{align*}
		\mathbb{E}\left[\left|\left| \hat{f}_{h}(X_i)-f(X_i) \right|\right|^2||\mathcal{E}_i||^2\right] &= O\left(\frac{1}{n^2h^{2d}}\right) + O\left(\frac{1}{n^2}\right) +  O\left(\frac{1}{nh^d}+\frac{1}{n}\right) + O(h^{2\nu}) = O\left(\frac{1}{nh^d}+h^{2\nu}\right),
	\end{align*}
	concluding the proof of (2). 
	
	The proof of (3) is similar. We have,
	\begin{align*}
		||\hat{f}_h(X_i)(\hat{m}_h(X_i) - m(X_i))||^2 &= \left|\left|\frac{1}{n}\sum\limits_{k=1}^n K_h(X_i-X_k) (Y_k - m(X_i))\right|\right|^2\\
		&=\frac{1}{n^2} \sum\limits_{k,l=1}^n \langle  K_h(X_i-X_k)(Y_k-m(X_i)),K_h(X_i-X_l)(Y_l-m(X_i))\\
		&= \frac{1}{n^2} \sum_{\substack{k=1 \\ k\neq i}}^{n} |K_h(X_i-X_k)|^2||Y_k-m(X_i)||^2 + \frac{1}{n^2}\frac{K^2(0)}{h^{2d}}||Y_i-m(X_i)||^2\\
		& \quad +  \frac{2}{n^2} \sum_{\substack{k=1 \\ k\neq i}}^{n} \left\langle  K_h(X_i-X_k)(Y_k-m(X_i)),\frac{K(0)}{h^d}(Y_i-m(X_i))\right\rangle\\
		& \quad +  \frac{1}{n^2} \sum_{\substack{k,l=1 \\ k,l\neq i \\ k\neq l }}^{n} \left\langle  K_h(X_i-X_k)(Y_k-m(X_i)),K_h(X_i-X_l)(Y_l-m(X_i))\right\rangle.
	\end{align*}
	Let $i,k,l$ be three different indices. Then, by Lemma \ref{Lm1}-(2), Lemma \ref{Lm1}-(4), (T1), the i.i.d. assumptions and the Cauchy-Schwarz inequality,
	\begin{align*}
		&\mathbb{E}[|K_h(X_i-X_k)|^2||Y_k-m(X_i)||^2]\le \frac{||K||_\infty}{h^d}\mathbb{E}[|K_h(X_i-X_k)| \ ||Y_k-m(X_i)||^2] = O\left(\frac{1}{h^d}\right),\\
		&\mathbb{E}[\left\langle  K_h(X_i-X_k)(Y_k-m(X_i)),(Y_i-m(X_i))\right\rangle]=\mathbb{E}[\left\langle  \mathbb{E}[K_h(X_i-X_k)(Y_k-m(X_i))|X_i,Y_i],(Y_i-m(X_i))\right\rangle]\\
		& \hspace{70.1mm} = \mathbb{E}[\left\langle  \mathbb{E}[K_h(X_i-X_k)(Y_k-m(X_i))|X_i],(Y_i-m(X_i))\right\rangle]\\
		& \hspace{70.1mm} = \mathbb{E}[\left\langle O_{L^2}(h^{\nu}),(Y_i-m(X_i))\right\rangle] = O(h^\nu),\\
		&\mathbb{E}[\left\langle  K_h(X_i-X_k)(Y_k-m(X_i)),K_h(X_i-X_l)(Y_l-m(X_i))\right\rangle]=\mathbb{E}[||\mathbb{E}[K_h(X_i-X_k)(m(X_k)-m(X_i))|X_i]||^2]\\
		&\hspace{89.9mm} = O(h^{2\nu}).
	\end{align*}
	Ultimately,
	$$\mathbb{E}\left[||\hat{f}_h(X_i)(\hat{m}_h(X_i) - m(X_i))||^2\right] = O\left(\frac{1}{nh^d}\right) + O\left(\frac{1}{n^2h^{2d}}\right) + O\left(\frac{h^\nu}{nh^d}\right) + O\left(h^{2\nu}\right) =  O\left(\frac{1}{nh^d} + h^{2\nu}\right).$$
	
	To prove (4), note that, by definition of $\hat{f}_h$,
	$$\frac{1}{n}\sum\limits_{k=1}^n K_h(X_i-X_k)(c(Z_i,\cdot)-c(Z_k,\cdot)) -\hat{f}_h(X_i)(c(i,\cdot)-b(i,\cdot)) = \frac{1}{n}\sum\limits_{k=1}^n K_h(X_i-X_k)(b(X_i,\cdot)-c(Z_k,\cdot)).$$
	Let us decompose and bound the square norm of the term in the right hand side of the last equality:
	\begin{align*}
		&\left|\left|\frac{1}{n}\sum\limits_{k=1}^n K_h(X_i-X_k)(b(X_i,\cdot)-c(Z_k,\cdot))\right|\right|^2\\
		=\ &\frac{1}{n^2}\sum\limits_{k=1}^n |K_h(X_i-X_k)|^2||b(X_i,\cdot)-c(Z_k,\cdot)||^2\\
		& + 	\frac{1}{n^2}\sum_{\substack{k,l=1 \\ k\neq l}}^{n} K_h(X_i-X_k) K_h(X_i-X_l)\langle b(X_i,\cdot)-c(Z_k,\cdot),b(X_i,\cdot)-c(Z_l,\cdot) \rangle\\
		=\ &\frac{1}{n^2}\sum_{\substack{k=1 \\ k\neq i}}^{n} |K_h(X_i-X_k)|^2||b(X_i,\cdot)-c(Z_k,\cdot)||^2 + \frac{1}{n^2}\frac{K^2(0)}{h^{2d}}||b(X_i,\cdot)-c(Z_i,\cdot)||^2\\
		& + 	\frac{2}{n^2}\frac{K(0)}{h^{d}}\sum_{\substack{k=1 \\ k\neq i}}^{n}K_h(X_i-X_k) \langle b(X_i,\cdot)-c(Z_k,\cdot),b(X_i,\cdot)-c(Z_i,\cdot) \rangle\\
		& + 	\frac{1}{n^2}\sum_{\substack{k,l=1 \\ k,l\neq i\\ k\neq l}}^{n} K_h(X_i-X_k) K_h(X_i-X_l)\langle b(X_i,\cdot)-c(Z_k,\cdot),b(X_i,\cdot)-c(Z_l,\cdot) \rangle\\
		\le\ &\frac{3}{n^2}\frac{||K||_\infty}{h^{d}}4||c||_\infty\sum_{\substack{k=1 \\ k\neq i}}^{n} |K_h(X_i-X_k)| + \frac{1}{n^2}\frac{||K||^2_\infty}{h^{2d}}4||c||_\infty\label{Lm2.2}\tag{Lm2.2}\\
		& + 	\frac{1}{n^2}\sum_{\substack{k,l=1 \\ k,l\neq i\\ k\neq l}}^{n} K_h(X_i-X_k) K_h(X_i-X_l)\langle b(X_i,\cdot)-c(Z_k,\cdot),b(X_i,\cdot)-c(Z_l,\cdot) \rangle.
	\end{align*}
	Let $i,k,l$ be three different indices. By Lemma \ref{Lm1}-(3),
	\begin{align*}
		\mathbb{E}\left[|K_h(X_i-X_k)| \ ||\mathcal{E}_i||^2\right] &= \mathbb{E}\left[\mathbb{E}[|K_h(X_i-X_k)| \ |\mathcal{E}_i,X_i]\ ||\mathcal{E}_i||^2\right]\\
		&= \mathbb{E}\left[\mathbb{E}[|K_h(X_i-X_k)| \ |X_i]\ ||\mathcal{E}_i||^2\right]\\
		&\le 2^d||f||_\infty||K||_\infty\mathbb{E}\left[||\mathcal{E}_i||^2\right]. 
	\end{align*}
	Moreover, by Lemma \ref{Lm1}-(7), the i.i.d. assumptions and the Cauchy-Schwarz inequality
	\begin{align*}
		&\mathbb{E}\left[K_h(X_i-X_k) K_h(X_i-X_l)\langle b(X_i,\cdot)-c(Z_k,\cdot),b(X_i,\cdot)-c(Z_l,\cdot) \rangle||\mathcal{E}_i||^2\right]\\
		=\ &\mathbb{E}\left[K_h(X_i-X_k) K_h(X_i-X_l)\langle b(X_i,\cdot)-b(X_k,\cdot),b(X_i,\cdot)-b(X_l,\cdot) \rangle||\mathcal{E}_i||^2\right]\\
		=\ &\mathbb{E}\left[||\mathbb{E}\left[K_h(X_i-X_k)(b(X_i,\cdot)-b(X_k,\cdot)) | X_i,\mathcal{E}_i\right]||^2||\mathcal{E}_i||^2\right]\\
		=\ &\mathbb{E}\left[||\mathbb{E}\left[K_h(X_i-X_k)(b(X_i,\cdot)-b(X_k,\cdot)) | X_i\right]||^2||\mathcal{E}_i||^2\right]\\
		\le\ &\mathbb{E}\left[||\mathbb{E}\left[K_h(X_i-X_k)(b(X_i,\cdot)-b(X_k,\cdot)) | X_i\right]||^4\right]^\frac{1}{2}\mathbb{E}\left[||\mathcal{E}_i||^4\right]^\frac{1}{2}\\
		=\ &o(1).
	\end{align*}
	Thus, it follows from (\ref{Lm2.2}) that
	\begin{align*}
		&\mathbb{E}\left[\left|\left|\left(\frac{1}{n}\sum\limits_{k=1}^n K_h(X_i-X_k)(c(Z_i,\cdot)-c(Z_k,\cdot)) -\hat{f}_h(X_i)(c(i,\cdot)-b(i,\cdot))\right)\otimes\mathcal{E}_i\right|\right|^2\right]\\
		=\ &O\left(\frac{1}{nh^d}\right) + O\left(\frac{1}{n^2h^{2d}}\right) + o(1)\\ 
		=\ &o(1).
	\end{align*}
\end{proof}

\section{Proofs of Theorems 1 and 2}

The following proposition studies the limiting behavior of $\mathbb{E}[\psi_n(i,k)]$, $\psi^1_n(i)$ and $\psi^2_n(i,j)$ under the regression model in (\ref{Sm.5}).

\begin{prop}
	\label{Pr2}
	Assume (A1)-(A3) and (T1)-(T5) hold for the regression model in (\ref{Sm.5}) and let $B$ be defined as in Lemma \ref{Lm1}. Given two i.i.d. copies $X$ and $X'$, and two indices $i\neq k$, it follows:
	\begin{enumerate}
		\item $\mathbb{E}[\psi_n(i,k)] = \mathbb{E}[c(Z,\cdot)\otimes f(X)(Y-m(X))] + O(h^{\nu})$.
		\item $\mathbb{E}[\psi_n(i,k)] = \mathbb{E}[c(Z,\cdot)\otimes \mathbb{E}[K_h(X-X')(m(X)-m(X'))|X]] =-h^\nu\mathbb{E}[c(Z,\cdot)\otimes B(X)] + o(h^{\nu})$ under $\mathcal{H}_0$.
		\item $\mathbb{E}[\psi_n(i,k)] = -h^\nu\mathbb{E}[c(Z,\cdot)\otimes B(X)] + o(h^{\nu}) + \mathbb{E}[c(Z,\cdot)\otimes f(X)\tilde{g}(Z)]/n^\gamma + O\left(h^\nu/n^\gamma\right)$ under $\{\mathcal{H}_{1n}\}_{n\in \mathbb{N}}$.
		\item $\psi^1_{n}(i) = O_{L^2}(1)$.
		\item $\psi^1_{n}(i) =\frac{1}{2}(c(Z_i,\cdot)-b(X_i,\cdot))\otimes f(X_i)\mathcal{E}_i + o_{L^{2}}(1)$ under $\mathcal{H}_0$.
		\item $\psi^1_{n}(i) =\frac{1}{2}(c(Z_i,\cdot)-b(X_i,\cdot))\otimes f(X_i)\mathcal{E}_i + o_{L^{2}}(1)$ under $\{\mathcal{H}_{1n}\}_{n\in \mathbb{N}}$.
		\item $\begin{aligned}[t]
			\psi^2_{n}(i,k) = O_{L^2}\left(1/{h^{\frac{d}{2}}}\right).
		\end{aligned}$
	\end{enumerate}
\end{prop}

\begin{proof}[\bfseries\textup{Proof of Proposition \ref{Pr2}}]
	
	By the symmetry of $\psi_n(i,k)$ and the identical distribution of $(Z_i,\mathcal{E}_i)$ and $(Z_k,\mathcal{E}_k)$ for all $i\neq k$ we have
	\begin{align*}
		\mathbb{E}[\psi_n(i,k)] &=  \mathbb{E}\left[\frac{1}{2}(c(Z_i,\cdot)-c(Z_k,\cdot))\otimes K_h(X_i-X_k)(Y_i-Y_k)\right]\\
		&= \mathbb{E}\left[c(Z_i,\cdot)\otimes K_h(X_i-X_k)(Y_i-Y_k)\right]\\
		 &= \mathbb{E}\left[\mathbb{E}\left[c(Z_i,\cdot)\otimes K_h(X_i-X_k)(Y_i-Y_k)|X_i,Z_i,Y_i\right]\right]\\
		&= \mathbb{E}\left[c(Z_i,\cdot)\otimes \mathbb{E}[K_h(X_i-X_k)(Y_i-Y_k)|X_i,Y_i]\right].
	\end{align*}
	By Lemma \ref{Lm1}-(1), $\mathbb{E}[K_h(X_i-X_k)(Y_i-Y_k)|X_i,Y_i] = f(X_i)(Y_i-m(X_i)) + O_{L^1}(h^{\nu})$. Therefore,
	$$\mathbb{E}[\psi_n(i,k)] = \mathbb{E}\left[c(Z_i,\cdot)\otimes f(X_i)(Y_i-m(X_i))\right] + h^\nu\mathbb{E}\left[c(Z_i,\cdot)\otimes  O_{L^1}(1) \right].$$
	Moreover, by the boundedness of $c$ in (A3),
	$$||\mathbb{E}\left[c(Z_i,\cdot)\otimes O_{L^1}(1) \right]||\le \mathbb{E}\left[||c(Z_i,\cdot)|| ||O_{L^1}(1)|| \right] = \mathbb{E}\left[\sqrt{c(Z_i,Z_i)} ||O_{L^1}(1)|| \right] = O(1),$$
	which implies that $h^\nu \mathbb{E}\left[c(Z_i,\cdot)\otimes O_{L^1}(1) \right] = O(h^\nu)$. Thus, (1) holds.
	
	To prove (2) note that, under $\mathcal{H}_0$,
	\begin{align*}
		 \mathbb{E}[\psi_n(i,k)] &= \mathbb{E}\left[\mathbb{E}\left[c(Z_i,\cdot)\otimes K_h(X_i-X_k)(Y_i-Y_k)|Z_i\right]\right]\\
		 &= \mathbb{E}\left[c(Z_i,\cdot)\otimes \mathbb{E}\left[K_h(X_i-X_k)(Y_i-Y_k)|Z_i\right]\right]\\
		 &= \mathbb{E}\left[c(Z_i,\cdot)\otimes \mathbb{E}\left[K_h(X_i-X_k)\mathbb{E}[Y_i-Y_k|Z_i,X_k]|Z_i\right]\right]\\
		 &= \mathbb{E}\left[c(Z_i,\cdot)\otimes \mathbb{E}\left[K_h(X_i-X_k)(m(X_i)-m(X_k))|Z_i\right]\right]\\
		 &= \mathbb{E}\left[c(Z_i,\cdot)\otimes \mathbb{E}\left[K_h(X_i-X_k)(m(X_i)-m(X_k))|X_i\right]\right].
	\end{align*}
	Then, by Lemma \ref{Lm1}-(2), $\mathbb{E}\left[K_h(X_i-X_k)(m(X_i)-m(X_k))|X_i\right] = -h^\nu B(X_i) + o_{L^2}(h^{\nu})$. Therefore, by the boundedness of $c$, it follows that
	$$\mathbb{E}[\psi_n(i,k)] = -h^\nu\mathbb{E}[c(Z,\cdot)\otimes B(X)] + o(h^{\nu}).$$

	The proof of (3) is similar. Under $\{\mathcal{H}_{1n}\}_{n\in \mathbb{N}}$ we have, by previous arguments, 
	\begin{align*}
		\mathbb{E}[\psi_n(i,k)] &= \mathbb{E}\left[c(Z_i,\cdot)\otimes \mathbb{E}\left[K_h(X_i-X_k)\mathbb{E}[Y_i-Y_k|Z_i,X_k]|Z_i\right]\right]\\
		&= \mathbb{E}\left[c(Z_i,\cdot)\otimes \mathbb{E}\left[K_h(X_i-X_k)\left(m(X_i)+\frac{\tilde{g}(Z_i)}{n^\gamma}-m(X_k)\right)|Z_i\right]\right]\\
		&= \mathbb{E}\left[c(Z_i,\cdot)\otimes \mathbb{E}\left[K_h(X_i-X_k)(m(X_i)-m(X_k))|X_i\right]\right]\\
		 &\quad + \frac{1}{n^\gamma}\mathbb{E}[c(Z_i,\cdot)\otimes \tilde{g}(Z_i)\mathbb{E}[K_h(X_i-X_k)|X_i]].
	\end{align*}
	By Lemma \ref{Lm1}-(2),	$\mathbb{E}\left[K_h(X_i-X_k)(m(X_i)-m(X_k))|X_i\right] = -h^\nu B(X_i) + o_{L^2}(h^{\nu})$ and by Lemma \ref{Lm1}-(5), $\mathbb{E}\left[K_h(X_i-X_k)|X_i\right] = f(X_i) + O_{L^4}(h^\nu)$. Combining this results with the Cauchy-Schwarz inequality and the bound $||c(Z_i,\cdot)||^2 \le ||c||_\infty$, it follows that
	$$\mathbb{E}[\psi_n(i,k)] =h^\nu\mathbb{E}[c(Z_i,\cdot)\otimes (-B(X_i))] + o(h^{\nu}) + \frac{1}{n^\gamma}\mathbb{E}[c(Z_i,\cdot)\otimes f(X_i)\tilde{g}(Z_i)] + O\left(\frac{h^\nu}{n^\gamma}\right),$$
	so (3) holds.
	
	 By (1), $\psi^1_n(i) = \mathbb{E}[\psi_n(i,k)|i] + O(1)$. Therefore, it suffices to show that $\mathbb{E}[\psi_n(i,k)|i] = O_{L^{2}}(1)$
	 to prove (4). Let us consider the following bound:
	 \begin{align*}
	 	||\mathbb{E}[\psi_n(i,k)|i]|| &\le  \mathbb{E}[||\psi_n(i,k)|| \ |i] = \frac{1}{2}\mathbb{E}[||c(Z_i,\cdot)-c(Z_k,\cdot)||\ || K_h(X_i-X_k)(Y_i-Y_k)|| \ |i] \\
	 	&\le \frac{1}{2}\mathbb{E}[\sqrt{||c(Z_i,\cdot)-c(Z_k,\cdot)||^2}|| \ |K_h(X_i-X_k)|(||Y_i||+||Y_k||)|i]\\
	 	&\le \frac{1}{2}\mathbb{E}[\sqrt{c(Z_i,Z_i)-2c(Z_i,Z_k)+c(Z_k,Z_k)} |K_h(X_i-X_k)|(||Y_i||+||Y_k||)|i]\\
	 	&\le  \mathbb{E}[ |K_h(X_i-X_k)| (||Y_i||+||Y_k||)|] =||Y_i|| \mathbb{E}[ |K_h(X_i-X_k)| \ |i] + \mathbb{E}[ |K_h(X_i-X_k)| \ ||Y_k|| \ |i],
	 \end{align*}
	 where we have used that $||c||_\infty \le 1$. On the one hand, by Lemma \ref{Lm1}-(3),
	 $$\mathbb{E}\left[||Y_i||^{2} \mathbb{E}[ |K_h(X_i-X_k)| \ |i]^{2}\right]\le \mathbb{E}\left[||Y_i||^{2}\right]2^{2d}||f||^{2}_\infty||K||^{2}_\infty = O(1).$$
	 On the other hand, by the Cauchy-Schwarz inequality based on $|K_h(X_i-X_k)| = \sqrt{K_h(X_i-X_k)}^2$ and Lemma \ref{Lm1}-(4) we have
	 \begin{align*}
	 	\mathbb{E}\left[\mathbb{E}[ |K_h(X_i-X_k)| \ ||Y_i|| \ |i]^{2}\right]&\le 	\mathbb{E}\left[\mathbb{E}\left[ |K_h(X_i-X_k)| \ ||Y_i||^{2}|i\right]\mathbb{E}[ |K_h(X_i-X_k)| \ |i]\right] \\
	 	&\le \mathbb{E}\left[\mathbb{E}\left[ |K_h(X_i-X_k)| \ ||Y_i||^{2}|i\right]\right]2^d||f||_\infty||K||_\infty\\
	 	&= \mathbb{E}\left[ |K_h(X_i-X_k)| \ ||Y_i||^{2}\right]2^d||f||_\infty||K||_\infty = O(1).
	 \end{align*}
	 Therefore, $\mathbb{E}[||\mathbb{E}[\psi_n(i,k)|i]||^{2}]=O(1)$ or, equivalently, $\mathbb{E}[\psi_n(i,k)|i] = O_{L^{2}}(1)$. 
	 
	 Given that, by definition, $\psi^1_n(i)= \mathbb{E}[\psi_n(i,k)|i] - \mathbb{E}[\psi_n(i,k)]$, and taking (2) into account, the proof of (5) follows if we show that
	\begin{equation}\label{Pr2.2}\tag{Pr2.2}
		\mathbb{E}[\psi_n(i,k)|i] = \frac{1}{2}(c(Z_i,\cdot)-b(X_i,\cdot))\otimes f(X_i)\mathcal{E}_i + o_{L^{2}}(1).
	\end{equation}
	However, under $\mathcal{H}_0$ we have
	\begin{align*}
		\mathbb{E}[\psi_n(i,k)|i] &= \frac{1}{2}\mathbb{E}\left[(c(Z_i,\cdot)-c(Z_k,\cdot))\otimes K_h(X_i-X_k)(Y_i-Y_k)|i\right]\\
		&= \frac{1}{2}\mathbb{E}\left[(c(Z_i,\cdot)-c(Z_k,\cdot))\otimes\mathbb{E}\left[ K_h(X_i-X_k)(Y_i-Y_k)|Z_i,Y_i,Z_k\right]|Z_i,Y_i\right]\\
		&= \frac{1}{2}\mathbb{E}\left[(c(Z_i,\cdot)-c(Z_k,\cdot))\otimes K_h(X_i-X_k)(Y_i-m(X_k))|Z_i,Y_i\right]\\
		&= \frac{1}{2}\mathbb{E}\left[\mathbb{E}\left[(c(Z_i,\cdot)-c(Z_k,\cdot))|Z_i,Y_i,X_k\right]\otimes K_h(X_i-X_k)(Y_i-m(X_k))|Z_i,Y_i\right]\\
		&= \frac{1}{2}\mathbb{E}\left[(c(Z_i,\cdot)-b(X_k,\cdot))\otimes K_h(X_i-X_k)(Y_i-m(X_k))|Z_i,Y_i\right]\\
		&= \frac{1}{2}(c(Z_i,\cdot)-b(X_i,\cdot))\otimes f(X_i)(Y_i-m(X_i))  + o_{L^{2}}(1),
	\end{align*}
	where the last equality follows from  Lemma \ref{Lm1}-(8) and the boundedness of $c$ (and $b$). 
	
	The proof of (6) is analogous since, given (3), it suffices to show that (\ref{Pr2.2}) also holds under $\{\mathcal{H}_{1n}\}_{n\in \mathbb{N}}$. By previous arguments based on Lemma \ref{Lm1}-(8) and the boundedness of $c$, it holds
	\begin{align*}
		\mathbb{E}[\psi_n(i,k)|i] &= \frac{1}{2}\mathbb{E}\left[(c(Z_i,\cdot)-c(Z_k,\cdot))\otimes\mathbb{E}\left[ K_h(X_i-X_k)(Y_i-Y_k)|Z_i,Y_i,Z_k\right]|Z_i,Y_i\right]\\
		&= \frac{1}{2}\mathbb{E}\left[(c(Z_i,\cdot)-c(Z_k,\cdot))\otimes K_h(X_i-X_k)\left(Y_i-m(X_k) - \frac{\tilde{g}(Z_k)}{n^\gamma}\right)|Z_i,Y_i\right]\\
		&= \frac{1}{2}(c(Z_i,\cdot)-b(X_i,\cdot))\otimes f(X_i)(Y_i-m(X_i))  + o_{L^{2}}(1)\\
		& \quad - \frac{1}{2n^\gamma}\mathbb{E}\left[(c(Z_i,\cdot)-c(Z_k,\cdot))\otimes K_h(X_i-X_k) \tilde{g}(Z_k)|Z_i,Y_i\right].
	\end{align*}
	Moreover, we can use Lemma \ref{Lm1}-(3), the Cauchy-Schwarz inequality and the boundedness of $c$ to obtain the bound,
	\begin{align*}
		&||\mathbb{E}\left[(c(Z_i,\cdot)-c(Z_k,\cdot))\otimes K_h(X_i-X_k) \tilde{g}(Z_k)|Z_i,Y_i\right]||^2\\
		\le \ &4\mathbb{E}\left[|K_h(X_i-X_k)| \ ||\tilde{g}(Z_k)|| \ |Z_i,Y_i\right]^2\\
		\le \ &4\mathbb{E}\left[|K_h(X_i-X_k)| \ ||\tilde{g}(Z_k)||^2 \ |Z_i\right]\mathbb{E}\left[|K_h(X_i-X_k)| \ | X_i\right]\\
		\le \ &2^{d+2}||f||_\infty||K||_\infty\mathbb{E}\left[|K_h(X_i-X_k)| \ ||\tilde{g}(Z_k)||^2 \ |X_i\right].
	\end{align*}
	Thus, taking expectations,
	\begin{align*}
		&\mathbb{E}\left[||\mathbb{E}\left[(c(Z_i,\cdot)-c(Z_k,\cdot))\otimes K_h(X_i-X_k) \tilde{g}(Z_k)|Z_i,Y_i\right]||^2\right]\\
		\le \ &2^{d+2}||f||_\infty||K||_\infty \mathbb{E}\left[|K_h(X_i-X_k)| \  ||\tilde{g}(Z_k)||^2 \right]\\
		= \ &2^{d+2}||f||_\infty||K||_\infty\mathbb{E}\left[\mathbb{E}\left[|K_h(X_i-X_k)| \ |Z_k\right] ||\tilde{g}(Z_k)||^2 \right]\\
		= \ &2^{d+2}||f||_\infty||K||_\infty \mathbb{E}\left[\mathbb{E}\left[|K_h(X_i-X_k)| \ |X_k\right] ||\tilde{g}(Z_k)||^2 \right]\\
		\le \ &2^{2d+2}||f||^2_\infty||K||^2_\infty \mathbb{E}\left[||\tilde{g}(Z_k)||^2 \right].
	\end{align*}
	Ultimately, we have
	$$ 	\mathbb{E}[\psi_n(i,k)|i] = \frac{1}{2}(c(Z_i,\cdot)-b(X_i,\cdot))\otimes f(X_i)(Y_i-m(X_i))  + o_{L^{2}}(1) + O_{L^{2}}\left(\frac{1}{n^\gamma}\right),$$
	and, given that $\gamma>0$ under $\{\mathcal{H}_{1n}\}_{n\in \mathbb{N}}$, (6) holds.
	
	By (1) and (4), we know that $\psi^1_n(i) + \psi^1_n(k) - \mathbb{E}[\psi_n(i,k)] = O_{L^2}(1)$.
	Therefore, to prove (7) it suffices to show that $\psi_n(i,k) = O_{L^{2}}\left(1/{h^d}\right)$. However, 
	$$||\psi_n(i,k)||\le \frac{1}{2}||c(Z_i,\cdot)-c(Z_k,\cdot)|| \ |K_h(X_i-X_k)| \ ||Y_i-Y_k||\le |K_h(X_i-X_k)| \ ||Y_i-Y_k||,$$
	and, by Lemma \ref{Lm1}-(4),
	\begin{equation*}
		\mathbb{E}\left[||\psi_n(i,k)||^2\right]\le \mathbb{E}\left[|K_h(X_i-X_k)|^2 ||Y_i-Y_k||^2\right] = O\left(\frac{1}{h^d}\right).
	\end{equation*}
\end{proof}

Proposition \ref{Pr2} suffices to characterize the asymptotic behavior of $V_n$ under the regression model in (\ref{Sm.5}). Thus, to prove Theorems 1 and 2, we are left to study the relation between $V_n$ and $U_n$.

\begin{prop}\label{Pr3}
	Under the regression model in (\ref{Sm.5}), assuming that (A1)-(A3) and (T1)-(T5) hold, we have
	$$n^4V_n - n(n-1)(n-2)(n-3)U_n = O_{L^1}(n^3).$$
	Moreover, under either $\mathcal{H}_0$ or $\{\mathcal{H}_{1n}\}_{n\in \mathbb{N}}$, 
	$$n^4V_n - n(n-1)(n-2)(n-3)U_n = n^3\mathbb{E}\left[||(c(Z,\cdot) - b(X,\cdot))\otimes f(X)\mathcal{E}||^2\right] + o_{L^1}(n^3).$$
\end{prop}
\begin{proof}[\bfseries\textup{Proof of Proposition 3}]
Let $\psi_n$ is defined as 
$$\psi_n((z,y),(z',y')) = \frac{1}{2}(c(z,\cdot)-c(z',\cdot))\otimes K_h(x-x')(y-y'),$$
for all $(z,y),(z',y')$ in the support of $(Z,Y)$. Then,
\begin{align*}
	n^4V_n &= \sum\limits_{i,j,k,l}\langle \psi_n((Z_i,Y_i),(Z_k,Y_k)),\psi_n((Z_j,Y_j),(Z_l,Y_l)) \rangle,\\
	n(n-1)(n-2)(n-3)U_n &= \sum\limits_{i\neq j\neq k\neq l}\langle \psi_n((Z_i,Y_i),(Z_k,Y_k)),\psi_n((Z_j,Y_j),(Z_l,Y_l)) \rangle.
\end{align*}	
By definition of $\psi_n$, it is clear that $\psi_n((Z_i,Y_i),(Z_i,Y_i)) = 0$ for all $i$. Therefore, by the inclusion-exclusion principle, the sum of $\langle \psi_n((Z_i,Y_i),(Z_k,Y_k)),\psi_n((Z_j,Y_j),(Z_l,Y_l)) \rangle$ in indices $i\neq j\neq k\neq l$ can be obtained from the sum in indices $i,j,k,l$ by:
\begin{enumerate}
	\item Substracting the sum in indices $i,j,k,l$ where there is, at least, one pair of equal indices.
	\item Adding back the sum in indices $i,j,k,l$ where there are two disjoint pairs of equal indices.
\end{enumerate}
Let $\sum\limits_{\substack{i,j,k,l\\ \ge 1 \text{Pair}}}$ denote the sum in indices $i,j,k,l$ where there is at least $1$ pair of indices. By the symmetry of $\psi_n$, it is clear that
\begin{align*}
	&\sum\limits_{\substack{i,j,k,l\\ \ge 1 \text{Pair}}}\langle \psi_n((Z_i,Y_i),(Z_k,Y_k)),\psi_n((Z_j,Y_j),(Z_l,Y_l)) \rangle\\
	=\ &\sum\limits_{i,k,l}\langle \psi_n((Z_i,Y_i),(Z_k,Y_k)),\psi_n((Z_i,Y_i),(Z_l,Y_l)) \rangle + \sum\limits_{i,j,k}\langle \psi_n((Z_i,Y_i),(Z_k,Y_k)),\psi_n((Z_j,Y_j),(Z_i,Y_i)) \rangle\\
	+& \sum\limits_{i,j,l}\langle \psi_n((Z_i,Y_i),(Z_j,Y_j)),\psi_n((Z_j,Y_j),(Z_l,Y_l)) \rangle +  \sum\limits_{i,j,k}\langle \psi_n((Z_i,Y_i),(Z_j,Y_j)),\psi_n((Z_j,Y_j),(Z_j,Y_j)) \rangle\\
	=\  &4\sum\limits_{i,k,l}\langle \psi_n((Z_i,Y_i),(Z_k,Y_k)),\psi_n((Z_i,Y_i),(Z_l,Y_l)) \rangle\\
	=\  &\sum\limits_{i,k,l}\left(c(Z_i,Z_i)-c(Z_i,Z_l) - c(Z_k,Z_i) + c(Z_k,Z_l)\right)K_h(X_i-X_k)K_h(X_i-X_l)\langle Y_i-Y_k, Y_i-Y_l\rangle\\
	=\  &\sum\limits_{i\neq k\neq l}\left(c(Z_i,Z_i)-c(Z_i,Z_l) - c(Z_k,Z_i) + c(Z_k,Z_l)\right)K_h(X_i-X_k)K_h(X_i-X_l)\langle Y_i-Y_k, Y_i-Y_l\rangle\\
	+& \sum\limits_{i,k}\left(c(Z_i,Z_i)- 2c(Z_i,Z_k) + c(Z_k,Z_k)\right)K^2_h(X_i-X_k)||Y_i-Y_k||^2.
\end{align*}
Similarly, letting $\sum\limits_{\substack{i,j,k,l\\ 2 \text{Pair}}}$ denote the sum in indices $i,j,k,l$ where there are $2$ pairs of disjoint indices, we have
\begin{align*}
	&\sum\limits_{\substack{i,j,k,l\\  2 \text{Pair}}}\langle \psi_n((Z_i,Y_i),(Z_k,Y_k)),\psi_n((Z_j,Y_j),(Z_l,Y_l)) \rangle\\
	=\ &\sum\limits_{i\neq k}\langle \psi_n((Z_i,Y_i),(Z_k,Y_k)),\psi_n((Z_i,Y_i),(Z_k,Y_k)) \rangle + \sum\limits_{i\neq k}\langle \psi_n((Z_i,Y_i),(Z_k,Y_k)),\psi_n((Z_k,Y_k),(Z_i,Y_i)) \rangle\\
	=\  &2\sum\limits_{i\neq k}|| \psi_n((Z_i,Y_i),(Z_k,Y_k)),\psi_n((Z_i,Y_i),(Z_k,Y_k)) ||^2\\
	=\  &\frac{1}{2}\sum\limits_{i,k}\left(c(Z_i,Z_i)- 2c(Z_i,Z_k) + c(Z_k,Z_k)\right)K^2_h(X_i-X_k)||Y_i-Y_k||^2.
\end{align*}
Thus,
\begin{align*}
	&n^4V_n - n(n-1)(n-2)(n-3)U_n\\
	=\ &\sum\limits_{i,k,l}\left(c(Z_i,Z_i)-c(Z_i,Z_l) - c(Z_k,Z_i) + c(Z_k,Z_l)\right)K_h(X_i-X_k)K_h(X_i-X_l)\langle Y_i-Y_k, Y_i-Y_l\rangle\\
	&+ \frac{1}{2}\sum\limits_{i,k}\left(c(Z_i,Z_i)- 2c(Z_i,Z_k) + c(Z_k,Z_k)\right)K^2_h(X_i-X_k)||Y_i-Y_k||^2\\
	=\ &\sum\limits_{i\neq k\neq l}\left(c(Z_i,Z_i)-c(Z_i,Z_l) - c(Z_k,Z_i) + c(Z_k,Z_l)\right)K_h(X_i-X_k)K_h(X_i-X_l)\langle Y_i-Y_k, Y_i-Y_l\rangle\\
	&+ \frac{3}{2}\sum\limits_{i,k}\left(c(Z_i,Z_i)- 2c(Z_i,Z_k) + c(Z_k,Z_k)\right)K^2_h(X_i-X_k)||Y_i-Y_k||^2.
\end{align*}
In the following we assume that, given three indices $i,k,l$, these indices are mutually distinct unless stated otherwise by a summation symbol. By the boundedness of $c$ and $K$, as well as Lemma \ref{Lm1}-(4), it holds that
\begin{align*}
	&\mathbb{E}\left[\left|\frac{3}{2}\sum\limits_{i,k}\left(c(Z_i,Z_i)- 2c(Z_i,Z_k) + c(Z_k,Z_k)\right)K^2_h(X_i-X_k)||Y_i-Y_k||^2\right| \right]\\
 \le\ &6n^2\mathbb{E}\left[K^2_h(X_i-X_k)||Y_i-Y_k||^2\right]\\
  =\ &O\left(\frac{n^2}{h^d}\right)\\
  =\ &o(n^3),
\end{align*}
 Similarly, by the Cauchy-Scwharz inequality, the sample independence and Lemma \ref{Lm1}-(3),
\begin{align*}
	&\mathbb{E}\left[\left|\sum\limits_{i\neq k\neq l}\left(c(Z_i,Z_i)-c(Z_i,Z_l) - c(Z_k,Z_i) + c(Z_k,Z_l)\right)K_h(X_i-X_k)K_h(X_i-X_l)\langle Y_i-Y_k, Y_i-Y_l\rangle\right| \right]\\
	\le\ &4n^3\mathbb{E}\left[|K_h(X_i-X_k)|\ |K_h(X_i-X_l)|\ ||Y_i-Y_k|| \ ||Y_i-Y_l||\right]\\
	=\ &4n^3\mathbb{E}\left[\mathbb{E}\left[|K_h(X_i-X_k)|\ ||Y_i-Y_k||\ | X_i,Y_i\right]^2\right]\\
	\le \ &4n^3\mathbb{E}\left[\mathbb{E}\left[|K_h(X_i-X_k)|\ | X_i\right]\mathbb{E}\left[|K_h(X_i-X_k)|\ ||Y_i-Y_k||^2 | X_i,Y_i\right]\right]\\
	\le\ &2^{d+2}||f||_\infty ||K||_\infty n^3\mathbb{E}\left[\mathbb{E}\left[|K_h(X_i-X_k)|\ ||Y_i-Y_k||^2 | X_i,Y_i\right]\right]\\
	=\ &2^{d+2}||f||_\infty ||K||_\infty n^3\mathbb{E}\left[|K_h(X_i-X_k)|\ ||Y_i-Y_k||^2 \right]\\
	=\ &O(n^3),
\end{align*}
and, therefore, $n^4V_n - n(n-1)(n-2)(n-3)U_n = O_{L^1}(n^3) + o_{L^1}(n^3) = O_{L^1}(n^3)$.

Now, let us assume that $\mathcal{H}_{1n}$ holds. On the one hand, by previous arguments,
$$\frac{1}{2}\sum\limits_{i,k}\left(c(Z_i,Z_i)- 2c(Z_i,Z_k) + c(Z_k,Z_k)\right)K^2_h(X_i-X_k)||Y_i-Y_k||^2 = o_{L^1}(n^3).$$
On the other hand,
\begin{align*}
	&\sum\limits_{i,k,l}\left(c(Z_i,Z_i)-c(Z_i,Z_l) - c(Z_k,Z_i) + c(Z_k,Z_l)\right)K_h(X_i-X_k)K_h(X_i-X_l)\langle Y_i-Y_k, Y_i-Y_l\rangle\\
	=\ &n^3 \frac{1}{n}\sum\limits_{i=1}^n\left|\left| \frac{1}{n}\sum\limits_{k=1}^n(c(Z_i,\cdot)-c(Z_k,\cdot))\otimes K_h(X_i-X_k)(Y_i-Y_k)\right|\right|^2,
\end{align*}
so, by the weak law of large numbers, it suffices to show that
\begin{equation}\label{Pr3.1}\tag{Pr3.1}
	\frac{1}{n}\sum\limits_{k=1}^n(c(Z_i,\cdot)-c(Z_k,\cdot))\otimes K_h(X_i-X_k)(Y_i-Y_k) = (c(Z_i,\cdot) - b(X_i,\cdot))\otimes f(X_i)\mathcal{E}_i +o_{L^2}(1).
\end{equation}
Let us consider the following decomposition:
\begin{align*}
	&\frac{1}{n}\sum\limits_{k=1}^n(c(Z_i,\cdot)-c(Z_k,\cdot))\otimes K_h(X_i-X_k)(Y_i-Y_k)\\
	=\ &\frac{1}{n}\sum\limits_{k=1}^n(c(Z_i,\cdot)-c(Z_k,\cdot))K_h(X_i-X_k)\otimes \mathcal{E}_i - \frac{1}{n}\sum\limits_{k=1}^n(c(Z_i,\cdot)-c(Z_k,\cdot))\otimes K_h(X_i-X_k)\mathcal{E}_k \label{Pr3.2}\tag{Pr3.2}\\
	&+\frac{1}{n}\sum\limits_{k=1}^n(c(Z_i,\cdot)-c(Z_k,\cdot))\otimes K_h(X_i-X_k)(m(X_i)-m(X_i))\\
	&+\frac{1}{n^{1+\gamma}}\sum\limits_{k=1}^n(c(Z_i,\cdot)-c(Z_k,\cdot))\otimes K_h(X_i-X_k)(\tilde{g}(Z_i) -\tilde{g}(Z_k)).
\end{align*}
By the Cauchy-Scwharz inequality, the boundedness of $c$, the sample independence and Lemma \ref{Lm1}-(3), the last term on the right hand side of (\ref{Pr3.2}) verifies
\begin{align*}
	&\mathbb{E}\left[\left|\left|\frac{1}{n^{1+\gamma}}\sum\limits_{k=1}^n(c(Z_i,\cdot)-c(Z_k,\cdot))\otimes K_h(X_i-X_k)(\tilde{g}(Z_i) -\tilde{g}(Z_k))\right|\right|^2\right]\\
	\le\ &\mathbb{E}\left[\left|\frac{2}{n^{1+\gamma}}\sum\limits_{k=1}^n |K_h(X_i-X_k)| \ ||\tilde{g}(Z_i) -\tilde{g}(Z_k)||\right|^2\right]\\
	\le\ &\frac{4}{n^{2+2\gamma}}\mathbb{E}\left[\sum\limits_{k=1}^n |K_h(X_i-X_k)|^2 ||\tilde{g}(Z_i) -\tilde{g}(Z_k)||^2\right]\\
	 &+ \frac{4}{n^{2+2\gamma}}\mathbb{E}\left[\sum\limits_{k\neq l} |K_h(X_i-X_k)| \ |K_h(X_i-X_l)| \ ||\tilde{g}(Z_i) -\tilde{g}(Z_k)||\ ||\tilde{g}(Z_i) -\tilde{g}(Z_l)||\right]\\
	 \le\ &\frac{16||K||_\infty}{n^{1+2\gamma}h^d}\mathbb{E}\left[ |K_h(X_i-X_k)|\ ||\tilde{g}(Z_i)||^2\right] + \frac{4}{n^{2\gamma}}\mathbb{E}\left[\mathbb{E}\left[|K_h(X_i-X_k)|\ ||\tilde{g}(Z_i) -\tilde{g}(Z_k)|| \ |Z_i\right]^2\right]\\
	 	\le\ &\frac{16||K||_\infty}{n^{1+2\gamma}h^d}\mathbb{E}\left[ \mathbb{E}\right[|K_h(X_i-X_k)|\ |X_i\left] ||\tilde{g}(Z_i)||^2\right]\\
	 	 &+ \frac{4}{n^{2\gamma}}\mathbb{E}\left[\mathbb{E}\left[|K_h(X_i-X_k)|\ |X_i\right] \mathbb{E}\left[|K_h(X_i-X_k)|\ ||\tilde{g}(Z_i) -\tilde{g}(Z_k)||^2 \ |Z_i\right]\right]\\
	 	\le\ &\frac{2^{d+4}||f||_\infty||K||^2_\infty}{n^{1+2\gamma}h^d}\mathbb{E}\left[ ||\tilde{g}(Z_i)||^2\right] + \frac{2^{d+2}||f||_\infty||K||_\infty}{n^{2\gamma}}\mathbb{E}\left[|K_h(X_i-X_k)|\ ||\tilde{g}(Z_i) -\tilde{g}(Z_k)||^2\right]\\
	 	\le\ &\frac{2^{d+4}||f||_\infty||K||^2_\infty}{n^{1+2\gamma}h^d}\mathbb{E}\left[ ||\tilde{g}(Z_i)||^2\right] + \frac{2^{d+4}||f||_\infty||K||_\infty}{n^{2\gamma}}\mathbb{E}\left[\mathbb{E}\left[|K_h(X_i-X_k)|\ |X_i\right] ||\tilde{g}(Z_i)||^2\right]\\
	 	\le\ &\frac{2^{d+4}||f||_\infty||K||^2_\infty}{n^{1+2\gamma}h^d}\mathbb{E}\left[ ||\tilde{g}(Z_i)||^2\right] + \frac{2^{2d+4}||f||^2_\infty||K||^2_\infty}{n^{2\gamma}}\mathbb{E}\left[||\tilde{g}(Z_i)||^2\right]\\
	 	=\ &O\left(\frac{1}{n^{1+2\gamma}h^d}\right) + O\left(\frac{1}{n^{2\gamma}}\right),
\end{align*}
and, thus, 
\begin{equation}\label{Pr3.3}\tag{Pr3.3}
	\frac{1}{n^{1+\gamma}}\sum\limits_{k=1}^n(c(Z_i,\cdot)-c(Z_k,\cdot))\otimes K_h(X_i-X_k)(\tilde{g}(Z_i) -\tilde{g}(Z_k)) = o_{L^2}(1).
\end{equation}
Similarly, by Lemma \ref{Lm1}-(6), we have the following bound for the second order moment of the third term on the right hand side of (\ref{Pr3.2}):
\begin{align*}
	&\mathbb{E}\left[\left|\left|\frac{1}{n}\sum\limits_{k=1}^n(c(Z_i,\cdot)-c(Z_k,\cdot))\otimes K_h(X_i-X_k)(m(X_i) -m(X_k))\right|\right|^2\right]\\
	\le\ &\frac{2^{d+4}||f||_\infty||K||^2_\infty}{nh^d}\mathbb{E}\left[ ||m(X_i)||^2\right]+ 4\mathbb{E}\left[\mathbb{E}\left[|K_h(X_i-X_k)|\ ||m(X_i) -m(X_k)|| \ |X_i\right]^2\right]\\
	=\ &O\left(\frac{1}{nh^d}\right) + O(h^2).
\end{align*}
Therefore,
\begin{equation}\label{Pr3.4}\tag{Pr3.4}
	\frac{1}{n}\sum\limits_{k=1}^n(c(Z_i,\cdot)-c(Z_k,\cdot))\otimes K_h(X_i-X_k)(m(X_i) -m(X_k)) = o_{L^2}(1).
\end{equation}
For the second term on the right hand side of (\ref{Pr3.2}), by similar arguments,  we can write
\begin{align*}
	&\mathbb{E}\left[\left|\left|\frac{1}{n}\sum\limits_{k=1}^n(c(Z_i,\cdot)-c(Z_k,\cdot))\otimes K_h(X_i-X_k)\mathcal{E}_k\right|\right|^2\right]\\
	=\ &\mathbb{E}\left[\left|\left|\frac{1}{n}\sum\limits_{\substack{1\le k \le n \\ k\neq i}}(c(Z_i,\cdot)-c(Z_k,\cdot))\otimes K_h(X_i-X_k)\mathcal{E}_k\right|\right|^2\right]\\
	=\ &\frac{1}{n^2}\sum\limits_{\substack{1\le k \le n \\ k\neq i}}\mathbb{E}\left[(c(Z_i,Z_i) - 2c(Z_i,Z_k) + c(Z_k,Z_k))K^2_h(X_i-X_k) ||\mathcal{E}_k||^2 \right]\\
	&+ \frac{1}{n^2}\sum\limits_{\substack{1\le k,l \le n \\ k,l\neq i \\ k\neq l}}\mathbb{E}\left[(c(Z_i,Z_i) - c(Z_k,Z_i) -c(Z_i,Z_l) + c(Z_k,Z_l))K_h(X_i-X_k)K_h(X_j-X_l) \langle \mathcal{E}_k,\mathcal{E}_l\rangle \right]\\
	\le\ &\frac{2^{d+4}||f||_\infty||K||^2_\infty}{nh^d}\mathbb{E}\left[ ||\mathcal{E}_k||^2\right],
\end{align*}
where the last inequality follows from 
$$ \mathbb{E}\left[\langle \mathcal{E}_k,\mathcal{E}_l\rangle | Z_k,Z_i,Z_l\right] = \left\langle \mathbb{E}\left[\mathcal{E}_k|Z_k\right],\mathbb{E}\left[\mathcal{E}_l|Z_l\right]\right\rangle = 0,$$
for all indices $i\neq k \neq l$, by the sample independence assumption. Thus,
\begin{equation}\label{Pr3.5}\tag{Pr3.5}
	\frac{1}{n}\sum\limits_{k=1}^n(c(Z_i,\cdot)-c(Z_k,\cdot))\otimes K_h(X_i-X_k)\mathcal{E}_k = o_{L^2}(1).
\end{equation}
Finally, by Lemma \ref{Lm2}-(2) and Lemma \ref{Lm2}-(4), we have
\begin{align*}
	 \frac{1}{n}\sum\limits_{k=1}^n(c(Z_i,\cdot)-c(Z_k,\cdot))K_h(X_i-X_k)\otimes \mathcal{E}_i &= (c(Z_i,\cdot)-b(X_i,\cdot))\otimes \frac{1}{n}\sum\limits_{k=1}^nK_h(X_i-X_k)\mathcal{E}_i + o_{L^2}(1)\\
	 &=	 (c(Z_i,\cdot)-b(X_i,\cdot))\otimes f(X_i)\mathcal{E}_i + o_{L^2}(1)\\
	 &\quad + \left(\frac{1}{n}\sum\limits_{k=1}^nK_h(X_i-X_k)-f(X_i)\right)(c(Z_i,\cdot)-b(X_i,\cdot))\otimes \mathcal{E}_i\\
	 &= (c(Z_i,\cdot)-b(X_i,\cdot))\otimes f(X_i)\mathcal{E}_i + o_{L^2}(1) .\label{Pr3.6}\tag{Pr3.6}
\end{align*}
Considering (\ref{Pr3.2}), (\ref{Pr3.3}), (\ref{Pr3.4}), (\ref{Pr3.5}) and (\ref{Pr3.6}), it is clear that (\ref{Pr3.1}) holds.

\end{proof}

We can finally prove Theorem 1 and Theorem 2 combining the results in Propositions \ref{Pr2} and \ref{Pr3}.

\begin{proof}[\bfseries\textup{Proof of Theorem 1}]
	By virtue of Proposition \ref{Pr3}, we have:
	\begin{itemize}
		\item $V_nn^3/((n-1)(n-2)(n-3)) - U_n = O_{\mathbb{P}}(1/n)$,
		\item $V_nn^3/((n-1)(n-2)(n-3)) - U_n = \mathbb{E}\left[||(c(Z,\cdot) - b(X,\cdot))\otimes f(X)\mathcal{E}||^2\right]/n+ o_{\mathbb{P}}(1/n)$ under $\mathcal{H}_0$.
	\end{itemize}
	Let $\tilde{V}_n$ denote $V_n(n^2/(n-1)^2)$ so that, according to the expression of $V_n$ in (\ref{Sm.2}),
	$$\tilde{V}_n = \left|\left| \frac{1}{{n\choose 2}}\sum\limits_{i<k} \psi_n((Z_i,Y_i),(Z_k,Y_k))  \right|\right|^2.$$
	 Then,
	\begin{equation*}
		\tilde{V}_n - V_n = \tilde{V}_n\left(\frac{n^2-(n-1)^2}{n^2}\right) = \tilde{V}_nO\left(\frac{1}{n}\right).
	\end{equation*}
	Moreover, 
	 $$\frac{n^3}{(n-1)(n-2)(n-3)}V_n-V_n = V_nO\left(\frac{1}{n}\right).$$ 
	Therefore, if $\tilde{V}_n = O_\mathbb{P}(1)$, which implies that $V_n= O_\mathbb{P}(1)$, we have
	\begin{align*}
		 U_n &= V_n +  \left(U_n - \frac{n^3}{(n-1)(n-2)(n-3)}V_n\right) + \left(\frac{n^3}{(n-1)(n-2)(n-3)}V_n-V_n\right)\\
		 &= V_n - \frac{\mathbb{E}\left[||(c(Z,\cdot) - b(X,\cdot))\otimes f(X)\mathcal{E}||^2\right]}{n} + O_{\mathbb{P}}\left(\frac{1}{n}\right)\\
		 &= \tilde{V}_n - \frac{\mathbb{E}\left[||(c(Z,\cdot) - b(X,\cdot))\otimes f(X)\mathcal{E}||^2\right]}{n} + O_{\mathbb{P}}\left(\frac{1}{n}\right).\label{Th1.1}\tag{Th1.1}
	\end{align*}
	Therefore, it suffices to show that $\tilde{V}_n = \text{KCMD}(f(X)(Y-m(X))|Z) + O(h^\nu) + O_\mathbb{P}\left(1/\sqrt{n}\right)$ to prove Theorem 1-(1). Similarly, if $\tilde{V}_n = o_\mathbb{P}(1)$, it follows that
	\begin{equation}\label{Th1.2}\tag{Th1.2}
		U_n = \tilde{V}_n - \frac{\mathbb{E}\left[||(c(Z,\cdot) - b(X,\cdot))\otimes f(X)\mathcal{E}||^2\right]}{n} + o_{\mathbb{P}}\left(\frac{1}{n}\right),
	\end{equation}
	and, thus, it suffices to establish the asymptotic behavior of $\tilde{V}_n$ under $\mathcal{H}_0$ to derive that of $U_n$.
	
	By virtue of the Hoeffding decomposition we have
	$$\frac{1}{{n\choose 2}}\sum\limits_{i< k} \psi_n(i,k) =  \mathbb{E}[\psi_n(i,k)] + \frac{2}{n}\sum\limits_{i=1}^n \psi^1_n(i)  + \frac{1}{{n \choose 2}}\sum\limits_{i<j}^n \psi^2_n(i,j).$$
	By Proposition \ref{Pr2}-(1), we know that $\mathbb{E}[\psi_n(i,k)] = \mathbb{E}[c(Z,\cdot)\otimes f(X)(Y-m(X))] + O(h^{\nu})$. Thus,
	\begin{equation}\label{Th1.3}\tag{Th1.3}
		\frac{1}{{n\choose 2}}\sum\limits_{i<k}\psi_n(i,k) = \mathbb{E}[c(Z,\cdot)\otimes f(X)(Y-m(X))] + O(h^{\nu}) + \frac{2}{n}\sum\limits_{i=1}^n \psi^1_n(i) +\frac{1}{{n \choose 2}}\sum\limits_{i<k}^n \psi^2_n(i,k).
	\end{equation}
	On the one hand, given the i.i.d. hyphotesis on $\{Z_i,Y_i\}_{i=1}^n$ and Proposition \ref{Pr2}-(3) we have
	\begin{align*}
		\mathbb{E}\left[\left|\left|\sqrt{n}\frac{1}{n}\sum\limits_{i=1}^n \psi^1_n(i)\right|\right|^2\right]
		&= \frac{1}{n}\sum\limits_{i,k}\mathbb{E}\left[\langle \psi^1_n(i), \psi^1_n(k) \rangle\right]
		= \mathbb{E}\left[|| \psi^1_n(0)||^2\right] = O(1). \label{Th1.4}\tag{Th1.4}
	\end{align*}
	On the other hand, taking into account the symmetry and degeneracy of the kernel $\psi^2_n(i,k)$,
	$$ \mathbb{E}\left[\left|\left|\frac{\sqrt{n}}{{n \choose 2}}\sum\limits_{i<k}^n \psi^2_n(i,k)\right|\right|^2\right]
	= \frac{4}{n(n-1)^2}\sum\limits_{i<k}\sum\limits_{j<l}\mathbb{E}\left[\langle \psi^2_n(i,k), \psi^2_n(j,l) \rangle\right] = \frac{4}{n(n-1)^2}\sum\limits_{i<k}\mathbb{E}\left[|| \psi^2_n(i,k) ||^2\right],$$
	since
	$$\mathbb{E}\left[\langle \psi^2_n(i_1,i_2), \psi^2_n(i_1,i_4) \rangle\right] = \mathbb{E}\left[\langle \psi^2_n(i_1,i_2), \psi^2_n(i_3,i_4) \rangle\right] = 0,$$
	for all indices $i_1\neq i_2\neq i_3 \neq i_4$. Therefore, by Proposition \ref{Pr2}-(4),
	\begin{equation}\label{Th1.5}\tag{Th1.5}
		 \mathbb{E}\left[\left|\left|\frac{\sqrt{n}}{{n \choose 2}}\sum\limits_{i<k}^n \psi^2_n(i,k)\right|\right|^2\right] = O\left(\frac{1}{nh^d}\right)=o(1).
	\end{equation}
	Combining (\ref{Th1.5}), (\ref{Th1.4}) and (\ref{Th1.5}), it follows that
	$$\frac{1}{{n\choose 2}}\sum\limits_{i<k}\psi_n(i,k) = \mathbb{E}[c(Z,\cdot)\otimes f(X)(Y-m(X))] + O(h^{\nu}) + O_\mathbb{P}\left(\frac{1}{\sqrt{n}}\right) + o_\mathbb{P}\left(\frac{1}{\sqrt{n}}\right),$$
	and, taking squared norms,
	$$\tilde{V}_n = \text{KCMD}(f(X)(Y-m(X))|Z) + O(h^\nu) + O_\mathbb{P}\left(\frac{1}{\sqrt{n}}\right),$$
	so Theorem 1-(1) holds by (\ref{Th1.1}). 
	
	To prove Theorem 1-(2) and Theorem 1-(3) recall that, under $\mathcal{H}_0$, we have
	$$\mathbb{E}[\psi_n(i,k)] = \mathbb{E}\left[c(Z,\cdot)\otimes \mathbb{E}[K_h(X-X')(m(X)-m(X'))|X]\right] = - h^\nu\mathbb{E}[c(Z,\cdot)\otimes B(X)] + o(h^{\nu}),$$ 
	by Proposition \ref{Pr2}-(2). Moreover, by Proposition \ref{Pr2}-(3), 
	\begin{align*}
		\sqrt{n}\frac{2}{n}\sum\limits_{i=1}^n \psi^1_n(i) &= \sqrt{n}\frac{1}{n}\sum\limits_{i=1}^n\left( (c(Z_i,\cdot)-b(X_i,\cdot))\otimes f(X_i)\mathcal{E}_i + o_{L^{2}}(1)_i\right)\\
		&= \frac{1}{\sqrt{n}}\sum\limits_{i=1}^n (c(Z_i,\cdot)-b(X_i,\cdot))\otimes f(X_i)\mathcal{E}_i + o_{L^{2}}(1).
	\end{align*}
	By the central limit theorem in Hilbert spaces (see e.g. Theorem 6.5.1 in \citeauthor{Grenander1963}, \citeyear{Grenander1963}) and Slutsky's Lemma,
	$$\sqrt{n}\frac{2}{n}\sum\limits_{i=1}^n \psi^1_n(i) \xrightarrow{\mathcal{D}}N_{{\mathcal{HS}}},$$
	where $N_{{\mathcal{HS}}}$ is a zero-mean Gaussian random variable taking values in ${\mathcal{HS}}$ with the same covariance operator as $(c(Z,\cdot)-b(X,\cdot))\otimes f(X)\mathcal{E}$. Then, by (\ref{Th1.5}), if $nh^{2\nu}\rightarrow{M}$ where $0\le M<\infty$,
	$$\frac{\sqrt{n}}{{n\choose 2}}\sum\limits_{i<k}\psi_n(i,k) =  -\sqrt{n}h^{\nu}\mathbb{E}[c(Z,\cdot)\otimes B(X)] + \sqrt{n}\frac{2}{n}\sum\limits_{i=1}^n \psi^1_n(i) + \sqrt{n} o_\mathbb{P}\left(\frac{1}{\sqrt{n}}\right)\xrightarrow{\mathcal{D}}-\sqrt{M}\mathbb{E}[c(Z,\cdot)\otimes B(X)]+N_{{\mathcal{HS}}},$$
	and, by the continuous mapping theorem, $n\tilde{V}_n\xrightarrow{\mathcal{D}}||-\sqrt{M}\mathbb{E}[c(Z,\cdot)\otimes B(X)]+N_{{\mathcal{HS}}}||^2$ (a limit which has the same probaiblity distribution as $||\sqrt{M}\mathbb{E}[c(Z,\cdot)\otimes B(X)]+N_{{\mathcal{HS}}}||^2$). If $nh^{2\nu}\rightarrow{\infty}$, we have,
	\begin{align*}
		\tilde{V}_n &= \left|\left|\mathbb{E}\left[c(Z,\cdot)\otimes \mathbb{E}[K_h(X-X')(m(X)-m(X'))|X]\right] +\frac{2}{n}\sum\limits_{i=1}^n \psi^1_n(i) + o_\mathbb{P}\left(\frac{1}{\sqrt{n}}\right)\right|\right|^2 \\
		&=  \text{KCMD}(\mathbb{E}[K_h(X-X')(m(X)-m(X'))|X]|Z) - 2h^\nu\left\langle \mathbb{E}[c(Z,\cdot)\otimes B(X)], \frac{2}{n}\sum\limits_{i=1}^n \psi^1_n(i)\right\rangle\\
		&\quad + o_\mathbb{P}\left(\frac{h^{\nu}}{\sqrt{n}}\right) + O_\mathbb{P}\left(\frac{1}{n}\right),
	\end{align*}
	since $\mu_{Bn} = \mathbb{E}[K_h(X-X')(m(X)-m(X'))|X]$ verifies $\mu_{Bn}= h^\nu(-B(X)+o_{L^1}(1))$. Therefore, it follows from the central limit theorem that
	$$\frac{\sqrt{n}}{h^\nu}(\tilde{V}_n -\mu_{Bn}) \xrightarrow{\mathcal{D}} -2\left\langle \mathbb{E}[c(Z,\cdot)\otimes B(X)], N_{{\mathcal{HS}}}\right\rangle,$$
	where $-2\left\langle \mathbb{E}[c(Z,\cdot)\otimes B(X)], N_{{\mathcal{HS}}}\right\rangle\sim N(0,4\sigma^2_B)$, with 
	\begin{align*}
		S &= \text{Var}\left( \left\langle \mathbb{E}[c(Z,\cdot)\otimes B(X)],(c(Z,\cdot)-b(X,\cdot))\otimes f(X)\mathcal{E}\right\rangle\right) \\
		&= \text{Var}\left( \mathbb{E}\left[(c(Z,Z')-b(X,Z'))\langle f(X)\mathcal{E},B(X')\rangle |Z,\mathcal{E}\right] \right).
	\end{align*}
	Finally, in both cases ($nh^{2\nu}\rightarrow{M}\ge 0$ and $nh^{2\nu}\rightarrow\infty$), $\tilde{V}_n = o_{\mathbb{P}}(1)$ and, by (\ref{Th1.2}), we have 
	\begin{itemize}
		\item $nU_n = n\tilde{V}_n - \mathbb{E}\left[\left|\left|N_{\mathcal{HS}}\right|\right|^2\right] + o_{\mathbb{P}}(1)$,
		\item $\sqrt{n}(U_n - \mu_{Bn})/h^\nu = \sqrt{n}(\tilde{V}_n - \mu_{Bn})/h^\nu + \sqrt{n}(U_n - \tilde{V}_n)/h^\nu =  \sqrt{n}(\tilde{V}_n - \mu_{Bn})/h^\nu + O_{\mathbb{P}}(1/\sqrt{n}h^\nu)$,
	\end{itemize}
	where we have implicitly used that $ \mathbb{E}\left[\left|\left|N_{\mathcal{HS}}\right|\right|^2\right] = \mathbb{E}\left[||(c(Z,\cdot) - b(X,\cdot))\otimes f(X)\mathcal{E}||^2\right]$, so Theorem 1-(2) and Theorem 1-(3) hold.
\end{proof}

\begin{proof}[\bfseries\textup{Proof of Theorem 2}]
	
	Let us consider, once again, $\tilde{V}_n = V_n(n^2/(n-1)^2)$ so that
	$$\tilde{V}_n = \left|\left|\frac{1}{{n\choose 2}}\sum\limits_{i<k}\psi_n(i,k)\right|\right|^2,$$
	as well as the Hoeffding decomposition 
	\begin{equation}\label{Th2.1}\tag{Th2.1}
		\frac{1}{{n\choose 2}}\sum\limits_{i< k} \psi_n(i,k) =  \mathbb{E}[\psi_n(i,k)] + \frac{2}{n}\sum\limits_{i=1}^n \psi^1_n(i)  + \frac{1}{{n \choose 2}}\sum\limits_{i<j}^n \psi^2_n(i,j).
	\end{equation}
	Given Proposition \ref{Pr2}-(6) and Proposition \ref{Pr2}-(7), by the same arguments used in the proof of Theorem 1, it follows that 
	$$\sqrt{n}\left(\frac{2}{n}\sum\limits_{i=1}^n \psi^1_n(i)  + \frac{1}{{n \choose 2}}\sum\limits_{i<j}^n \psi^2_n(i,j)\right)\xrightarrow{\mathcal{D}}N_{\mathcal{HS}},$$
	where $N_{\mathcal{HS}}$ is a $\mathcal{HS}$-valued Gaussian random variable with zero mean and the same covariance operator as $(c(Z_i,\cdot)-b(X_i,\cdot))\otimes f(X_i)\mathcal{E}_i$. Thus, the change in behavior of $\tilde{V}_n$ under $\{\mathcal{H}_{1n}\}_{n\in \mathbb{N}}$ with respect to that of $\tilde{V}_n$ under $\mathcal{H}_{0}$ depends, exclusively, on the asymptotic deviation of $\mathbb{E}[\psi_n(i,k)]$ with the exponent $\gamma>0$. Particularly, by Proposition \ref{Pr2}-(3), we have
	\begin{align*}
		\sqrt{n}\mathbb{E}[\psi_n(i,k)] &= \sqrt{n}\left(-h^\nu\mathbb{E}[c(Z_i,\cdot)\otimes B(X_i)] + o(h^{\nu}) + \frac{1}{n^\gamma}\mathbb{E}[c(Z_i,\cdot)\otimes f(X_i)\tilde{g}(Z_i)] + O\left(\frac{h^\nu}{n^\gamma}\right)\right)\\
		&= -\sqrt{n}h^\nu\mathbb{E}[c(Z_i,\cdot)\otimes B(X_i)] + \frac{\sqrt{n}}{n^\gamma}\mathbb{E}[c(Z_i,\cdot)\otimes f(X_i)\tilde{g}(Z_i)] + o\left(\sqrt{n}h^\nu + \frac{\sqrt{n}}{n^\gamma}\right)\label{Th2.2}\tag{Th2.2} .
	\end{align*}	 
	Assuming $nh^{2\nu}\rightarrow{M}\ge 0$ and $\gamma\ge1/2$, it holds
	$$\lim\limits_{n\rightarrow \infty}\sqrt{n}\mathbb{E}[\psi_n(i,k)] = -\sqrt{M}\mathbb{E}[c(Z_i,\cdot)\otimes B(X_i)] + \ind{\gamma = \frac{1}{2}}\mathbb{E}[c(Z_i,\cdot)\otimes f(X_i)\tilde{g}(Z_i)].$$
	Therefore, 
	$$\frac{\sqrt{n}}{{n\choose 2}}\sum\limits_{i< k} \psi_n(i,k) \xrightarrow{D} -\sqrt{M}\mathbb{E}[c(Z_i,\cdot)\otimes B(X_i)] + \ind{\gamma = \frac{1}{2}}\mathbb{E}[c(Z_i,\cdot)\otimes f(X_i)\tilde{g}(Z_i)] +N_{\mathcal{HS}},$$
	and, by the continuous mapping theorem,
	$$ n\tilde{V}_n = \left|\left| \frac{\sqrt{n}}{{n\choose 2}}\sum\limits_{i< k} \psi_n(i,k) \right|\right|^2\xrightarrow{D} \left|\left|\mathbb{E}[c(Z_i,\cdot)\otimes \left(\ind{\gamma = \frac{1}{2}}f(X_i)\tilde{g}(Z_i) - \sqrt{M}B(X_i))\right) +N_{\mathcal{HS}}\right|\right|^2.$$
	If $nh^{2\nu}\rightarrow{M}\ge 0$ and $\gamma<1/2$, then
	\begin{align*}
		n\tilde{V}_n &= \left|\left|\frac{\sqrt{n}}{n^\gamma}\mathbb{E}[c(Z_i,\cdot)\otimes f(X_i)\tilde{g}(Z_i)] + o\left(\sqrt{n}h^\nu + \frac{\sqrt{n}}{n^\gamma}\right) +O_{\mathbb{P}}\left(1\right)\right|\right|^2\\
		&= n^{1-2\gamma}\text{KCMD}(f(X_i)\tilde{g}(Z_i)|Z_i) + o_{\mathbb{P}}(n^{1-2\gamma}) \xrightarrow{\mathbb{P}}\infty.
	\end{align*}
	The case of $nh^{2\nu}\rightarrow\infty$ is more complicated. By Proposition \ref{Pr2}, computing the square norm of all terms in the Hoeffding decomposition in (\ref{Th2.1}) once expanded $\mathbb{E}[\psi_n(i,k)]$ as in (\ref{Th2.2}), we have
	\begin{align*}
		\tilde{V}_n &= \text{KCMD}(\mathbb{E}[K_h(X-X')(m(X)-m(X'))|X]|Z) + \frac{1}{n^{2\gamma}}\text{KCMD}(f(X)\tilde{g}(Z)|Z) + o\left(\frac{1}{n^{2\gamma}}\right)\\
		&\quad -\frac{2h^\nu}{n^\gamma} \mathbb{E}\left[c(Z,Z')\left\langle B(X), f(X')\tilde{g}(Z')\right\rangle\right] + o\left(\frac{h^\nu}{n^{\gamma}}\right) - 2h^\nu\left\langle \mathbb{E}[c(Z,\cdot)\otimes B(X)], \frac{2}{n}\sum\limits_{i=1}^n \psi^1_n(i)\right\rangle\\
		&\quad + o_\mathbb{P}\left(\frac{h^\nu}{\sqrt{n}}\right) + \frac{2}{n^\gamma}\left\langle \mathbb{E}[c(Z,\cdot)\otimes f(X)\tilde{g}(Z)], \frac{2}{n}\sum\limits_{i=1}^n \psi^1_n(i)\right\rangle + o_\mathbb{P}\left(\frac{1}{\sqrt{n}n^\gamma}\right) + O_{\mathbb{P}}\left(\frac{1}{n}\right).
	\end{align*}
	Let $\mu_{Bn} = \text{KCMD}(\mathbb{E}[K_h(X-X')(m(X)-m(X'))|X]|Z)$. If $\gamma >1/2$, then $\sqrt{n}/n^\gamma \rightarrow 0$ and $n^\gamma h^\nu = \sqrt{n}h^\nu n^\gamma/\sqrt{n}\rightarrow \infty$. Therefore, by the continuous mapping theorem and Slutsky's theorem,
	\begin{align*}
		\frac{\sqrt{n}}{h^\nu}(\tilde{V}_n - \mu_{Bn})&= O\left(\frac{\sqrt{n}}{ n^{2\gamma} h^\nu}\right) + O\left(\frac{\sqrt{n}}{n^{\gamma}}\right) - 2\left\langle \mathbb{E}[c(Z,\cdot)\otimes B(X)], \frac{2}{\sqrt{n}}\sum\limits_{i=1}^n \psi^1_n(i)\right\rangle + o_\mathbb{P}\left(1\right)\\
		 &\quad + O_\mathbb{P}\left(\frac{1}{ n^\gamma h^\nu}\right) + O_{\mathbb{P}}\left(\frac{1}{\sqrt{n}h^\nu}\right)\\
		&=- 2\left\langle \mathbb{E}[c(Z,\cdot)\otimes B(X)], \frac{2}{\sqrt{n}}\sum\limits_{i=1}^n \psi^1_n(i)\right\rangle  + o_\mathbb{P}\left(1\right) \xrightarrow{\mathcal{D}} - 2\left\langle \mathbb{E}[c(Z,\cdot)\otimes B(X)], N_{\mathcal{HS}}\right\rangle,
	\end{align*}
	which is the same limiting distribution obtained under $\mathcal{H}_0$ in the proof of Theorem 1 when $nh^{2\nu}\rightarrow\infty$. If  $\gamma =1/2$, then $n^\gamma h^\nu \rightarrow \infty$ but $\sqrt{n}/n^\gamma =1$. In such case we have,
	\begin{align*}
		\frac{\sqrt{n}}{h^\nu}(\tilde{V}_n - \mu_{Bn})	&=-2\mathbb{E}\left[c(Z,Z')\left\langle B(X), f(X')\tilde{g}(Z')\right\rangle\right]- 2\left\langle \mathbb{E}[c(Z,\cdot)\otimes B(X)], \frac{2}{\sqrt{n}}\sum\limits_{i=1}^n \psi^1_n(i)\right\rangle + o_\mathbb{P}\left(1\right)\\
		&\xrightarrow{\mathcal{D}} -2\mathbb{E}\left[c(Z,Z')\left\langle B(X), f(X')\tilde{g}(Z')\right\rangle\right] - 2\left\langle \mathbb{E}[c(Z,\cdot)\otimes B(X)], N_{\mathcal{HS}}\right\rangle.
	\end{align*}
	The asymptotic distribution in this case consists in a location shif with respect to the limit under the null. Finally, if $\gamma<1/2$ we can distinguish three different scenarios.
	\begin{itemize}
		\item If $n^\gamma h^\nu\rightarrow 0$, then $\sqrt{n}/(n^{2\gamma}h^\nu)$ dominates $\sqrt{n}/n^\gamma$, $1/(n^\gamma h^\nu)$ and $1/(\sqrt{n}h^\nu)$. Therefore,
		$$\frac{\sqrt{n}}{h^\nu}(\tilde{V}_n -\mu_{Bn}) = \frac{\sqrt{n}}{h^\nu n^{2\gamma}}\text{KCMD}(f(X)\tilde{g}(Z)|Z) + o_\mathbb{P}\left(\frac{\sqrt{n}}{h^\nu n^{2\gamma}}\right)\xrightarrow{\mathbb{P}}\infty,$$
		since $\text{KCMD}(f(X)\tilde{g}(Z)|Z)>0$.
		\item If $n^\gamma h^\nu\rightarrow \sqrt{\raisebox{0.1ex}{\(\tilde{M}\)}}> 0$, then $\sqrt{n}/n^\gamma$ and $\sqrt{n}/(n^{2\gamma} h^\nu )$ have the same asymptotic order, wich dominates both $1/( n^\gamma h^\nu)$ and $1/(\sqrt{n}h^\nu )$. In that case,
		$$\frac{\sqrt{n}}{h^\nu}(\tilde{V}_n - \mu_{Bn})
			=\frac{\sqrt{n}}{n^\gamma}\left(\frac{1}{\sqrt{\raisebox{0.1ex}{\(\tilde{M}\)}}}\text{KCMD}(f(X)\tilde{g}(Z)|Z)-2\mathbb{E}\left[c(Z,Z')\left\langle B(X), f(X')\tilde{g}(Z')\right\rangle\right]\right) + o_\mathbb{P}\left(\frac{\sqrt{n}}{n^\gamma}\right).$$
		\item If $n^\gamma h^\nu\rightarrow \infty$, then $\sqrt{n}/n^\gamma$ dominates $\sqrt{n}/( n^{2\gamma}h^\nu)$, $1/(n^\gamma h^\nu )$ and $1/(\sqrt{n}h^\nu)$. Thus, 
		$$\frac{\sqrt{n}}{h^\nu}(\tilde{V}_n - \mu_{Bn})= -\frac{2\sqrt{n}}{n^\gamma}\mathbb{E}\left[c(Z,Z')\left\langle B(X), f(X')\tilde{g}(Z')\right\rangle\right] + o_\mathbb{P}\left(\frac{\sqrt{n}}{n^\gamma}\right).$$
	\end{itemize}
	In the last two scenarios, the test statistic can tend in probability to $\infty$, $-\infty$ or even have a proper limit in distribution depending on the sign of $\mathbb{E}\left[c(Z,Z')\left\langle B(X), f(X')\tilde{g}(Z')\right\rangle\right]$, the relationship with $\text{KCMD}(f(X)\tilde{g}(Z)|Z)$ and the behavior of lower order terms. 
	
	Finally, since $\gamma>0$, we have $\tilde{V}_n = o_{\mathbb{P}}(1)$ in all scenarios. By Proposition \ref{Pr3}, following the same arguments used in the proof of Theorem 1, it holds that
	$$	U_n = \tilde{V}_n - \frac{\mathbb{E}\left[\left|\left|N_{\mathcal{HS}}\right|\right|^2\right]}{n} + o_{\mathbb{P}}\left(\frac{1}{n}\right).$$
	Therefore, the distribution of $U_n$ and $\tilde{V}_n$ coincide if $nh^{2\nu} \rightarrow\infty$ whereas, if $nh^{2\nu} \rightarrow M\ge 0$, 
	$$n(U_n -\tilde{V}_n) \xrightarrow{\mathbb{P}} -\mathbb{E}\left[\left|\left|N_{\mathcal{HS}}\right|\right|^2\right],$$ 
	so Theorem 2 holds by virtue of Slutsky's theorem.
\end{proof}

\section{Bootstrap asymptotics: background}

\begin{definition}
	\label{Def4}
	
	Let $S$ be a normed space and let $\mathcal{A}_n^*$ and $\mathcal{A}$ be an $S$-valued bootstrap statistic based on a sample $\{\eta_n\}_{n=1}^\infty$, which takes values in some measurable space $E$, and an independent $S$-valued random variable, respectively. We say that
	$\mathcal{A}_n^*$ converges weakly to $\mathcal{A}$ almost surely, and we denote it by $\mathcal{A}_n^* \xrightarrow{\mathcal{D}^*} \mathcal{A}$ a.s. if, conditionally on $\{\eta_n\}_{n=1}^\infty$, $\mathcal{A}_n^*$ converges weakly to $\mathcal{A}$ for almost all sample realizations of $\{\eta_n\}_{n=1}^\infty$. We also say that $\mathcal{A}_n^*$ converges in probability $\mathbb{P^*}$ to $0$ almost surely, and we denote it by $A_n^* = o_{\mathbb{P}^*}(1)$ a.s. if $\mathbb{P}^*(||\mathcal{A}_n^*||>M) \rightarrow{0}$ a.s. for all $M>0$. Moreover, we say that $\mathcal{A}_n^*$ converges weakly to $\mathcal{A}$ (resp. in probability $\mathbb{P^*}$ to $0$) in probability $\mathbb{P}$, and we denote it by $\mathcal{A}_n^* \xrightarrow{\mathcal{D}^*} \mathcal{A}$ in $\mathbb{P}$ (resp. $A_n^* = o_{\mathbb{P}^*}(1)$ in $\mathbb{P}$) if, for any subsequence $\{n_k\}_{k\in \mathbb{N}}$ there is a further subsequence $\{n_{k_l}\}_{l\in \mathbb{N}}$ such that 
	$\mathcal{A}_{n_{k_l}}^* \xrightarrow{\mathcal{D}^*} \mathcal{A}$ a.s. (resp. $\mathcal{A}_{n_{k_l}}^* \xrightarrow{\mathbb{P}^*} 0$ a.s.).
\end{definition}

Note that usual results regarding weak convergence (convergence in distribution if $\mathcal{S}$ is an Euclidean space) and convergence in probability, such as the continuous mapping theorem and Slutsky's theorem, trivially hold almost surely in the sense implied by Definition \ref{Def4}. By arguments based on subsequences, they also hold for weak convergence and convergence in probability $\mathbb{P}^*$, in probability $\mathbb{P}$. The following lemma justifies the sufficiency of the latter type of convergence for bootstrap-calibrated tests.

\begin{lemma}
	\label{Lm3}
	Let $\mathcal{A}_n^*$ and $\mathcal{A}$ be as in Definition \ref{Def4}, with $\mathcal{S}=\mathbb{R}$. Denote the cumulative dsitribution functions of $\mathcal{A}_n^*$ and $\mathcal{A}$ by $F_n^*$ and $F$, respectively. Assume that $\mathcal{A}_n^* \xrightarrow{\mathcal{D}^*} \mathcal{A}$ in $\mathbb{P}$ and also that $F$ is continuous. Then,
	$$\sup\limits_{x\in \mathbb{R}}|F_n^*(x) - F(x)| \xrightarrow{\mathbb{P}} 0.$$
	Furthermore, for a significance level $\alpha \in (0,1)$, denote the $(1-\alpha)$th quantiles of the distributions of $\mathcal{A}_n^*$ and $\mathcal{A}$ by $Q_n^*(1-\alpha)$ and $Q(1-\alpha)$, respectively. We have
	$$Q_n^*(1-\alpha) \xrightarrow{\mathbb{P}} Q(1-\alpha).$$
	In addition, let $\tilde{\mathcal{A}}_n$ be another sequence of random variables with a continous limit in distribution, $\tilde{\mathcal{A}}$. Then,
	$$\mathbb{P}(\tilde{\mathcal{A}}_n>Q_n^*(1-\alpha)) \rightarrow \mathbb{P}(\tilde{\mathcal{A}}>Q(1-\alpha)).$$ 
	If $\tilde{\mathcal{A}}_n\xrightarrow{\mathbb{P}}\infty$ and $Q_n^*(1-\alpha) = O_{\mathbb{P}}(1)$, we have
	$$\mathbb{P}(\tilde{\mathcal{A}}_n>Q_n^*(1-\alpha)) \rightarrow 1.$$ 
	Finally, denoting by $\mathbb{E}^*$ the expectation operator with respect to $\mathbb{P}^*$, it holds that
	$$\mathbb{E}^*[|\mathcal{A}_n^*|] = O_{\mathbb{P}}(1) \implies Q_n^*(1-\alpha) = O_{\mathbb{P}}(1).$$
\end{lemma}

\begin{proof}[\bfseries\textup{Proof of Lemma \ref{Lm3}}]

	By the definition of weak convergence in probability $\mathbb{P}$, for any subsequence $\{n_k\}_{k\in \mathbb{N}}$ there is a further subsequence $\{n_{k_l}\}_{l\in \mathbb{N}}$ such that 
	$\mathcal{A}_{n_{k_l}}^* \xrightarrow{\mathcal{D}^*} \mathcal{A}$ a.s.; that is, $F_{n_{k_l}}^*(x)-F(x) \rightarrow 0$ a.s. for all $x\in \mathbb{R}$ continuity point of $F$. Since $F$ is continuous, the convergence of distribution functions is, in fact, uniform:
	$$\sup\limits_{x\in \mathbb{R}}|F_{n_{k_l}}^*(x) - F(x)| \xrightarrow{a.s.} 0.$$
	As $\{n_k\}_{k\in \mathbb{N}}$ is arbitrary, it holds that
	$$\sup\limits_{x\in \mathbb{R}}|F_n^*(x) - F(x)| \xrightarrow{\mathbb{P}} 0.$$
	
	Let $M>0$. By the definition of $Q(1-\alpha)$, $\mathbb{P}(\mathcal{A}\le Q(1-\alpha)-M)<1-\alpha$. Let $\epsilon_1 = 1-\alpha - F(Q(1-\alpha)-M)>0$. It is immediate that
	\begin{align*}
		|\mathbb{P}^*(\mathcal{A}_n^*\le Q(1-\alpha) - M) - \mathbb{P}(\mathcal{A}\le Q(1-\alpha)-M)|<\epsilon_1 &\implies \mathbb{P}^*(\mathcal{A}_n^*\le Q(1-\alpha) - M) < 1-\alpha\\
		&\implies Q_n^*(1-\alpha) > Q(1-\alpha) - M.
	\end{align*}
	Analogously, let $\epsilon_2 = \alpha - F(Q(1-\alpha)+M)>0$. Then, 
	\begin{align*}
	|\mathbb{P}^*(\mathcal{A}_n^*> Q(1-\alpha) + M) - \mathbb{P}(\mathcal{A}> Q(1-\alpha)+M)|<\epsilon_2  &\implies \mathbb{P}^*(\mathcal{A}_n^*> Q(1-\alpha) + M) < \alpha\\
		&\implies Q_n^*(1-\alpha) < Q(1-\alpha) + M.
	\end{align*}
	Note that
	$$|\mathbb{P}^*(\mathcal{A}_n^*> Q(1-\alpha) + M) - \mathbb{P}(\mathcal{A}> Q(1-\alpha)+M)| = |\mathbb{P}^*(\mathcal{A}_n^*\le Q(1-\alpha) + M) - \mathbb{P}(\mathcal{A}\le Q(1-\alpha)+M)|, $$
	and, thus, if $\sup\limits_{x\in \mathbb{R}} |F^*_n(x) - F(x)| \le \epsilon = \min\{\epsilon_1,\epsilon_2\}$, then $Q_n^*(1-\alpha) > Q(1-\alpha) - M$ and $Q_n^*(1-\alpha) < Q(1-\alpha) + M$ hold simultaneously. By the definition of convergence in distribution in probability $\mathbb{P}$, for all $\varsigma>0$ there exists some $n_0\in\mathbb{N}$ such that for all $n\ge n_0$,
	$$\mathbb{P}(Q(1-\alpha) - M  < Q_n^*(1-\alpha) < Q(1-\alpha) + M)
		\ge\ \mathbb{P}\left(\sup\limits_{x\in \mathbb{R}} |F^*_n(x) - F(x)| \le \epsilon\right) > 1-\varsigma,$$
	since $\mathbb{P}\left(\sup\limits_{x\in \mathbb{R}} |F^*_n(x) - F(x)| > \epsilon\right) < \varsigma$ for $n$ large enough.
	This implies that, for any arbitrary $M>0$ and $\varsigma>0$, if $n$ is sufficiently large,
	$$\mathbb{P}(|Q(1-\alpha) - Q_n^*(1-\alpha)|<M) > 1-\varsigma,$$
	or, equivalently, $Q_n^*(1-\alpha)\xrightarrow{\mathbb{P}}Q(1-\alpha)$.
	
	Now, given the sequence $\{\tilde{\mathcal{A}}_n\}_{n=1}^\infty$ and its limit $\tilde{\mathcal{A}}$, it follows that
	\begin{align*}
	&\mathbb{P}(\tilde{\mathcal{A}}_n>Q_n^*(1-\alpha)) - \mathbb{P}(\tilde{\mathcal{A}}>Q(1-\alpha))\\
	=\ &\mathbb{P}(\tilde{\mathcal{A}}_n>Q_n^*(1-\alpha)) - \mathbb{P}(\tilde{\mathcal{A}}_n>Q(1-\alpha)) + \mathbb{P}(\tilde{\mathcal{A}}_n>Q(1-\alpha)) - \mathbb{P}(\tilde{\mathcal{A}}>Q(1-\alpha)).
	\end{align*}
	The convergence in distribution of $\tilde{\mathcal{A}}_n$ to $\tilde{\mathcal{A}}$, the latter being a continuous random variable, implies that 
	$$\mathbb{P}(\tilde{\mathcal{A}}_n>Q(1-\alpha)) - \mathbb{P}(\tilde{\mathcal{A}}>Q(1-\alpha))\rightarrow{0}.$$
	Moreover, 
	\begin{align*}
		&\mathbb{P}(\tilde{\mathcal{A}}_n>Q_n^*(1-\alpha)) - \mathbb{P}(\tilde{\mathcal{A}}_n>Q(1-\alpha)) \\
		=\ &\mathbb{P}(\tilde{\mathcal{A}}_n>Q_n^*(1-\alpha),\tilde{\mathcal{A}}_n\le Q(1-\alpha)) - \mathbb{P}(\tilde{\mathcal{A}}_n>Q(1-\alpha),\tilde{\mathcal{A}}_n\le Q_n^*(1-\alpha)).
	\end{align*}
	Observe that, based on previous arguments, we have
	$$\mathbb{P}(\tilde{\mathcal{A}}_n>Q_n^*(1-\alpha),\tilde{\mathcal{A}}_n\le Q(1-\alpha)) \le  \mathbb{P}(Q_n^*(1-\alpha)< Q(1-\alpha)) \rightarrow 0,$$
	Analogously,
	$$\mathbb{P}(\tilde{\mathcal{A}}_n>Q(1-\alpha),\tilde{\mathcal{A}}_n\le Q_n^*(1-\alpha)) \le  \mathbb{P}(Q_n^*(1-\alpha)> Q(1-\alpha)) \rightarrow 0,$$
	so $\mathbb{P}(\tilde{\mathcal{A}}_n>Q_n^*(1-\alpha)) \rightarrow \mathbb{P}(\tilde{\mathcal{A}}>Q(1-\alpha))$.
	
	Let us now assume that $\tilde{\mathcal{A}}_n \xrightarrow{\mathbb{P}}\infty$ and $Q_n^*(1-\alpha) = O_{\mathbb{P}}(1)$. For all $M>0$ it is satisfied that
	\begin{align*}
		\mathbb{P}(\tilde{\mathcal{A}}_n\le Q_n^*(1-\alpha)) &\le \mathbb{P}(\tilde{\mathcal{A}}_n\le Q_n^*(1-\alpha), Q_n^*(1-\alpha)\le M) + \mathbb{P}(\tilde{\mathcal{A}}_n\le Q_n^*(1-\alpha), Q_n^*(1-\alpha)>M) \\
		&\le  \mathbb{P}(\tilde{\mathcal{A}}_n\le  M) +  \mathbb{P}(Q_n^*(1-\alpha)>M).
	\end{align*}
	Given an arbitrary $\varsigma>0$, if $n$ is sufficiently large, by the convergence in probability of $\tilde{\mathcal{A}}_n$ to infinity and the bound in probability of $Q_n^*(1-\alpha)$, there will be a certain $M>0$ such that
	$$	\mathbb{P}(\tilde{\mathcal{A}}_n\le Q_n^*(1-\alpha)) \le  \mathbb{P}(\tilde{\mathcal{A}}_n\le  M) +  \mathbb{P}(Q_n^*(1-\alpha)>M) \le \varsigma,$$
	and, in conclusion, $\mathbb{P}(\tilde{\mathcal{A}}_n\le Q_n^*(1-\alpha)) \rightarrow{0}.$
	
	To prove the last result, let $M_{n}(1-\alpha)= \mathbb{E}^*[|\mathcal{A}_n^*|]/\alpha$. By virtue of Markov's inequality, it follows that
	\begin{align*}
		&\mathbb{P^*}(|\mathcal{A}_n^*|>M_{n}(1-\alpha))\le \alpha\\
		 \iff\ &\mathbb{P^*}(|\mathcal{A}_n^*|\le M_{n}(1-\alpha)) = \mathbb{P^*}(\mathcal{A}_n^* \le M_{n}(1-\alpha)) - \mathbb{P^*}(\mathcal{A}_n^* < -M_{n}(1-\alpha))\ge 1-\alpha.
	\end{align*}
	On the one hand,
	$$\mathbb{P^*}(\mathcal{A}_n^*\le M_{n}(1-\alpha)) \ge  \mathbb{P^*}(|\mathcal{A}_n^*|\le M_{n}(1-\alpha))\ge 1-\alpha.$$
	Then, by definition of $Q_n^*(1-\alpha)$, we have that $Q_n^*(1-\alpha)\le M_{n}(1-\alpha)$. On the other hand,
	$$-\mathbb{P^*}(\mathcal{A}_n^*< -M_{n}(1-\alpha)) \ge 1-\alpha - \mathbb{P^*}(\mathcal{A}_n^* \le M_{n}(1-\alpha)) \ge -\alpha \implies \mathbb{P^*}(\mathcal{A}_n^* < -M_{n}(1-\alpha)) \le \alpha$$
	and thus, $-M_{n}(1-\alpha)\le Q_{n}^*(\alpha)$. Equivalently, since $\alpha\in(0,1)$ is arbitrary, $-M_{n}(\alpha)\le Q_n^*(1-\alpha)$. Therefore,
	$$ |Q_n^*(1-\alpha)| \le \max\left\{M_{n}(1-\alpha),M_{n}(\alpha)\right\} =  \max\left\{\frac{1}{\alpha},\frac{1}{1-\alpha}\right\}\mathbb{E}^*[|\mathcal{A}_n^*|] = O_{\mathbb{P}}(1)\implies Q_n^*(1-\alpha)=O_{\mathbb{P}}(1).$$
\end{proof}

The following central limit theorem for multiplier-based bootstrap statistic is a key result in the proofs of Theorems 3 and 4. We did not find a satisfactory proof for this result in the literature, so we provide a relatively simple one.

\begin{Theorem}	
	\label{Th5}
	Let $\{r_i\}_{i=1}^n$ be an i.i.d. sample of real-valued random variables drawn from a distribution with zero mean, unit variance and finite $(2+\delta)$th order moment, for some $\delta>0$. Let $\{S_i\}_{i=1}^n$ be an i.i.d sample of random variables taking values in a separable Hilbert space $\mathcal{S}$, which is also independent of $\{r_i\}_{i=1}^n$, and such that $\mathbb{E}[||S_i||^{2+\delta}]<\infty$. Then,
	conditionally on $\{S_i\}_{i=1}^n$, the bootstrap statistic $\mathcal{A}_n^* =  \sum\limits_{i=1}^n r_iS_i/\sqrt{n}$ verifies
	$$\mathcal{A}_n^*\xrightarrow{\mathcal{D}^*}N_{{\mathcal{S}}} \ \text{a.s.},$$
	where $N_{{\mathcal{S}}}$ is a zero-mean $\mathcal{S}$-valued Gaussian random variable with the same covariance operator as $S_i$ for all $i$. Moreover
	$$ ||\mathcal{A}_n^*||^2 \xrightarrow{\mathcal{D}^*} ||N_{{\mathcal{S}}}||^2 \ \text{a.s.} $$
	In addition,
	$$\frac{1}{(n-1)}\sum\limits_{i\neq j} r_ir_j\langle S_i,S_j\rangle \xrightarrow{\mathcal{D}^*} ||N_{{\mathcal{S}}}||^2 - \mathbb{E}\left[ \left|\left|N_{\mathcal{S}}\right|\right|^2\right]\ \text{a.s.} $$
\end{Theorem}

\begin{proof}[\bfseries\textup{Proof of Theorem \ref{Th5}}]
	
	Let us first verify that, for all $0\neq s\in \mathcal{S}$, 
	\begin{equation}\label{Th5.1}\tag{Th5.1}
		\sum\limits_{i=1}^{{n}} \left\langle s,\frac{1}{\sqrt{n}} r_iS_i \right\rangle\xrightarrow{\mathcal{D}^*} N(0,\sigma^2_s) \ \text{a.s.},
	\end{equation}
	for some $\sigma^2_s\ge 0$. Using the lineality of the inner product,
	$$\sum\limits_{i=1}^{{n}} \left\langle s,\frac{1}{\sqrt{n}} r_iS_i \right\rangle = \sum\limits_{i=1}^{{n}} r_i\frac{1}{\sqrt{n}}\left\langle s, S_i \right\rangle = \sum\limits_{i=1}^{{n}}r_iw_{in},$$
	where $w_{in}$ stands for $\left\langle s,S_i \right\rangle /\sqrt{n}$. By the zero mean of the boostrap multipliers, 
	$$\mathbb{E}^*\left[\sum\limits_{i=1}^{{n}}r_iw_{in}\right] =0,$$ 
	where $\mathbb{E}^*$ denotes conditional expectation given $\{S_i\}_{i=1}^n$. Moreover, by the strong law of large of large numbers and the bootstrap multipliers independence and unit variance,
	$$\text{Var}^*\left(\sum\limits_{i=1}^{{n}}r_iw_{in}\right) = \frac{1}{{n}}\sum\limits_{i=1}^{{n}} \left\langle s, S_i\right\rangle^2 \xrightarrow{a.s.} \mathbb{E}\left[\left\langle s, S_i \right\rangle^2\right],$$
	where $\text{Var}^*$ denotes conditional variance given $\{S_i\}_{i=1}^n$, since, by the Cauchy-Schwarz inequality,
	$$\mathbb{E}\left[\left\langle s, S_i \right\rangle^2\right] \le ||s||^2\mathbb{E}\left[||S_i||^2\right]<\infty.$$
	Therefore, if $\mathbb{E}\left[\left\langle s, S_i \right\rangle^2\right] = 0$ we have, by Chebyshev's inequality, that $\sum\limits_{i=1}^{n}r_iw_{in} = o_{\mathbb{P}^*}(1)$ a.s. Let us now assume that $\mathbb{E}\left[\left\langle s, S_i \right\rangle^2\right] >0$. Given $\delta>0$,
	$$\mathbb{E}\left[\left\langle s, S_i \right\rangle^{2+\delta}\right] \le ||s||^{2+\delta}\mathbb{E}\left[||S_i||^{2+\delta}\right]<\infty$$
	and, thus, for any $j = 1,\ldots, n$,
	$$\sum\limits_{i=1}^{{n}}\mathbb{E}^*[r_i^{2+\delta}]w^{2+\delta}_{ik} = \mathbb{E}^*[r_j^{2+\delta}]\frac{1}{{n}^{1+\frac{\delta}{2}}} \sum\limits_{i=1}^{{n}} \left\langle s, S_i \right\rangle^{2+\delta}\xrightarrow{a.s.} 0.$$
	Therefore, the Lyapunov condition is verified for the triangular array $\{\{r_iw_{in}\}_{i=1}^{n}\}_{n=1}^\infty$:
	$$\frac{1}{\text{Var}^*\left(\sum\limits_{i=1}^{{n}}r_iw_{in}\right)^{2+\delta}}\sum\limits_{i=1}^{{n}}\mathbb{E}^*\left[|r_iw_{in}|^{2+\delta}\right] \xrightarrow{a.s.}0.$$
	In particular, Lindeberg condition is also verified and, ultimately, (\ref{Th5.1}) holds with $\sigma^2_s = \mathbb{E}\left[\left\langle s, S_i \right\rangle^2\right].$
	
	Note that, by the independence, zero mean and unit variance of the bootstrap multipliers, the covariance operator of $\sum\limits_{i=1}^n r_iS_i/\sqrt{n}$, $\Gamma_n$, verifies
	$$\Gamma_n= \mathbb{E}^*\left[\left(\frac{1}{\sqrt{n}}\sum\limits_{i=1}^n r_iS_i\right) \otimes \left(\frac{1}{\sqrt{n}}\sum\limits_{i=1}^n r_iS_i\right)\right] = \frac{1}{n}\sum\limits_{i=1}^n S_i \otimes S_i.$$
	and, thus, for any $s\in \mathcal{S}$ we have, by the strong law of large numbers,
	\begin{equation}\label{Th5.2}\tag{Th5.2}
		\left\langle \Gamma(s),s\right\rangle = \frac{1}{n}\sum\limits_{i=1}^n \langle S_i,s\rangle^2 \xrightarrow{a.s.}\mathbb{E}\left[\langle S_i,s\rangle^2\right] =  \left\langle \mathbb{E}\left[S_i \otimes S_i\right](s),s\right\rangle.
	\end{equation}
	Moreover, as $\text{trace}(S_i\otimes S_i) = ||S_i||^2$ and the trace is a linear opreator, it holds that
	\begin{equation}\label{Th5.3}\tag{Th5.3}
		\text{trace}\left(\Gamma_n\right) = \frac{1}{n}\sum\limits_{i=1}^n ||S_i||^2 \xrightarrow{a.s.} \mathbb{E}[||S_i||^2] = \text{trace}\left(\mathbb{E}[S_i\otimes S_i]\right).
	\end{equation}
	Since (\ref{Th5.1}), (\ref{Th5.2}) and (\ref{Th5.3}) hold with $\sigma^2_s = \mathbb{E}\left[\left\langle S_i,s \right\rangle^2\right] = \left\langle\mathbb{E}\left[S_i \otimes S_i\right](s),s\right\rangle$ for all $s\neq 0$, it follows from Lemma 3.1 in \cite{Chen1998} (see also Remark 3.3 in that document) that
	$$\sum\limits_{i=1}^{n} \frac{1}{\sqrt{n}} r_iS_i \xrightarrow{\mathcal{D}^*} N_{\mathcal{S}} \ \text{a.s.},$$
	where $N_{\mathcal{S}}$ is Gaussian, zero-mean and has the same covariance operator as $S_i$ for all $i$. Then, by the continuous mapping theorem,
	$$ \left|\left|\sum\limits_{i=1}^{n} \frac{1}{\sqrt{n}} r_iS_i\right|\right|^2 \xrightarrow{\mathcal{D}^*}  \left|\left|N_{\mathcal{S}}\right|\right|^2 \ \text{a.s.}.$$
	Furthermore, note that
	$$ \frac{1}{(n-1)}\sum\limits_{i\neq j} r_ir_j\langle S_i,S_j\rangle =\frac{n}{(n-1)} \left(\left|\left|\sum\limits_{i=1}^{n} \frac{1}{\sqrt{n}} r_iS_i\right|\right|^2 - \frac{1}{n}\sum\limits_{i=1}^n r_i^2||S_i||^2\right).$$
	 It follows from the strong law of large numbers that $ \sum\limits_{i=1}^n ||S_i||^{2+\delta}/n \rightarrow \mathbb{E}\left[ \left|\left|S_i\right|\right|^{2+\delta}\right] \ \text{a.s.}$, which implies that 
	$\sup\limits_{n\in \mathbb{N}}\frac{1}{{n}}\sum\limits_{i=1}^{{n}}|| S_i||^{2+\delta}
	< \infty \ \text{a.s.}$
	By the strong law of large numbers for weighted averages of independent random variables (see e.g. Theorem 3.4 in \citeauthor{Baxter2004}, \citeyear{Baxter2004}) we have, $\mathbb{P}$-almost surely; that is, for almost all sequences $\{S_i\}_{i=1}^\infty$, that $\sum\limits_{i=1}^n (r_i^2-1)||S_i||^2/n$ converges to zero $\mathbb{P}^*$-almost surely; that is, for almost all sequences $\{r_i\}_{i=1}^\infty$, given a particular sequence $\{S_i\}_{i=1}^\infty$. In turn, $\sum\limits_{i=1}^n (r_i^2-1)||S_i||^2/n = o_{\mathbb{P}^*}(1)$ a.s. and, ultimately,
	$$ \frac{1}{n}\sum\limits_{i=1}^n r_i^2||S_i||^2 - \mathbb{E}\left[ \left|\left|N_{\mathcal{S}}\right|\right|^2\right] =\frac{1}{n}\sum\limits_{i=1}^n (r_i^2-1)||S_i||^2 +\left(\frac{1}{n}\sum\limits_{i=1}^n ||S_i||^2- \mathbb{E}\left[ \left|\left|S_i\right|\right|^2\right]\right)  = o_{\mathbb{P}^*}(1) \ \text{a.s.}$$
	Therefore, by Slutsky's theorem,
	\begin{align*}
		\frac{1}{(n-1)}\sum\limits_{i\neq j} r_ir_j\langle S_i,S_j\rangle 
		\xrightarrow{\mathcal{D}^*}  \left|\left|N_{\mathcal{S}}\right|\right|^2 - \mathbb{E}\left[ \left|\left|N_{\mathcal{S}}\right|\right|^2\right] \ \text{a.s.}
	\end{align*}
\end{proof}

\section{Proofs of Theorems 3 and 4}

Recall that $nU_n^{*}$ is defined as
\begin{equation}\label{Sm.6}\tag{Sm.6}
	nU_n^{*} = \frac{1}{n-1}\sum\limits_{i\neq j} r_ir_jD_{ij}\left\langle \tilde{\hat{\mathcal{E}}}_i , \tilde{\hat{\mathcal{E}}}_j \right\rangle ,
\end{equation}
where $\{r_i\}_{i=1}^n$ is a collection of i.i.d random variables, independent of $\{(Z_i,Y_i)\}_{i=1}^n$, with zero mean, unit variance and finite $(2+\delta)$th order moment, $D_{ij}$ verifies
$$D_{ij} = \frac{1}{n^2}\sum\limits_{k,l=1}^n K_h(X_i-X_k)K_h(X_j-X_l)\left(c(Z_i,Z_j)-c(Z_i,Z_l) - c(Z_k,Z_j) + c(Z_k,Z_l)\right),$$
for all $i,j = 1,\ldots, n$, 
$$\tilde{\hat{\mathcal{E}}}_i = Y_i - \hat{m}_{\tilde{h}}(X_i)\ind{\hat{f}'_{h}(X_i)\le |\hat{f}_{\tilde{h}}(X_i)|\kappa_n} + \hat{m}'_{h}(X_i)\ind{\hat{f}'_{h}(X_i)>|\hat{f}_{\tilde{h}}(X_i)|\kappa_n},$$
where $\hat{f}_{\tilde{h}}$, $\hat{f}'_{h}$ and $\hat{m}_{\tilde{h}}$ are defined as in Lemma $\ref{Lm2}$, 
$$\hat{m}'_{h}(x) =  \frac{1}{n}\sum\limits_{k=1}^n\frac{|K_h(x-X_k)|}{\hat{f}'_{h}(x)}Y_k,$$
and $\kappa_n$ and $\tilde{h}$ are chosen so that $\tilde{h}\rightarrow0$, $n\tilde{h}^d\rightarrow\infty$, $\kappa_n \tilde{h}^\nu \rightarrow0$ and $\kappa_n^2/(n\tilde{h}^d)\rightarrow 0$.

Before establishing the asymptotic behavior of $nU_n^*$, let us prove that the limit distributions of $nU_n$ under $\mathcal{H}_0$ and $\{\mathcal{H}_{1n}\}_{n\in \mathbb{N}}$, when $nh^{2\nu}\rightarrow 0$, are absolutely continuous.

\begin{lemma}
	\label{Lm4}
	Let $N_{{\mathcal{HS}}}$ be a $\mathcal{HS}$-valued zero-mean Gaussian random variable with the same covariance operator as $(c(Z,\cdot)-b(X,\cdot))\otimes f(X)\mathcal{E}$ and let $B$ be defined as in Lemma \ref{Lm1}. Then, both $||\sqrt{M}\mathbb{E}[c(Z,\cdot)\otimes B(X)]+N_{{\mathcal{HS}}}||^2$ and $||\mathbb{E}[c(Z,\cdot)\otimes (\sqrt{M}B(X)-f(X)\tilde{g}(Z))]+N_{{\mathcal{HS}}}||^2$ are absolutely continuous if $\mathbb{P}(\mathcal{E}\neq 0)>0$ and $\sigma(W)\nsubseteq\sigma(X)$.
\end{lemma}

\begin{proof}[\bfseries\textup{Proof of Lemma \ref{Lm4}}]
	
	The proof is the same in both cases so we focus on $||\sqrt{M}\mathbb{E}[c(Z,\cdot)\otimes B(X)]+N_{{\mathcal{HS}}}||^2$. Let $\{(\lambda_j,\phi_j)\}_{j=1}^\infty$ denote the eigensystem of the covariance operator of $N_{{\mathcal{HS}}}$, which coincides with the covariance operator of $(c(Z,\cdot) - b(X,\cdot))\otimes f(X)\mathcal{E}$. Assume that the eigenvalues $\{\lambda_j\}_{j=1}^\infty$ are sorted in decreasing order. Since $\mathbb{P}(\mathcal{E}\neq 0)>0$ and $\sigma(W)\nsubseteq\sigma(X)$, $N_{\mathcal{HS}}$ is not constant and, thus, $\lambda_1>0$. Then, the following Karhunnen-Loève representation holds almost surely:
	$$ ||\sqrt{M}\mathbb{E}[c(Z,\cdot)\otimes B(X)]+N_{{\mathcal{HS}}}||^2 = \sum\limits_{j=1}^\infty \langle \sqrt{M}\mathbb{E}[c(Z,\cdot)\otimes B(X)]+N_{{\mathcal{HS}}},\phi_j \rangle^2.$$
	Furthermore, the variables $\langle \sqrt{M}\mathbb{E}[c(Z,\cdot)\otimes B(X)]+N_{{\mathcal{HS}}},\phi_j \rangle$ are Gaussian, independent, with variance $\lambda_j$ and mean $\langle \sqrt{M}\mathbb{E}[c(Z,\cdot)\otimes B(X)],\phi_j \rangle$. Thus, we can decompose $||\sqrt{M}\mathbb{E}[c(Z,\cdot)\otimes B(X)]+N_{{\mathcal{HS}}}||^2$ as the sum of $\langle \sqrt{M}\mathbb{E}[c(Z,\cdot)\otimes B(X)]+N_{{\mathcal{HS}}},\phi_1 \rangle^2$, wich is a scaled noncentral chi-squared random variable (and therefore, absolutely continuous) and $||\sqrt{M}\mathbb{E}[c(Z,\cdot)\otimes B(X)]+N_{{\mathcal{HS}}}||^2 - \sqrt{M}\mathbb{E}[c(Z,\cdot)\otimes B(X)]+N_{{\mathcal{HS}}},\phi_1 \rangle^2$. 
	Then, $||\sqrt{M}\mathbb{E}[c(Z,\cdot)\otimes B(X)]+N_{{\mathcal{HS}}}||^2$ is absolutely continuous as the sum of two independent random variables where, at least one of them, is absolutely continuous.
\end{proof}

\begin{proof}[\bfseries\textup{Proof of Theorem 3}]

	Let us first show that $nU_n^*$, defined as in (\ref{Sm.6}), verifies $\mathbb{E}^*[|nU_n^*|] = O_{\mathbb{P}}(1)$. Let $nV_n^*$ be defined as
	\begin{align*}
		nV_n^* &= (n-1)U_n^* + \frac{1}{n}\sum\limits_{i=1}^n r_i^2 D_{ii} ||\tilde{\hat{\mathcal{E}}}_i||^2\\
		&=\frac{1}{n}\sum\limits_{i,j} r_ir_jD_{ij}\left\langle \tilde{\hat{\mathcal{E}}}_i , \tilde{\hat{\mathcal{E}}}_j \right\rangle \\
		&=\left|\left| \frac{1}{n}\sum\limits_{i=1}^n r_i \left(\frac{1}{n}\sum\limits_{k=1}^n K_h(X_i-X_k)(c(Z_i,\cdot)-c(Z_k,\cdot))\right)\otimes\tilde{\hat{\mathcal{E}}}_i  \right|\right|^2,
	\end{align*}
	where
	\begin{align*}
		D_{ij} &= \frac{1}{n^2}\sum\limits_{k,l=1}^n K_h(X_i-X_k)K_h(X_j-X_l)\left(c(Z_i,Z_j)-c(Z_i,Z_l) - c(Z_k,Z_j) + c(Z_k,Z_l)\right)\\
		&=\left\langle \frac{1}{n}\sum\limits_{k=1}^n K_h(X_i-X_k)(c(Z_i,\cdot)-c(Z_k,\cdot)),\frac{1}{n}\sum\limits_{l=1}^n K_h(X_j-X_l)(c(Z_j,\cdot)-c(Z_l,\cdot)) \right\rangle.
	\end{align*}
	By the boundedness of $c$, the zero mean, unit variance and the independence of the bootstrap multipliers, it holds
	\begin{align*}
		\mathbb{E}^*[nV^{*}_n] &= \frac{1}{n}\sum\limits_{i=1}^n \left|\left| \left(\frac{1}{n}\sum\limits_{k=1}^n K_h(X_i-X_k)(c(Z_i,\cdot)-c(Z_k,\cdot))\right)\otimes\tilde{\hat{\mathcal{E}}}_i  \right|\right|^2\\
		 &\le \frac{4}{n}\sum\limits_{i=1}^n \left| \frac{1}{n}\sum\limits_{k=1}^n |K_h(X_i-X_k)| \ ||\tilde{\hat{\mathcal{E}}}_i||  \right|^2\\
		 &= \frac{4}{n}\sum\limits_{i=1}^n \left|\left| \hat{f}'_{h}(X_i)\tilde{\hat{\mathcal{E}}}_i  \right|\right|^2\\
		  &\le \frac{8}{n}\sum\limits_{i=1}^n\left(\left|\left| \hat{f}'_{h}(X_i)(Y_i-m(X_i))  \right|\right|^2 + \left|\left| \hat{f}'_{h}(X_i)(\tilde{\hat{\mathcal{E}}}_i-(Y_i-m(X_i)))  \right|\right|^2\right).
	\end{align*}
	 On the one hand, $(\hat{f}'_{h}(X_i)-f(X_i))(Y_i-m(X_i)) = O_{L^2}(1)$ by Lemma \ref{Lm2}-(1) and $f(X_i)(Y_i-m(X_i)) = O_{L^2}(1)$ by (T1) and (T5). Thus, $\hat{f}'_{h}(X_i)(Y_i-m(X_i)) = O_{L^2}(1)$. On the other hand, by the definition of $\tilde{\hat{\mathcal{E}}}$ and $\hat{m}_{h,\tilde{h}}$,
	\begin{align*}
		||\hat{f}'_{h}(X_i)(\tilde{\hat{\mathcal{E}}}_i-(Y_i-m(X_i)))||^2  &= ||\hat{f}'_{h}(X_i)(\hat{m}_{h,\tilde{h}}(X_i) - m(X_i))||^2\\
		 &\le 2||\hat{f}'_{h}(X_i)(\hat{m}'_{h}(X_i)-m(X_i))||^2+ 2\kappa^2_n||\hat{f}_{\tilde{h}}(X_i)(\hat{m}_{\tilde{h}}(X_i)-m(X_i))||^2.
	\end{align*}
	By Lemma \ref{Lm2}-(3),
	$$\mathbb{E}\left[\kappa^2_n||\hat{f}_{\tilde{h}}(X_i)(\hat{m}_{\tilde{h}}(X_i)-m(X_i))||^2\right] = O\left(\frac{\kappa_n^2}{n\tilde{h}^d}\right)+ O\left(\kappa_n^2\tilde{h}^{2\nu}\right)=o(1).$$
	Moreover, define $K'$ as $K'(x) = |K(x)|/\int |K(x)|dx$. Is is clear that $K'$ is a second order kernel (the corresponding $\nu$ is $2$). Thus,
	$$\hat{f}'_{h}(X_i)(\hat{m}'_{h}(X_i)-m(X_i)) = \frac{1}{n}\sum\limits_{k=1}^n |K_h(X_i-X_k)| (Y_k - m(X_i)) = \frac{\int |K(x)|dx}{n}\sum\limits_{k=1}^n K'_h(X_i-X_k) (Y_k - m(X_i)),$$
	and we can apply Lemma \ref{Lm2}-(3) again to obtain
	$$\mathbb{E}\left[||\hat{f}'_{h}(X_i)(\hat{m}'_{h}(X_i)-m(X_i))||^2\right] = O\left(\frac{1}{nh^d}\right)+ O\left(h^4\right)=o(1).$$
	Therefore, $\mathbb{E}^*[nV^{*}_n] = O_{L^1}(1)$. In addition, by previous arguments,
	$$ \mathbb{E}^*\left[\frac{1}{n}\sum\limits_{i=1}^n r_i^2 D_{ii} ||\tilde{\hat{\mathcal{E}}}_i||^2\right] = \frac{1}{n}\sum\limits_{i=1}^n|D_{ii}| \ ||\tilde{\hat{\mathcal{E}}}_i||^2\le  \frac{1}{n}\sum\limits_{i=1}^n ||\hat{f}'_h(X_i) \tilde{\hat{\mathcal{E}}}_i||^2= O_{L^1}(1),$$
	which implies that $\mathbb{E}^*\left[|(n-1)U^{*}_n|\right] = O_{L^1}(1)$ and, in turn, that $\mathbb{E}^*\left[|nU_n^*|\right] = O_{\mathbb{P}}(1)$. Then, by Lemma \ref{Lm3}, $Q_n^*(1-\alpha) = O_{\mathbb{P}}(1)$, where $Q_n^*(1-\alpha)$ denotes the $(1-\alpha)$th quantile of the conditional distribution of $nU_n^{*}$. Moreover, under $\mathcal{H}_1$ we have $nU_n\xrightarrow{\mathbb{P}}\infty$ by Theorem 1 and, thus, 
	$$ \mathbb{P}(nU_n>Q_n^*(1-\alpha)) \rightarrow 1,$$
	under (fixed) alternative hypothesis.
	
	Let us now show that, under $\mathcal{H}_0$,
	\begin{align*}
		nV_n^{*} &= \left|\left| \frac{1}{\sqrt{n}}\sum\limits_{i=1}^n r_i \left(\frac{1}{n}\sum\limits_{k=1}^n K_h(X_i-X_k)(c(Z_i,\cdot)-c(Z_k,\cdot))\right)\otimes(\mathcal{E}_i+(\tilde{\hat{\mathcal{E}}}_i-\mathcal{E}_i )) \right|\right|^2\\
		&= \left|\left| \frac{1}{\sqrt{n}}\sum\limits_{i=1}^n r_i \left(\frac{1}{n}\sum\limits_{k=1}^n K_h(X_i-X_k)(c(Z_i,\cdot)-c(Z_k,\cdot))\right)\otimes\mathcal{E}_i \right|\right|^2 + o_{\mathbb{P}^*}(1)  \ \text{in }\mathbb{P}. \label{Th3.1}\tag{Th3.1}
	\end{align*}
	By virtue of the Markov inequality, it is sufficient to prove that
	$$ \mathbb{E}^*\left[\left|\left| \frac{1}{\sqrt{n}}\sum\limits_{i=1}^n r_i \left(\frac{1}{n}\sum\limits_{k=1}^n K_h(X_i-X_k)(c(Z_i,\cdot)-c(Z_k,\cdot))\right)\otimes(\tilde{\hat{\mathcal{E}}}_i-\mathcal{E}_i ) \right|\right|^2\right] = o_{\mathbb{P}}(1),$$
	and also that
	$$\mathbb{E}^*\left[\left|\left| \frac{1}{\sqrt{n}}\sum\limits_{i=1}^n r_i \left(\frac{1}{n}\sum\limits_{k=1}^n K_h(X_i-X_k)(c(Z_i,\cdot)-c(Z_k,\cdot))\right)\otimes\mathcal{E}_i  \right|\right|^2\right] = O_{\mathbb{P}}(1).$$
	By previous arguments and Lemma \ref{Lm2}-(3), taking into account that $Y_i -m(X_i) = \mathcal{E}_i$ under the null hypothesis,
	\begin{align*}
		&\mathbb{E}^*\left[\left|\left| \frac{1}{\sqrt{n}}\sum\limits_{i=1}^n r_i \left(\frac{1}{n}\sum\limits_{k=1}^n K_h(X_i-X_k)(c(Z_i,\cdot)-c(Z_k,\cdot))\right)\otimes(\tilde{\hat{\mathcal{E}}}_i-\mathcal{E}_i ) \right|\right|^2\right]\\
		=\ &\frac{1}{n}\sum\limits_{i=1}^n \left|\left| \left(\frac{1}{n}\sum\limits_{k=1}^n K_h(X_i-X_k)(c(Z_i,\cdot)-c(Z_k,\cdot))\right)\otimes(\tilde{\hat{\mathcal{E}}}_i-\mathcal{E}_i )  \right|\right|^2\\
		\le\ &\frac{4}{n}\sum\limits_{i=1}^n \left|\left| \hat{f}'_{h}(X_i)(\tilde{\hat{\mathcal{E}}}_i-\mathcal{E}_i )  \right|\right|^2\\
		=\ &\frac{4}{n}\sum\limits_{i=1}^n \left|\left|\hat{f}'_{h}(X_i)(\hat{m}_{h,\tilde{h}}(X_i) - m(X_i))  \right|\right|^2\\
		 =\ &o_{L^1}(1).
	\end{align*}
	Similarly, since $\mathcal{E} = O_{L^2}(1)$, it follows from Lemma \ref{Lm2}-(1) that
	\begin{align*}
		&\mathbb{E}^*\left[\left|\left| \frac{1}{\sqrt{n}}\sum\limits_{i=1}^n r_i \left(\frac{1}{n}\sum\limits_{k=1}^n K_h(X_i-X_k)(c(Z_i,\cdot)-c(Z_k,\cdot))\right)\otimes\mathcal{E}_i \right|\right|^2\right]	\le \frac{4}{n}\sum\limits_{i=1}^n \left|\left| \hat{f}'_{h}(X_i)\mathcal{E}_i  \right|\right|^2 = O_{L^1}(1),
	\end{align*}
	and, therefore, (\ref{Th3.1}) holds. In addition, by Lemma \ref{Lm2}-(4), 
	\begin{align}\label{Th3.2}\tag{Th3.2}
		 \frac{1}{n}\sum\limits_{i=1}^n \left|\left| \left(\frac{1}{n}\sum\limits_{k=1}^n K_h(X_i-X_k)(c(Z_i,\cdot)-c(Z_k,\cdot)) -\hat{f}_h(X_i)(c(Z_i,\cdot)-b(X_i,\cdot))\right)\otimes\mathcal{E}_i\right|\right|^2 = o_{L^1}(1).
	\end{align}
	and, moreover, by Lemma \ref{Lm2}-(1),
		\begin{align}\label{Th3.3}\tag{Th3.3}
		\frac{1}{n}\sum\limits_{i=1}^n \left|\left| \hat{f}_h(X_i)(c(Z_i,\cdot)-b(X_i,\cdot))\otimes\mathcal{E}_i\right|\right|^2 \le \frac{4}{n}\sum\limits_{i=1}^n \left|\left| \hat{f}_h(X_i)\mathcal{E}_i\right|\right|^2 = O_{L^1}(1).
		\end{align}
	Finally, by Lemma \ref{Lm2}-(2),
	\begin{align}\label{Th3.4}\tag{Th3.4}
		\frac{1}{n}\sum\limits_{i=1}^n \left|\left| (\hat{f}_h(X_i)-f(X_i))(c(Z_i,\cdot)-b(X_i,\cdot))\otimes\mathcal{E}_i\right|\right|^2 \le \frac{4}{n}\sum\limits_{i=1}^n \left|\left| (\hat{f}_h(X_i)-f(X_i))\mathcal{E}_i\right|\right|^2 = o_{L^1}(1),
	\end{align}
	and, by the boundedness of $f$, 
	\begin{align}\label{Th3.5}\tag{Th3.5}
		\frac{1}{n}\sum\limits_{i=1}^n \left|\left| f(X_i)(c(Z_i,\cdot)-b(X_i,\cdot))\otimes\mathcal{E}_i\right|\right|^2 \le \frac{4||f||^2_\infty}{n}\sum\limits_{i=1}^n \left|\left|\mathcal{E}_i\right|\right|^2 = O_{L^1}(1),
	\end{align}
	By previous arguments, it follows from (\ref{Th3.1}), (\ref{Th3.2}), (\ref{Th3.3}), (\ref{Th3.4}) and (\ref{Th3.5}) that
	$$nV_n^{*} = \left|\left| \frac{1}{\sqrt{n}}\sum\limits_{i=1}^n r_i(c(Z_i,\cdot)-b(X_i,\cdot))\otimes f(X_i)\mathcal{E}_i \right|\right|^2 + o_{\mathbb{P}^*}(1)  \ \text{in }\mathbb{P},$$
	and, analogously,
	\begin{align*}
		 \frac{1}{n}\sum\limits_{i=1}^n r_i^2 D_{ii} ||\tilde{\hat{\mathcal{E}}}_i||^2 &=  \frac{1}{n}\sum\limits_{i=1}^n r_i^2 \left|\left|  \frac{1}{n}\sum\limits_{k=1}^n K_h(X_i-X_k)(c(Z_i,\cdot)-c(Z_k,\cdot))\otimes \tilde{\hat{\mathcal{E}}}_i\right|\right|^2\\
		 &= \frac{1}{n}\sum\limits_{i=1}^n r_i^2 ||(c(Z_i,\cdot)-b(X_i,\cdot))\otimes f(X_i)\mathcal{E}_i||^2+ o_{\mathbb{P}^*}(1)  \ \text{in }\mathbb{P}.
	\end{align*}
	Therefore,
	$$ (n-1)U_n^{*} = \frac{1}{n}\sum\limits_{i\neq j} r_ir_j\left\langle (c(Z_i,\cdot)-b(X_i,\cdot))\otimes f(X_i)\mathcal{E}_i , (c(Z_j,\cdot)-b(X_j,\cdot))\otimes f(X_j)\mathcal{E}_j \right\rangle + o_{\mathbb{P}^*}(1)  \ \text{in }\mathbb{P}.$$
	Taking into account that $||f(X_i)(c(Z_i,\cdot)-b(X_i,\cdot))\otimes\mathcal{E}_i||\le 2||f||_\infty ||\mathcal{E}_i||$ and that, by (T1), $\mathbb{E}[||\mathcal{E}||^4]<\infty$, along with the separability of $\mathcal{HS}$ induced by the separability of both $\mathcal{Y}$ and the RKHS $\mathcal{C}$, which follows from the continuity of kernel $c$ and the separability of $\mathcal{Z}$ (see e.g. Theorem 7 in \citeauthor{Hein2004}, \citeyear{Hein2004}), we can apply Theorem \ref{Th5} along with Slutsky's theorem, to obtain
	$$ nU_n^{*} \xrightarrow{\mathcal{D}^*} ||N_{\mathcal{HS}}||^2 - \mathbb{E}\left[||N_{\mathcal{HS}}||^2\right] \ \text{in }\mathbb{P},$$
	where $N_{\mathcal{HS}}$ is a $\mathcal{HS}$-valued Gaussian random variable with zero mean and the same covariance operator as $(c(Z_i,\cdot)-b(X_i,\cdot))\otimes f(X_i)\mathcal{E}_i$.
	
	Finally, denote by $F$ the cumulative distribution function of $||N_{\mathcal{HS}}||^2 - \mathbb{E}\left[||N_{\mathcal{HS}}||^2\right]$. Then,
	\begin{align*}
		&\sup\limits_{x\in \mathbb{R}}|\mathbb{P}^*(nU^*_n\le x) - \mathbb{P}(nU_n\le x)|\\
		 \le\ &\sup\limits_{x\in \mathbb{R}}|\mathbb{P}^*(nU^*_n\le x) - F(x)| + \sup\limits_{x\in \mathbb{R}}|F(x) - \mathbb{P}(nU_n\le x)|,
	\end{align*}
	and, thus, the rest of the proof trivially follows from Lemmas \ref{Lm3} and \ref{Lm4}, Theorem 1-(3) and the uniform convergence of cumulative distribution functions to a continous limit. 
\end{proof}

\begin{proof}[\bfseries\textup{Proof of Theorem 4}]
	We are going to show that, under $\{\mathcal{H}_{1n}\}_{n\in \mathbb{N}}$ and independently of $\gamma>0$,
	$$\frac{1}{n}\sum\limits_{i=1}^n \left|\left| \left(\frac{1}{n}\sum\limits_{k=1}^n K_h(X_i-X_k)(c(Z_i,\cdot)-c(Z_k,\cdot))\right)\otimes(\tilde{\hat{\mathcal{E}}}_i-\mathcal{E}_i )  \right|\right|^2 = o_{L^1}(1).$$
	In such case, by the same arguments used in the proof of Theorem 3, as well as Slutsky's theorem and Theorem \ref{Th5}, we would have
	$$ nU_n^{*} \xrightarrow{\mathcal{D}^*} ||N_{\mathcal{HS}}||^2 - \mathbb{E}\left[||N_{\mathcal{HS}}||^2\right] \ \text{in }\mathbb{P},$$
	and, by Lemmas \ref{Lm3} and \ref{Lm4} and Theorem 2-(2), Theorem 4 would follow.
	
	Under $\{\mathcal{H}_{1n}\}_{n\in \mathbb{N}}$, we have $\tilde{\hat{\mathcal{E}}}_i - \mathcal{E}_i = Y_i - \hat{m}_{h,\tilde{h}}(X_i) - \mathcal{E}_i = m(X_i)- \hat{m}_{h,\tilde{h}}(X_i)  +\tilde{g}(Z_i)/n^\gamma$. Thus, by the boundedness of $c$ and the definition of $\hat{m}_{h,\tilde{h}}(X_i)$, we have
	\begin{align*}
	&\frac{1}{n}\sum\limits_{i=1}^n \left|\left| \left(\frac{1}{n}\sum\limits_{k=1}^n K_h(X_i-X_k)(c(Z_i,\cdot)-c(Z_k,\cdot))\right)\otimes(\tilde{\hat{\mathcal{E}}}_i-\mathcal{E}_i )  \right|\right|^2 \\
		 \le\ &\frac{4}{n}\sum\limits_{i=1}^n \left|\left| \hat{f}'_{h}(X_i)(\tilde{\hat{\mathcal{E}}}_i-\mathcal{E}_i )  \right|\right|^2\\
		  \le\ &\frac{8}{n}\sum\limits_{i=1}^n \left|\left| \hat{f}'_{h}(X_i)(m(X_i)-\hat{m}_{h,\tilde{h}}(X_i))  \right|\right|^2 + \frac{8}{n^{1+\gamma}}\sum\limits_{i=1}^n \left|\left| \hat{f}'_{h}(X_i)\tilde{g}(Z_i) \right|\right|^2\\
		  \le\ &\frac{16}{n}\sum\limits_{i=1}^n \left|\left|\hat{f}'_{\tilde{h}}(X_i)(\hat{m}'_{\tilde{h}}(X_i)-m(X_i))\right|\right|^2 + \frac{16\kappa^2_n}{n}\sum\limits_{i=1}^n \left|\left|\hat{f}_{\tilde{h}}(X_i)(\hat{m}_{\tilde{h}}(X_i)-m(X_i))\right|\right|^2\\
		  & + \frac{8}{n^{1+\gamma}}\sum\limits_{i=1}^n \left|\left| \hat{f}'_{h}(X_i)\tilde{g}(Z_i) \right|\right|^2\\
		 =\ &o_{L^1}(1) + o_{L^1}(1) + o\left(\frac{1}{n^\gamma}\right)O_{L^1}(1) = o_{L^1}(1). 
	\end{align*}
	where the last equality follows from Lemma \ref{Lm2}-(3) (see the proof of Theorem 3) and also from Lemma \ref{Lm2}-(1), since $\tilde{g}(Z) = O_{L^2}(1)$, for any $\gamma>0$.
\end{proof}	

\newpage

\bibliography{biblio}
\bibliographystyle{apalike}